\documentclass[preprint]{imsart}
\pdfoutput=1
\usepackage[top=2.8cm, bottom=2.8cm, left=2.8cm, right=2.8cm]{geometry}
\usepackage[utf8]{inputenc}
\usepackage{xr}

\usepackage{tikz}

\usepackage{amsthm,amsmath,amssymb, booktabs}
\RequirePackage{threeparttable}
\usepackage{parskip}
\usepackage[authoryear]{natbib}
\usepackage{hypernat}
\usepackage{marvosym}
\usepackage[hang,small,bf]{caption}
\usepackage{mathtools}
\usepackage{courier}
\usepackage{rotating}
\usepackage{hyperref}
\hypersetup{
  colorlinks,%
  citecolor=blue,%
  filecolor=black,%
  linkcolor=blue,%
  urlcolor=blue
}
\usepackage{hypernat}
\usepackage[none]{hyphenat}
\usepackage{enumerate}
\usepackage{color}
\usepackage{bbm}
\usepackage{thmtools}
\usepackage{xcolor}
\usepackage{subcaption}
\usepackage{textcomp}
\usepackage{graphicx}
\usepackage{float}
\usepackage{algpseudocode}
\usepackage{algorithm}

\RequirePackage{adjustbox}
\RequirePackage{blindtext}
\newcommand{\indep}{\rotatebox[origin=c]{90}{$\models$}}
\RequirePackage{color}  
\numberwithin{equation}{section}

\newtheoremstyle{exampstyle}
{8pt} 
{8pt} 
{\it} 
{} 
{\bfseries} 
{.} 
{.5em} 
{} 

\theoremstyle{exampstyle}

\newtheorem{theorem}{Theorem}[section]

\newtheorem{lemma}{Lemma}
\newtheorem{corollary}[theorem]{Corollary}

\startlocaldefs

\newcommand{\eat}[1]{}

\renewcommand{\bar}[1]{\overline{#1}}
\renewcommand{\hat}[1]{\widehat{#1}}
\renewcommand{\tilde}[1]{\widetilde{#1}}

\theoremstyle{plain}

\def\beq{\begin{equation}}
\def\eeq{\end{equation}}
\def\ba{\begin{enumerate}[(a)]}
\def\bei{\begin{enumerate}[(i)]}
\def\be{\begin{enumerate}[(1)]}
\def\ee{\end{enumerate}}
\def\bi{\begin{itemize}}
\def\ei{\end{itemize}}
\def\beg{\begin{eg}}
\def\eeg{\end{eg}}
\def\bd{\begin{defn}}
\def\ed{\end{defn}}
\def\bt{\begin{thm}}
\def\et{\end{thm}}
\def\bl{\begin{lemma}}
\def\el{\end{lemma}}
\def\bfac{\begin{fact}}
\def\efac{\end{fact}}

\def\bc{\begin{corollary}}
\def\ec{\end{corollary}}
\def\bp{\begin{prop}}
\def\ep{\end{prop}}
\def\bo{\begin{observe}}
\def\eo{\end{observe}}
\def\bas{\begin{assumption}}
\def\eas{\end{assumption}}

\endlocaldefs

\begin{document}

\begin{frontmatter}
\title{Robust and Efficient Bayesian Inference for Non-Probability Samples}


\runtitle{Robust and Efficient Bayesian Inference for Non-Probability Samples}

\begin{aug}
\author[A]{\fnms{Ali} \snm{Rafei}},  
\author[A]{\fnms{Michael R.} \snm{Elliott}\thanks{\textit{Corresponding author; address: 426 Thompson St. Ann Arbor, MI 48109. Rm 4068 ISR, email: \href{mailto:mrelliot@umich.edu}{mrelliot@umich.edu}.}}},
\and
\author[B]{\fnms{Carol A. C.} \snm{Flannagan}} 

\address[A]{Survey Methodology Program, University of Michigan}
\vspace{-10pt}
\address[B]{University of Michigan Transportation Research Institute}
\vspace{10pt}
\end{aug}

\begin{abstract}
The declining response rates in probability surveys along with the widespread availability of unstructured data has led to growing research into non-probability samples. Existing robust approaches are not well-developed for non-Gaussian outcomes and may perform poorly in presence of influential pseudo-weights. Furthermore, their variance estimator lacks a unified framework and rely often on asymptotic theory. To address these gaps, we propose an alternative Bayesian approach using a partially linear Gaussian process regression that utilizes a prediction model with a flexible function of the pseudo-inclusion probabilities to impute the outcome variable for the reference survey. By efficiency, we mean not only computational scalability but also superiority with respect to variance. We also show that Gaussian process regression behaves as a kernel matching technique based on the estimated propensity scores, which yields double robustness and lowers sensitivity to influential pseudo-weights. Using the simulated posterior predictive distribution, one can directly quantify the uncertainty of the proposed estimator and derive associated $95\%$ credible intervals. We assess the repeated sampling properties of our method in two simulation studies. The application of this study deals with modeling count data with varying exposures under a non-probability sample setting.
\end{abstract}


\end{frontmatter}


\section{Introduction}\label{S:1}
The declining response rates in probability surveys increasingly challenge the validity of this long-standing touchstone for finite population inference \citep{neyman1934two, groves2011three, johnson2017big, miller2017there}. According to a report by Pew Research Center, the average response rate in telephone surveys has dropped by 75\% over the past two decades \citep{keeter2017low}. Unless additional effort and cost is expended, a similar trend is expected to hold for in-person household surveys. Researchers speculate that multiple factors, including the rising response burden from a multitude of surveys with lengthy and sophisticated instruments, busier-than-ever lifestyles, and increased privacy concerns, contribute to this downward trend \citep{williams2018trends, brick2013explaining}. It is perhaps because of this issue that pollsters increasingly fail to predict the outcome of the political elections in the U.S. \citep{forsberg2020polls, vittert2020predicting}. Although the emergence of responsive and adaptive survey designs has paved novel routes to maximize the response propensity by design \citep{groves2006responsive, brick2017responsive, tourangeau2017adaptive}, these approaches may not remain efficient forever as the cost of refusal conversion continues to rise \citep{luiten2020survey}.

At the same time, use of non-probability sample ``Big Data'' is gaining popularity increasingly in various domains. These massive unstructured data are often accumulated naturally as a byproduct of human activities. Examples include but not limited to administrative and transactional records, social media, and sensor data \citep{johnson2017big}. Being cheaper and faster to collect than probability sample data has led to growing interest in using this wide range of data for producing official statistics \citep{groves2011three, berkesewicz2018overview}. However, the organic nature of their data-generating process, mainly due to self-selection, makes big-data-based inferences prone to selection bias \citep{johnson2017big, kreuter201412}. For a biased sample with an unknown selection mechanism, larger data volume even amplifies the relative contribution of bias to total error, which extremely reduces the effective sample size \citep{keiding2016perils, meng2018statistical}. This heightens the need for novel approaches that optimally calibrate such large-scale unstructured data for finite population inference.\par

As the motivating application, we are interested in estimating crash rates per distance unit driven for a subpopulation of American drivers. The current estimates are based on a ratio of the annual total police-reported crashes and annual total miles driven obtained from the General Estimates System (GES) \citep{national2014national} and the American Driving Survey (ADS), respectively \citep{tefft2017rates}. The denominator, however, can be widely subject to systematic measurement error as it relies on respondents' guess/estimate about their annual driven miles over the telephone \citep{prieger2004empirical, kim2019american}. In contrast, naturalistic driving studies (NDS) are capable of capturing these two quantities simultaneously for each participant by monitoring traffic incidents and kinematic indices continuously via a series of in-vehicle sensors and cameras \citep{guo2009modeling}. However, because of the high administrative and technical costs, participants of NDS are usually recruited via convenience samples from limited geographical areas. Therefore, naive inference based on such a non-probabilistic sample may suffer from selection bias \citep{antin2015naturalistic, rafei2021thesis}.\par

Consider a two-sample setup where a well-designed probability sample is available with a common set of auxiliary variables, also known as a ``reference survey''. Under certain assumptions, two general classes of adjustment methods can be followed: (1) \emph{quasi-randomization} (QR), where the unknown selection probabilities are estimated for units of the non-probability sample \citep{lee2006propensity, lee2009estimation, valliant2011estimating}, and (2) \emph{prediction modeling} (PM), where the analytic variable is predicted for units of the reference survey \citep{rivers2007sampling, kim2012combining, wang2015forecasting, kim2018combining}. In either case, design-based approaches can then be utilized to compute point and interval estimates. To further protect against model misspecification, \cite{chen2019doubly} reconciles the QR approach with the PM method using the idea of augmented inverse propensity weighting (AIPW) \citep{robins1994estimation}. This method is doubly robust (DR) in the sense that the estimator is consistent if either model holds \citep{scharfstein1999adjusting}.\par

Although the QR, PM, and AIPW methods all involve modeling, the ultimate form of their estimator is design-based and therefore, are subject to the general drawbacks of the design-based methods. For instance, a presence of outlying pseudo-weights may yield extremely inefficient estimates, especially if the sample size is small \citep{chen2017approaches}. Furthermore, design-based approaches lack a unified framework for quantifying all sources of uncertainty in the point estimates \citep{zangeneh2012model}. Existing methods rely on multiple assumptions about the design of the two samples and the distribution of the outcome variable that hold often asymptotically \citep{chen2019doubly, kim2018combining}. As another limitation, the AIPW method proposed by \cite{chen2019doubly} requires the sets of variables governing the selection mechanism and response surface to be identical, while this may not be the case in many situations. Furthermore, a unique solution may not necessarily exist for the joint estimating equation systems associated with the two underlying models.\par

To minimize these limitations, \cite{zheng2003penalized} propose an alternative class of inferential methods for the probability surveys with a probability proportional-to-size (PPS) design, which is called Penalized Spline of Propensity Prediction (PSPP). Unlike the previously discussed methods, PSPP is fully model-based, in which the outcome variable is predicted for the entire non-sampled units of the population. This method borrows the idea of Linear-in-Weight Prediction (LWP), in which estimated pseudo-weights are specified as a predictor in the outcome model \citep{zhang2009extensions}. \cite{bang2005doubly} show that an LWP estimator is equivalent to an AIPW estimator in terms of double robustness. In situations where auxiliary variables are missing for the non-sampled units, \cite{little2007bayesian} recommend synthesizing multiple populations via finite population Bayesian bootstrapping (FPBB). \cite{an2008robust} extend this approach to a missing data imputation scenario, where measures of size are replaced by the estimated propensity scores (PS) of being observed, and demonstrate its DR property in a simulation study.\par

As a likelihood-based method, the PSPP can be implemented under a fully Bayesian framework. This allows for direct estimation of the variance by simulating the posterior predictive distribution of the population parameters. \cite{zangeneh2015bayesian} expands the Bayesian PSPP under a PPS design for situations where the totals of the measures of size are known from external data and where there is evidence of heteroscedasticity with respect to the estimated PS. Further extensions to probability samples with unequal selection probabilities are proposed by \cite{chen2012bayesian}. The PSPP is also suitable for situations where the design of the reference survey is complex. For instance, \cite{zhou2016synthetic} develop a synthetic population approach based on a multi-stage cluster survey by undoing the sampling steps through a weighted P\'olya posterior distribution. More recently, \cite{tan2018robust} compare PSPP with AIPW to make inference for incomplete data, where PS is predicted using Bayesian Additive Regression Trees (BART) and find that the former outperforms in terms of the mean squares error.\par

While the use of a more flexible non-parametric function of the estimated PS improves when influential pseudo-weights are present \citep{zhang2011comparative}, the theoretical rationale for using a penalized spline model among a wider class of smoothers is not quite clear. \cite{saarela2016bayesian} argue that the convergence of the posterior sampling to any well-defined joint distribution of the outcome and PS may be hard to achieve. Alternatively, in this study, we propose to use Gaussian Process (GP) priors to link the PS to the outcome conditional mean \citep{si2015bayesian}. GP is a powerful non-parametric Bayesian tool for functional regression that assigns prior distributions over multidimensional non-linear functions. Because of its flexibility and generalizability, GP is gaining popularity in statistics and machine learning \citep{neal1997monte, oakley2004probabilistic, williams2006gaussian, kaufman2010bayesian, yi2011penalized, shi2011gaussian, wang2019gaussian}. 

While the correspondence between splines and GP has long been understood \citep{kimeldorf1970correspondence, seeger2000relationships}, the latter can exploit a kernel with infinite basis functions \citep{williams2006gaussian}. In this regard, GP may outdo the spline in terms of flexibility while depending on no arbitrary tuning parameters. In a regular spline regression, one has to determine the polynomial order as well as the frequency and location of the knots empirically. More importantly, \cite{huang2019gpmatch} demonstrate that a stationary isotropic covariance matrix in GP behaves as a non-parametric matching technique using the estimated PS as a measure of similarity. In a more \emph{ad hoc} manner, \cite{rivers2007sampling} suggests matching units of a web non-probability survey to those from a parallel reference survey. Very recently, a kernel weighting approach has been proposed by \cite{wang2020improving}, where the weighted estimator is proved to be consistent under a weak exchangeability condition. To further weaken the modeling assumptions, \cite{kern2020boosted} propose to use algorithmic tree-based methods, including random forests and gradient tree boosting, for estimating the PS in kernel weighting.\par

To eliminate the need to account for the sampling weights in the PS model, one possible solution is to generate synthetic populations using the weights in the reference survey \citep{dong2014nonparametric, zangeneh2015bayesian, an2008robust}. However, fitting Bayesian joint models on a synthetic population can be computationally demanding if not impossible \citep{mercer2018selection}. In addition, one has to rely on Rubin's combining rules to derive the final point and interval estimates. Therefore, direct simulation of the posterior predictive distribution is not possible for the unknown population quantity in this approach. To augment the prediction model in the GPPP estimator while avoiding this technique, we propose an alternative approach that is inspired by \cite{si2015bayesian}. In this method, we limit the prediction of outcome to the reference survey, instead of the whole non-sampled units of the population. Having the sampling weights and other design features known in the reference survey, we utilize the posterior predictive draws of the outcome to derive the final point and variance estimates based on Rubin's combining rules \citep{rubin2004multiple}.\par

The rest of the article is organized as follows: In Section~\ref{S:2}, we start by describing the proposed method through rigorous mathematical notations. Section~\ref{S:3} assesses the repeated sampling properties of the proposed method and compares its performance with the AIPW estimator through a simulation study. Section~\ref{S:4} involves application on the sensor-based data from Strategic Highway Research Program 2 (SHRP2) where we deal with modeling count data with varying exposures. All the statistical analyses have been performed using R/Stan with annotated codes accessible publicly at \href{https://github.com/arafei/GPPP}{https://github.com/arafei/GPPP}. Finally, Section~\ref{S:5} reviews the strengths and weaknesses of the study in more detail and suggests some future research directions. Supplemental information, including proofs, additional theory, and preliminary descriptive results, are provided in Appendix~\ref{S:6}.\par


\section{Methods}\label{S:2}
\subsection{Notation}\label{S:2.1}
Consider $U$ to be a finite population of unknown size $N$. For each $i=1, ..., N$, we denote $y_i$ to be the realized values of a scalar outcome variable, $Y$, in $U$, and $x_i=[x_{i1}, x_{i2}, ... , x_{ip}]^T$ the values of a $p$-dimensional set of relevant auxiliary variables, $X$. Let $S_A$ be a non-probability sample selected from $U$ with $y_i$ being observed, and $n_A$ being the sample size. The main objective of descriptive inference is to learn about an unknown finite population quantity that is a function of $Y$, e.g. the population mean, $Q(y)= \bar y_U=\sum_{i=1}^Ny_i/N$. Suppose $\delta^A_{i}=I(i\in S_A)$ represents the inclusion indicator variable of $S_A$ for $i\in U$, and $x_i$ is the vector of variables associated with the selection mechanism of $S_A$. The unknown $\pi^A_i$ for $i\in S_A$ is the biggest challenge to the analyst, and unbiased inference about $Q(y)$ cannot be drawn without imposing a set of strong conditions. Therefore, we consider the following assumptions:
\begin{enumerate}
  \item[\textbf{C1.}] \textbf{Positivity}---The non-probability sample $S_A$ actually does have a probabilistic sampling mechanism, albeit unknown. That means $p(\delta^A_i=1|x_i)>0$ for all possible values of $x_i$ in $U$.
  \item[\textbf{C2.}] \textbf{Ignorability}---the selection mechanism of $S_A$ is fully governed by $x$, which implies that $Y\indep\delta^A|X$. Then, for $i\in U$, the pseudo-inclusion probability associated with $S_A$ is defined as $\pi^A_i=p(\delta^A_i=1|x_i)$.
  \item[\textbf{C3.}] \textbf{Independence}---units in $S_A$ are selected independently given $x$, i.e. $\delta^A_i\indep\delta^A_j|x_i,x_j$ for $i\neq j\in U$. Note that this assumption is made to avoid unnecessary complications; otherwise extensions can be given for situations where $S_A$ is a clustered sample.
\end{enumerate}

Now, suppose $S_R$ is a parallel ``reference survey'' of size $n_R$, for which the same set of covariates, $X$, has been measured. Note that in a non-probability sample setting, $y_i$ has to be unobserved for $i\in S_R$; otherwise, inference could be directly drawn based on $S_R$. Also, let $\delta^R_i=I(i\in S_R)$ denotes the inclusion indicator variable associated with $S_R$ for $i\in U$. Units of $S_R$ may be selected independently or through a stratified multistage clustered sampling design. Being a full probability sample implies that the selection mechanism in $S_R$ is ignorable given its design features, i.e. $p(\delta^R_i| y_i, d_i)=p(\delta^R_i|d_i)$ for $i\in U$, where $d_i$ denotes a set of associated design variables. Thus, one can define the inclusion probabilities in $S_R$ as $\pi^R_i=p(\delta^R_i=1|d_i)$ for $i\in U$. This study deals with reference surveys whose sampling design involves an independent selection of population units but with unequal inclusion probabilities. A common example of such a design is PPS, under which $d$ is regarded as a scalar variable representing measures of size in $U$, and $\pi^R_i\propto n^R\bar d/N\bar D$ according to \cite{zangeneh2015bayesian}. 

Normally, probability surveys come with a set of sampling weights in their public-use data which are inversely proportional to the selection probabilities, i.e. $w^R_i\propto 1/\pi^R_i$. While $w_i^R$'s may comprise of post-survey adjustments for non-response and non-coverage errors \citep{valliant2013practical}, not all the auxiliary information used for the construction of weights are necessarily supplied to the analyst. In addition, public-use survey data may lack a detailed guideline on how the sampling weights are calculated. To simplify the problem under such situations, \cite{si2015bayesian} assume that weights with identical values represent a unique post-stratum in the population. Therefore, one can identify $d$ as the indicator of $J$ unique post-strata in $U$, and consider $w^R_j\propto N_j/n^R_j$ $(j=1, 2, ... , J)$, where $N_j$ and $n^R_j$ are the $j$-th post-stratum size in $U$ and $S_R$, respectively. For instance, in RDD telephone surveys or mail surveys, whose design involves equiprobability sampling, the inequality in weights may arise exclusively from non-response adjustment or post-stratification.\par


Now, we combine the two samples and define $S_C=S_A\cup S_R$ with $n_C=n_A+n_R$ being the total sample size. While $X$ and $D$ may overlap or correlate, in addition to the aforementioned conditions, we assume that, conditional on $[X, D]$, $S_R$ and $S_A$ are selected independently, i.e. $\delta^A\indep\delta^R|X,D$. We denote this condition as \textbf{C4}. Considering \textbf{C1-C4}, the joint density of $y_i$, $\delta^A_i$ and $\delta^R_i$ can be factorized as below:
\begin{equation}\label{eq:2.1}
p(y_i, \delta^A_i, \delta^R_i| x_i, d_i; \theta, \beta)=p(y_i|x_i, d_i; \theta)p(\delta^A_i|x_i; \beta)p(\delta^R_i|d_i), \hspace{4mm} \forall i\in U
\end{equation}
where $\eta=(\theta, \beta)$ are some unknown parameters indexing the conditional distribution of $Y|X, D$ and $\delta^A|X$, respectively. The conditional density $p(y_i|x_i; \theta)$ denotes the underlying model that governs the response surface structure of a superpopulation from which $U$ has been selected. Also, $p(\delta^A_i|x_i;\beta)$ and $p(\delta^R_i|d_i)$ denote the randomization distributions associated with the design of $S_A$ and $S_R$, respectively. Note that the latter does not depend on any unknown parameter as $S_R$ is a probability sample with a known sampling design. A QR approach involves modeling $p(\delta^A_i|x_i; \beta)$, whereas a PM approach deals with modeling $p(y_i|x_i, d_i; \theta)$. In the following subsection, we describe a fully model-based strategy for Bayesian inference based on non-probability samples by jointly modeling the PS and the outcome variable.\par 





\subsection{Bayesian model-based inference}\label{S:2.2}
As discussed in the introduction, under a fully model-based approach, the inference paradigm is viewed as imputing the unobserved outcome, $Y$, for the non-sampled units of the population with respect to $S_A$, i.e. $\bar{S_A}=U-S_A$. Apparently, one can directly estimate the population unknown quantity as soon as $y_i$ is known for $i\in U$. For the population mean, such an estimator, also known as a \emph{prediction} estimator, is given by
\begin{equation}\label{eq:2.2}
\begin{aligned}
\hat{\bar y}_U &= \left(\sum_{i\in S_A}y_i + \sum_{i\in \bar S_A}\hat y_i\right)/N\\
&= \left(\sum_{i\in S_A}\left(y_i-\hat y_i\right) + \hat y_U\right)/N
\end{aligned}
\end{equation}
where $\hat y_i$ is the prediction of $y_i$ for $i\in U$, and $\hat y_U=\sum_{i\in U}\hat y_i$. The last expression in Eq.~\ref{eq:2.2} is also known as a ``generalized difference estimator'' \citep{wu2001model}.\par

A fully Bayesian approach specifies a model to the joint distribution of $(y_i, \delta^A_i)$ across the units of $U$, which can be formulated by
\begin{equation}\label{eq:2.3}
p(y_i, \delta^A_i | x_i, d_i; \theta, \beta)=p(y_i|x_i, d_i, \delta^A_i; \theta)p(\delta^A_i|x_i; \beta), \hspace{4mm} i\in U
\end{equation}
For brevity, to show that a variable, say $x$, is indexed across units of $U$, $S_A$, $S_R$, or $S_C$, we denote them by $x_U$, $x_A$, $x_R$, or $x_C$, respectively. Then, the likelihood function for $(\theta, \beta)$ based on the observed data is given by
\begin{equation}\label{eq:2.4}
L(\beta,\theta|y_A, \delta_U^A, x_U, d_U)\propto p(y_A,\delta_U^A|x_U, d_U,\theta,\beta)=\int p(y_U,\delta_U^A|x_U, d_U,\theta,\beta)dy_U
\end{equation}
Under a Bayesian approach, model parameters are assigned prior distributions $p(\theta,\beta|x_U, d_U)$, and analytical inference is drawn based on the posterior distribution as below:
\begin{equation}\label{eq:2.5}
p(\beta,\theta|y_A, \delta_U^A, x_U, d_U)\propto p(\theta,\beta|x_U, d_U)L(\beta,\theta|y_A, \delta_U^A, x_U, d_U)
\end{equation}
Note that in a Bayesian setting, to preserve the ignorable assumption, \textbf{C2}, in $S_A$, it is essential to specify independent priors, i.e. $p(\theta,\beta|x_U, d_U)=p(\theta|x_U, d_U)p(\beta|x_U, d_U)$ \citep{little2007bayesian}.

Descriptive inference about $\bar y_U$ requires deriving the posterior predictive distribution conditional on the observed data, which is given by
\begin{equation}\label{eq:2.6}
p(\bar y_U|y_A, \delta^A_U, x_U, d_U)=\int \int p(\bar y_U|y_A, \delta^A_U, x_U, d_U, \theta, 
\beta)p(\theta,\beta|y_A, \delta^A_U, x_U, d_U)d\theta d\beta
\end{equation}
We will further expand this formula in the following subsections and show how one can jointly estimate $(y_U, \pi^A_U)$ in deriving the posterior predictive distribution of $\bar y_U$.\par

Obviously, estimating $\hat y_U$ in Eq.~\ref{eq:2.2} requires $(X, D)$ to be observed for the entire population units, while it is not in practice, and most often, the measurement of auxiliary information is confined to the pooled sample, $S_C$. One way to tackle this issue is to generate a finite set of synthetic populations, say $M$, by undoing the sampling mechanism in $S_R$, which can be performed non-parametrically through the idea of finite population Bayesian bootstrapping (FPBB) \citep{little2007bayesian, dong2014nonparametric}. Given a synthetic population, attempts are then made to imputed the outcome variable for the entire non-sampled units. \par

However, considering the limited resources of high-performance computing, it is computationally expensive, if not infeasible, to fit Bayesian joint models repeatedly on large synthetic populations and simulate the posterior predictive distribution for all population units based on a custom posterior sampler \citep{mercer2018selection, savitsky2016bayesian}. In addition, the two-step algorithm proposed by \cite{zangeneh2015bayesian} may not be fully implementable on the existing Bayesian platforms such as Stan \citep{carpenter2017stan}, and therefore, the authors propose to combine the estimates across synthetic populations through Rubin's combining rules \citep{rubin1976inference}. This may not be ideal especially when the posterior predictive distribution of the target population quantity tends to be heavily skewed and consequently, a symmetric confidence interval may fail to properly approximate the direct credible intervals of the posterior predictive distribution.\par

\subsection{Proposed computationally tractable method}\label{S:2.3}
To reduce the computational burden and to be able to directly simulate the posterior predictive distribution of $\bar y_U$ via a unified algorithm that is implementable in Stan, we limit the imputation of the outcome, $y_i$, to units of the combined survey, i.e. $i\in S_C$, and use the following estimator, as defined by \cite{si2015bayesian}, to multiply impute $\hat y_U$ in Eq.~\ref{eq:2.2} $M$ times as below:
\begin{equation}\label{eq:2.7}
\begin{aligned}
\hat y^{(m)}_U &= \sum_{j=1}^{J}N^{(m)}_j\hat{\bar y}^{(m)}_j\\
&=\sum_{j=1}^{J}\sum_{i=1}^{n_{j}}\frac{N^{(m)}_{j[i]}}{n^R_{j[i]}} \hat y^{(m)}_{j[i]}
\end{aligned}
\end{equation}
where $\left[N_{j[i]}^{(m)}, \hat y^{(m)}_{j[i]}\right]$ is the $m$-th draw of the joint posterior predictive distribution of the $j$-th post-stratum size and $i$-th outcome within $j$-th post-stratum. Therefore, the $m$-th posterior predictive draw of $\bar y_U$ is given by 
\begin{equation}\label{eq:2.8}
\hat{\bar y}^{(m)}_U =\left(\sum_{i=1}^{n_A}\left(y_i - \hat y^{(m)}_i \right) + \hat y^{(m)}_U\right)/N
\end{equation}
To this end, we are interested in modeling the joint distribution of $(y_i, \delta^A_i, n^R_i)$ for $i\in S_C$ as below:
\begin{equation}\label{eq:2.9}
p(y_A, \delta^A_C, w_R, n_R|x_C, d_C, \theta, \beta, \xi)=p(n_R|w_R, \xi)p(y_A, \delta^A_C, w_R|x_C, d_C, \theta, \beta)
\end{equation}
where $n_R=[n^R_1, n^R_2, ..., n^R_J]^T$ is the sizes of post-strata in $S_R$, and $\xi$ is a $J$-dimensional vector of parameters associated with non-parametric modeling of $n_R|w_R$.
While we thoroughly discuss each components of the rightmost expression of Eq.~\ref{eq:2.9} later, one can derive the final estimate of $\bar y_U$ by
\begin{equation}\label{eq:2.10}
\hat{\bar y}_U = \frac{1}{M}\sum_{m=1}^M\hat{\bar y}_U^{(m)}
\end{equation}
and the associated $100(1-\alpha)\%$ credible interval can be constructed by sorting $(\hat{\bar y}^{(1)}_U, \hat{\bar y}^{(2)}_U, ... , \hat{\bar y}^{(M)}_U)$ ascendingly, and finding the $\alpha/2$ and $1-\alpha/2$ percentiles of this ordered sequence that correspond to lower and upper limits of the credible interval, respectively.

\subsubsection{Finite population Bayesian bootstrapping for modeling $p(n_R|w_R, \xi)$}\label{S:2.3.1}
We begin by modeling $p(n^R|w^R, \xi^R)$ non-parametrically via Bayesian bootstrapping (BB), with the aim to simulate the posterior predictive distribution of the $N_j$'s. The idea of BB operates quite similar to the regular bootstrap approach \citep{efron1981nonparametric}, except for the fact that BB simulates the posterior predictive distribution of a given population parameter instead of the sample distribution of the statistic estimating that parameter \citep{rubin1981bayesian}. In a finite population Bayesian bootstrap (FPBB) setting, the goal is to derive the posterior predictive distribution of the post-strata sizes for the non-sampled population units, i.e. $\bar S_R$. Although FPBB imposes no parametric assumptions, it is assumed that all the existing post-strata in $U$ are limited to those observed in the collected sample (exchangeability).\par 

Under a simple random sample, \cite{ghosh1983estimation} propose to use a Pol\'ya Urn Scheme, in which a Dirichlet-multinomial conjugate model is considered to expand the sample to the population. \cite{cohen1997bayesian} generalizes this approach to a weighted sample with independent draws, and the attributed Pol\'ya posterior distribution for the non-sampled units of $U$ given the observed sampling weights is formulated by \cite{dong2014nonparametric}. \cite{little2007bayesian} propose a modified FPBB method to generate synthetic populations based on the samples with a PPS design. Further extension based on a constrained BB is provided by \cite{zangeneh2015bayesian} for situations where totals are known for auxiliary variables at the population level. \par


In the present article, we modify the FPBB method proposed by \cite{little2007bayesian} by letting $v^R=\{v^R_1, v^R_2, ... , v^R_J\}$ represent the set of $J$ distinct values of the sampling weights in $S_R$, and $\xi^R=\{\xi^R_1, \xi^R_2, ... , \xi^R_J\}$ denote the vector of conditional probabilities that $p(w^R=v^R_j|\delta^R=1)=\xi^R_j$ for $j=1, 2, ... , J$, where $\sum_{j=1}^J\xi^R_j=1$. Now, suppose $n^R_j$ and $r^R_j$ are the frequencies of $w^R$ taking the value $v^R_j$ in $S_R$ and $\bar S_R$, respectively, for $j=1, 2, ... , J$. It is clear that $\sum_{j=1}^Kn^R_j=n_R$, and $\sum_{j=1}^Kr^R_j=N-n_R$. Considering a Dirichlet prior, i.e. $\xi^R\sim Dirichlet(\alpha_{J\times 1})$, $\alpha\in {\rm I\!R}^{J>0}$, with a multinomial likelihood function of $p(n^R_1, n^R_2, ... , n^R_J|\xi)\propto\prod_{j=1}^J(\xi^R_j)^{n^R_j}$, the posterior distribution of $\xi^R$ is given by $(\xi^R|n^R_1, n^R_2, ... , n^R_J)\sim Dirichlet(n^R_1+\alpha_1-1, n^R_2+\alpha_2-1, ... , n^R_J+\alpha_J-1)$. Using Bayes' rule, \cite{little2007bayesian} show that
\begin{equation}\label{eq:2.11}
\begin{aligned}
\xi_j^{\bar R}&=p(w^R_i=v^R_j|\delta_i^R=0)\\
&=p(\delta_i^R=0|w_i^R=v^R_j)\frac{p(w_i^R=v^R_j)}{p(\delta_i^R=0)}\\
&=p(\delta_i^R=0|w_i^R=v^R_j)\frac{p(w_i^R=v^R_j|\delta_i^R=0)p(\delta_i^R=0)+p(w_i^R=v^R_j|\delta_i^R=1)p(\delta_i^R=1)}{p(\delta_i^R=0)}\\
&=p(\delta_i^R=0|w_i^R=v^R_j)\bigg\{\xi^{\bar R} + \xi^R\frac{p(\delta_i^R=1)}{p(\delta_i^R=0)}\bigg\}
\end{aligned}  
\end{equation}
Since $p(\delta_i^R=0|w_i^R=v^R_j)=1-\pi^R_j$, and $p(\delta_i^R=1)/p(\delta_i^R=0)$ can be treated as a normalizing constant,
\begin{equation}\label{eq:2.12}
\xi_j^{\bar R}\propto\xi_j^R\frac{1-\pi^R_j}{\pi^R_j}
\end{equation}
After normalizing $\xi^{\bar R}$ such that $\sum_{j=1}^J \xi_j^{\bar R}=1$, the posterior predictive distribution of $r^R$ is given by
\begin{equation}\label{eq:2.13}
p(r^R_1, r^R_2, ... , r^R_J|n^R_1, n^R_2, ... , n^R_J, \xi^R)={N-n_R\choose r_1, r_2, ... , r_J} \prod_{j=1}^J\left[c\xi_j^R(1-\pi^R_j)/\pi^R_j\right]^{r^R_j}
\end{equation}
where $c$ is the normalizing constant. The $m$-th posterior predictive draw of the size of post-stratum $j$ in the population is $N_j^{(m)}=n^R_j+r_j^{R(m)}$, $(m=1, 2, ... , M)$.\par

\subsubsection{Modeling the joint distribution of $(y_i, \delta_i^A)$ given the combined sample}\label{S:2.3.2}
As discussed earlier, the goal of PM in this study is to model $p(y_i|x_i; \theta)$ in order to obtain the posterior predictive distribution of $y_i$ for $i\in S_R$, i.e. $p(y_R|y_A, x_C)\propto\int p(y_R|y_A, x_C; \theta)p(\theta|y_A, x_C)d\theta$. Although $\theta$ is a parameter defined in $U$, the ignorable assumption guarantees a consistent estimate of $\theta$ by fitting $p(y|x; \theta)$ on $S_A$, because
\begin{equation}\label{eq:2.14}
\begin{aligned}
p(y_A|x_A;\theta)&=p(y_U|\delta^A_U=1, x_U, d_U; \theta)\\
&=\frac{p(\delta^A_U=1|y_U, x_U; \theta)}{p(\delta^A_U=1|x_U; \theta)}p(y_U|x_U, d_U; \theta)\\
&=p(y_U|x_U, d_U; \theta)
\end{aligned}
\end{equation}
If $\pi^A_i$ was known for $i\in S_C$, one could augment the PM by incorporating $\pi^A_i$ as a predictor into the PM, e.g. $p(y_i|x_i, f(\pi^A_i); \theta)$. A robust estimator is achieved by choosing a flexible $f(.)$, as detailed later.\par


While a non-probability sample is characterized by its unknown selection mechanism, given the conditions \textbf{C1-C4}, $\pi^A_i$ can be estimated by modeling $p(\delta^A_U|x_U;\beta)$. Assuming that $S_A$ is selected by a Poisson sampling, one can formulate the likelihood of $\beta$ given $\delta^A_U$ as:
\begin{equation}\label{eq:2.15}
L(\beta|\delta_U^A, x_U)= \prod_{i=1}^N p(\delta^A_i=1|x_i, \beta)^{\delta^A_i}\left[1-p(\delta^A_i=1|x_i, \beta)\right]^{1-\delta^A_i}
\end{equation}
Under a logistic regression model,
\begin{equation}\label{eq:2.16}
\pi^A_i=p(\delta^A_i=1|x_i;\beta)=\frac{exp\{x^T_i\beta\}}{1+exp\{x^T_i\beta\}}
\end{equation}
By assigning appropriate prior distributions to $\beta$, one can simulate the posterior distribution of $\pi^A_i$ for $i\in U$ through the Hamiltonian Monte Carlo (HMC) algorithm. 

One major issue with Eq.~\ref{eq:2.15} is that the observed $(\delta^A_i, x_i)$ is restricted to $S_C$. Although there exist several approaches restricting the estimation of $\beta$ to $S_C$ \citep{Valliant2018, Elliott2017Inference, chen2019doubly, wang2020adjusted}, the majority rely on a pseudo-maximum likelihood estimation (PMLE) idea to account for unequal $w^R_i$'s, which necessitates solving a set of estimating equations. A corresponding method in a Bayesian setting is called pseudo-Bayesian. While such a method guarantees consistency in point estimates, the uncertainty tends to be underestimated in the posterior distribution of parameters \citep{savitsky2016bayesian, gunawan2020bayesian, williams2021uncertainty}. To avoid this problem, we employ a two-step pseudo-weighting approach proposed by \cite{Elliott2017Inference}. Assuming that $p(\delta_i^A+\delta_i^R=2)\approx 0$, i.e. $S_A$ and $S_R$ have no overlap, one can show that
\begin{equation}\label{eq:2.17}
  p(\delta^A_U=1|x_U; \beta)= p(\delta^R_U=1|x_U;\gamma)\frac{p(\delta^A_C=1|x_C;\phi)}{1-p(\delta^A_C=1|x_C;\phi)}
\end{equation}
where $\beta=(\gamma,\phi)^T$ is the associated model parameters. \cite{rafei2020big} call this approach propensity-adjusted probability prediction (PAPP) and prove the asymptotic properties of a pseudo-weighted estimate based on this method including consistency and variance estimation. As can be seen, this approach reduces the modeling of $p(\delta^A_U=1|x_U)$ to the modeling of $p(\delta^A_C=1|x_C)$ with an additional step, which is modeling $p(\delta^R_U|x_U)$. Treating $\pi^R_i$ as a random variable for $i\in S_A$ conditional on $x_i$, one can estimate this probability by regressing the $\pi^R_i$'s on the $x_i$'s in $U$ \citep{pfeffermann2009inference}, because
\begin{equation}\label{eq:2.18}
\begin{aligned}
p(\delta^R_U=1|x_U;\gamma) &=\int_0^1 p(\delta^R_U=1| \pi^R_U, x_U;\gamma) p(\pi^R_U|x_U;\gamma)d\pi^R_U\\
&=\int_0^1\pi^R_Up(\pi^R_U|x_U;\gamma)d\pi^R_U\\
&=E(\pi^R_U|x_U;\gamma)
\end{aligned}
\end{equation}
\cite{pfeffermann1999parametric} demonstrate that $E(\pi^R_U|x_U)=E^{-1}(w_R|x_R)$ where $w_R$ are the sampling weights in $S_R$. Since $\pi^R_i$ is only observed in $S_R$, then, the sample estimator of $\pi^R_i$ is given by
\begin{equation}\label{eq:2.19}
  p(\delta^A_C=1|x_C; \gamma, \phi)= E^{-1}(w_R|x_C;\gamma)\frac{p(\delta^A_C=1|x_C;\phi)}{1-p(\delta^A_C=1|x_C;\phi)}
\end{equation}
$E(w_R|x)$ is modeled using a GLM with a $log$ link function, as the distribution of the $w^R_i$'s tends to be right-skewed in the actual survey data. In addition, we know that the sampling weights are usually a multiplicative factor of selection probabilities$\times$non-response adjustment$\times$post-stratification. Therefore, given the posterior distribution of $p(\gamma,\beta|x_C, w_R)$, one can obtain the posterior distribution of $\pi^A_i$ for $i\in S_C$ by 
\begin{equation}\label{eq:2.20}
  p(\delta^A_C=1|x_C; \gamma, \phi)= exp\big\{x^T_C(\phi-\gamma)\big\}
\end{equation}
Note that modeling $w^R_i$ is not required if $w^R_i$ is known for $i\in S_A$ \citep{rafei2021robust}. The joint distribution of $(y_A, \delta^A_C, w_R)$ can be written as:
\begin{equation}\label{eq:2.21}
  p(y_A, \delta^A_C, w_R|x_C)=\int p(y_R, y_A|f(\pi^A[x_C, \delta^A_C,w_R;\gamma,\phi]), x_C;\theta)p(\delta^A_C|x_C;\phi) p(w_R|x_R;\gamma) dy_R
\end{equation}
where $\pi^A[x_C, \delta^A_C,w_R;\gamma,\phi]=exp\big\{x^T_C(\phi-\gamma)\big\}$ according to Eq.~\ref{eq:2.20}. The corresponding posterior predictive distribution of $y_R$ is given by
\begin{equation}\label{eq:2.22}
\begin{aligned}
  p(y_R|y_A, \delta^A_C, w_R, x_C, \delta^A_C,\pi^R_R) &=\int\int\int p(y_R|y_A, f(\pi^A[x_C, \delta^A_C,\pi^R_R;\gamma,\phi]), x_C;\theta)\\
  &\times p(\phi|\delta^A_C,x_C) p(\gamma|w_R, x_R) d\theta d\phi d\gamma
\end{aligned}
\end{equation}
Although \cite{zigler2016central} argues that such a factorization of the joint distribution of $(y_i, \delta^A_i, w^R)$ does not correspond to a valid use of the Bayes' theorem, for certain reasons, it has been advocated by several studies. First, \cite{little2004model} highlights the fact that Bayesian joint modeling can result in better repeated sampling properties. It has been well-understood that the performance of the alternative two-step Bayesian methods with respect to frequentist properties depends on the choice of priors \citep{kaplan2012two}. Furthermore, having both $\pi^A_i$ and $x_i$ as predictors in the PM cuts the notorious feedback between the QR and PM models, which leads to incorrect estimation of the PS posterior distribution \citep{zigler2013model}.\par

However, what matters most in this study is the double robustness property that the likelihood factorization in Eq.~\ref{eq:2.21} offers. For instance, by choosing a parametric form $f(\pi^A_i)=\theta^*/\pi^A_i$, where $\theta^*$ is an unknown scalar parameter, this factorization leads to a linear-in-weight Prediction (LWP) model. \cite{scharfstein1999adjusting} and \cite{bang2005doubly} identified the correspondence between LWP and AIPW estimators. In the causal inference context, this has been termed a clever covariate by \cite{rose2008simple} as it characterizes the correct relationship between the propensity scores and the outcome model. In the context of item-missing data imputation, \cite{little2004robust} suggest that the use of a more flexible non-parametric function can improve the efficiency of the adjusted estimator, especially when there are extreme values in the estimated PS. The authors propose to use a penalized spline model, which is piecewise continuous polynomials of the estimated PS, paired with a mathematical penalization to find the best fit of PM to the data \citep{ruppert2003semiparametric, fahrmeir2011bayesian}. Alternatively, \cite{mccandless2009bayesian} suggest categorizing propensity scores into quantiles and using them as dummy variables to augment the PM.\par

In the current study, we extend the PSPP idea to a non-probability sample setting while using Gaussian process (GP) regression instead of a penalized spline model. As a flexible non-parametric Bayesian approach, GP can automatically capture non-linear associations as well as multi-way interactions \citep{rusmassen2005gaussian, neal1997monte}. Having $\pi^A_i=p(\delta^A_i=1|x_i,w^R; \gamma, \phi)$ estimated for $i\in S_C$, for a continuous outcome variable, we fit a semiparametric model on $S_A$ as below: 
\begin{equation}\label{eq:2.23}
y_i|x_i, d_i, \hat\pi_i, \theta = \theta_0+\sum_{j=1}^p\theta_jx_{ij}+\sum_{j=p+1}^{p+q}\theta_jd_{ij} +f\left(\hat\pi^A_i\right)+\epsilon_i
\end{equation}
where $\theta$ denotes a $(p+q+1)$-dimensional vector of the PM parameters, and $\epsilon_i \sim N(0, \sigma^2)$ with $\sigma^2$ being unknown. Eq.~\ref{eq:2.23} involves two parts: a linear regression parameterized by $\theta$ and a GP denoted by $f(.)$.\par

A GP $\{f(u):u\in R^N\}$ is a set of random variables, any finite number of which jointly follow a multivariate Gaussian distribution. In a full-ranked GP, $f(.)$ is \textit{a priori} defined by its mean and covariance functions as below:
\begin{equation}\label{eq:2.24}
f(u)\sim GP\left(\mu(u), K\left(u, u'\right)\right)
\end{equation}
where $\mu(u)$ is the mean vector and $K(u, u')$ is the covariance matrix. The latter encompasses all our prior beliefs about the functional association between $x$ and $y$, including continuity, smoothness, periodicity and scale properties \citep{riutort2020practical}. For notational simplicity, we set $\mu(u)=0$, though it is not necessary. It is worth noting that the LWP model can be viewed as a specific type of GP with a dot product covariance matrix as $\alpha^2[1+((\pi^A_i)^T\pi^A_j)^{-1}]$ if the regression coefficient is specified a prior of $N(0, \alpha^2)$ \citep{rusmassen2005gaussian}. 
While literature suggests a variety of covariance functions for GP, the most common type is the \textit{squared exponential} (SE) covariance matrix whose elements take the following form:
\begin{equation}\label{eq:2.25}
k(u,u')=\alpha^2exp\bigg\{-\frac{||u - u'||^2}{2\rho}\bigg\}
\end{equation}
where $\rho$ is called a \emph{length-scale} parameter, and $\alpha$ is known as the \emph{marginal standard error}. One can show that the SE covariance structure represents a kernel with an infinite number of basis functions \citep{rusmassen2005gaussian}.\par

From a weight-space viewpoint, \cite{huang2019gpmatch} show that with a stationary isotropic kernel, where $K(\pi^A_i, \pi^A_j)=f(||\pi^A_i-\pi^A_j||)$, GP acts as a non-parametric matching technique. \cite{wang2020improving} prove the \emph{consistency} of a kernel-weighted estimator under certain regularity conditions. Refer to Appendix~\ref{S:6.1} to see the connection between GP and kernel weighting. In our non-probability sample setting, one can view it as matching units of $S_A$ to units of $S_R$ based on the estimated propensity scores, $\pi^A_i$'s \citep{rivers2007sampling}. Further theoretical properties of kernel optimal matching, such as consistency, can be found in~\cite{kallus2018more}. Although the SE covariance has desirable properties, empirical results show that it is not a strong fit for the real-world data as it is infinitely differentiable \citep{rusmassen2005gaussian}. Therefore, 
we propose to use a Mat\'ern kernel added to an inhomogeneous standardized polynomial kernel of order $p$ as below:
\begin{equation}\label{eq:2.26}
\begin{aligned}
K(u_i,u_j)=\alpha^2\frac{2^{1-\nu}}{\Gamma(\nu)}\left(\sqrt{2\nu}\frac{||u_i-u_j||}{\rho}\right)^\nu &K_\nu \left(\sqrt{2\nu}\frac{||u_i-u_j||}{\rho}\right)\\
&+\left(\frac{\tau^2+u_i^Tu_j}{\sqrt{\tau^2+u_i^Tu_i}\sqrt{\tau^2+u_j^Tu_j}}\right)^p
\end{aligned}
\end{equation}
where $\Gamma(.)$ denotes the gamma function, and $K_\nu(.)$ is a modified Bessel function of the second kind. This combination of two kernels ensures capturing both local variations and long-range discrepancies in the estimated propensity scores \citep{vegetabile2018methods}. Note that for $\nu\rightarrow \infty$, Mat\'ern covariance will converge to the SE covariance, and the sum of two valid kernels is still a valid kernel.\par

In this chapter, we set $\nu=3/2$ and $p=1$ throughout the simulation and empirical studies, which yields the following covariance function:
\begin{equation}\label{eq:2.27}
K(u_i,u_j)=\alpha^2\left(1+\frac{\sqrt{3}||u_i-u_j||}{\rho}\right) exp\left(-\frac{\sqrt{3}||u_i-u_j||}{\rho}\right)+\frac{\tau^2+u_iu_j}{\sqrt{\tau^2+u^2_i}\sqrt{\tau^2+u_j^2}}
\end{equation}
In addition, we propose to use a $log$ transformation of the $\hat\pi^A_i$'s as GP input, i.e. $u_i=log(\hat\pi^A_i)$. This is because the input of GP will become a linear combination, i.e. $x_C^T(\phi-\gamma)$, and given normal priors assigned to $\beta$, this linear combination is expected to follow a Gaussian distribution \citep{si2015bayesian}. 

Fully Bayesian inference using GP comes with computational issues even for a moderate $n_A$ as one has to invert the covariance matrix at each posterior sampling step that needs $O(n_A^3)$ computations. The problem becomes even more severe when the joint posterior distribution of $(\pi^A_i, y_i)$ has to be simulated. we propose to use a low-ranked sparse GP based on the Laplace eigenvectors approximation \citep{solin2020hilbert, riutort2020practical}. Such a method reduces the computational complexity up to $O(n_Al^2)$ where $l<<n_A$ is the reduced rank of the covariance matrix.\par

Under a standard Bayesian framework, a set of independent prior distributions are assigned to the model parameters, and conditional on the observed data through a joint likelihood function, the associated posterior distributions are obtained. To this end, we use the ``black box'' solver Stan \citep{carpenter2017stan}, which employs an HMC technique to simulate the posterior predictive distribution of the parameters. In the following, we show the structure of our proposed method in Stan.\\

\noindent 
STEP 1: Priors\\\vspace{-5mm}
\begin{equation*}
\begin{aligned}
\hspace{10mm}\theta,\gamma,\phi &\sim t\text{-}student(3, 0, 1)\\
\lambda, \alpha, \sigma & \sim t\text{-}student^+(3, 0, 1)\\
\rho &\sim GIG(0, 1, 2)\\
\xi^R & \sim Dirichlet(1, 1, ... , 1)
\end{aligned}
\end{equation*}
STEP 2: Joint likelihood\\\vspace{-5mm}
\begin{equation*}
\begin{aligned}
\hspace{40mm}w^R|x_R,\gamma,\lambda &\sim N\left(exp\big\{x^T_R\gamma\big\}, \lambda^2\right)\\
\delta^A_C|x_C,\phi &\sim Bernoulli\left(logit^{-1}\{x^T_C\phi\}\right)\\
y_A|x_A,d_A,\theta, \sigma &\sim Normal\left(\theta_0+x^T_A\theta_1+d^T_A\theta_2+f\left(x^T_A(\phi-\gamma), \alpha, \rho, \tau\right), \sigma^2\right)\\
n^R|\xi^R &\sim Multinomial(n_R, \xi)
\end{aligned}
\end{equation*}\\
\noindent 
STEP 3: Posteriors\\\vspace{-5mm}
\begin{equation*}\label{eq:2.27}
\begin{aligned}
\hspace{35mm}\hat y_R|y_A, x_R, d_R,\theta, \sigma &\sim Normal\left(\theta_0+x^T_R\theta_1+d^T_R\theta_2+f\left(x^T_R(\phi-\gamma), \alpha, \rho, \tau\right), \sigma^2\right)\\
\hat N |\pi^R, \xi^R &\sim Multinomial\left(N-n_R, c\xi^R(1-\pi^R)/\pi^R\right)\\
\hat{\bar y}_U&=\bigg\{\sum_{j=1}^J \frac{\hat N_{j}}{n^R_{j}}\sum_{i=1}^{n^R_j}\hat y_{j[i]} + \sum_{i=1}^{n_A}\{y_i-\hat y_i\}\bigg\}/N
\end{aligned}
\end{equation*}
where $t\text{-}student^+$ denotes a half $t\text{-}student$ and $GIG$ stands for the Generalized Inverse Gaussian distribution, which is recommended in Stan User's Guide (Stan Development Team, 2019) for the length-scale parameter of a partially linear GP regression. Also, $f(.)$ denotes a low-ranked GP approximation with $l=10$ and a boundary condition factor of $c=1.25$, where the covariance function is given by Eq.~\ref{eq:2.27}. Throughout the analysis, we simulate the posterior predictive distribution of $\hat{\bar y}_U$ in Stan using $M=500$ HMC draws after discarding the first $500$ draws as the burn-in period.\par


\section{Simulation study}\label{S:3}
Two simulations are presented in this section, in which we compare the performance of our proposed GPPP method with those of LWP, AIPW, and PAPP with respect to the bias magnitude, efficiency, and accuracy of the variance estimator. All of the competing methods are DR, except for the PAPP method, which is an inverse PS weighted estimate of the observed $y_i$ for $i\in S_A$ with PS estimated from Eq.~\ref{eq:2.17}. The GPPP and LWP methods are fully implemented under a Bayesian setting, whereas AIPW and PAPP estimates are obtained under a frequentist method \citep{rafei2021robust}. Therefore, for the earlier class of methods, we are able to compute 95\% credible intervals (95\% CIs) while for the latter, a bootstrap method with $B=100$ replications is employed to estimate the variance and 95\% confidence intervals.\par

Various scenarios are considered with different assumptions about the functional form of the relationship among variables. For both studies, $S_A$ and $S_R$ are given a random selection mechanism with unequal inclusion probabilities. Note that units of both samples are selected independently with no clustering or stratification. Once $S_A$ and $S_R$ are drawn from $U$, we assume that $\pi^A_i$ for $i\in S_C$ and $y_j$ for $j\in S_R$ are unobserved, and the aim is to adjust for the selection bias in $S_A$ based on the combined sample, $S_C$. The simulation is then iterated $K=216$ times (which is a multiple of $36$ as conducted parallel computing using $36$ cores), where the bias-adjusted point estimates, SE and associated 95\% credible/confidence interval (CI) for $\bar y_U$ are estimated in each iteration.\par

To evaluate the repeated sampling properties of the competing method, relative bias (rBias), relative root mean square error (rMSE), the nominal coverage rate of 95\% CIs (crCI), relative length of 95\%CIs (rlCI) and SE ratio (rSE) are calculated as below:
\begin{align}
rbias\left(\hat{\bar y}_{U}\right) &=100 \times\frac{1}{K}\sum_{k=1}^K \left(\hat{\bar y}^{(k)}_{U}-\bar y_U\right) /\bar y_U\\
rMSE\left(\hat{\bar y}_{U}\right) &=100 \times\sqrt{\frac{1}{K}\sum_{k=1}^K\left(\hat{\bar y}^{(k)}_{U}-\bar y_U\right)^2} /\bar y_U\\
crCI\left(\hat{\bar y}_{U}\right) &=100 \times \frac{1}{K}\sum_{k=1}^K I\left(\big|\hat{\bar y}^{(k)}_{U} - \bar y_U\big| <z_{0.975}\sqrt{var\left(\hat{\bar y}^{(k)}_{U}\right)}\right)\\
rlCI\left(\hat{\bar y}_{U}\right) & = 100\times \frac{2}{K}\sum_{k=1}^K z_{0.975}\sqrt{var\left(\hat{\bar y}^{(k)}_{U}\right)}\\
rSE\left(\hat{\bar y}_{U}\right) &= \frac{1}{K}\sum_{k=1}^K \sqrt{var(\hat{\bar y}^{(k)}_{U})}/\sqrt{\frac{1}{K-1}\sum_{k=1}^K \left(\hat{\bar y}^{(k)}_{U}-\bar{\bar y}_{U}\right)^2}
\end{align}
where $\hat{\bar y}^{(k)}_{U}$ denotes the adjusted sample mean from iteration $k$, $\bar{\bar y}_{U}=\sum_{k=1}^K \hat{\bar y}^{(k)}_{U}/K$, $\bar y_U$ is the finite population true mean, and $var(.)$ represents the variance estimate of the adjusted mean based on the sample. Finally, to test the DR property of the proposed methods, we investigate different scenarios regarding whether models for QR and PM are correctly specified or not.\par

\subsection{Simulation I}\label{S:3.1} 
\subsubsection{Design}{S:3.1.1}
The design of our first study is based on the simulation implemented in \cite{chen2019doubly}. Consider a finite population of size $N=10^5$ with $z=\{z_1, z_2, z_3, z_4\}$ being a set of auxiliary variables generated as follows:
\begin{equation}\label{eq:3.1}
z_1\sim Ber(p=0.5) \hspace{15mm} z_2\sim U(0, 2) \hspace{15mm} z_3\sim Eup(\mu=1) \hspace{15mm} z_4\sim \chi^2_{(4)}
\end{equation}
and $x=\{x_1, x_2, x_3, x_4\}$ is subsequently defined as a linear function of $z$ as below:
\begin{equation}\label{eq:3.2}
x_1=z_1 \hspace{10mm} x_2=z_2+0.3z_1 \hspace{10mm} x_3=z_3+0.2(x_1 + x_2) \hspace{10mm} x_4=z_4+0.1(x_1+x_2+x_3)
\end{equation}
Given $x$, a continuous outcome variable $y$ is constructed by
\begin{equation}\label{eq:4.34}
y_i=2+x_{1i}+x_{2i}+x_{3i}+x_{4i}+\sigma\epsilon_i
\end{equation}
where $\epsilon_i\sim N(0, 1)$, and $\sigma$ is defined such that the correlation between $y_i$ and $\sum_{k=1}^4x_{ki}$ equals $\rho=0.8$. Further, associated with the design of $S_A$, a set of selection probabilities are assigned to the population units through the following logistic model:
\begin{equation}\label{eq:3.4}
log\left(\frac{\pi^A_i}{1-\pi^A_i}\right)=\gamma_0+0.1x_{1i}+0.2x_{2i}+0.1x_{3i}+0.2x_{4i}
\end{equation}
where $\gamma_0$ is determined such that $\sum_{i=1}^{N}\pi^A_i=n_A$. For the selection probabilities in $S_R$, we assume that $\pi^R_i\propto \gamma_1 + z_{3i}$, where $\gamma_1$ is obtained such that $max\{\pi^R_i\}/min\{\pi^R_i\}=50$. It is important to note that in this simulation study $\pi^R_i$ is assumed to be known for $i\in S_A$ as $z_3$ is observed in $S_A$.\par

Using these measures of size, we repeatedly draw pairs of samples corresponding to $S_A$ and $S_R$ from $U$ through a Poisson sampling design. The simulation is then repeated for different pairs of expected sample sizes, i.e. $(n_A, n_R)=(500, 500)$, $(n_A, n_R)=(1,000, 500)$ and also $(n_A, n_R)=(500, 1,000)$. (Note that the actual sample size is a random variable under a Poisson sampling design.) Both $Y$ and $\pi^A$ are associated with a linear combination of $X$ in this simulation study. Finally, in order to misspecify a model, we omit $x_4$ from the predictors of the working model.\par 

\subsubsection{Results}\label{S:3.1.2}
Table~\ref{tab:1} summarizes the numerical results of the first simulation study across different sample size scenarios for $\rho=0.8$. As illustrated, naive estimates of the population mean are biased in both $S_R$ and $S_A$ while weighting fully corrects for the bias in both samples. For the non-robust method, PAPP, estimates are unbiased as long as the QR model is correct. The DR methods produce unbiased estimates when either the QR model or PM holds, though there is evidence of residual bias for the LWP method when the QR model holds but the PM is misspecified. In terms of rMSE, all the methods perform similarly, except for the LWP method with correct and incorrect models specified the QR and PM, respectively, which shows higher degrees of rMSE compared to the alternative methods.\par 

\renewcommand{\tabcolsep}{3.5pt}
\begin{table}[hbt!]
\caption{Comparing the performance of the bias adjustment methods in the first simulation study for $\rho=0.8$}\label{tab:1}
\begin{threeparttable}
\tiny{\begin{tabular}{l l l l l l l l l l l l l l l l l l l}
\toprule
& \multicolumn{5}{c}{\textbf{$n_A=500, \hspace{2mm} n_R=500$}} & & \multicolumn{5}{c}{\textbf{$n_A=1,000, \hspace{2mm} n_R=500$}} & & \multicolumn{5}{c}{\textbf{$n_A=500, \hspace{2mm} n_R=1,000$}}\\\cline{2-6}\cline{8-12}\cline{14-18}
\textbf{Measure} & rBias & rMSE & crCI & rlCI & rSE & & rBias & rMSE & crCI & rlCI & rSE & & rBias & rMSE & crCI & rlCI & rSE \\
\midrule
\multicolumn{10}{l}{\textbf{Probability sample ($S_R$)}}  &    &    &  &  \\
\hline
\hspace{1mm}UW   & 8.866 & 9.093 & 0.926 & 0.724 & 0.982 &    & 8.866 & 9.093 & 0.926 & 0.724 & 0.982 &    & 8.819 & 8.942 & 0.000 & 0.513 & 0.953\\
\hspace{1mm}FW   & 0.150 & 2.322 & 93.981 & 0.844 & 0.998 &    & 0.150 & 2.322 & 93.981 & 0.844 & 0.998 &    & 0.030 & 1.692 & 94.907 & 0.598 & 0.969\\
\hline
\multicolumn{10}{l}{\textbf{Non-probability sample ($S_A$)}}   &  &      &    &  &   \\
\hline
\hspace{1mm}UW   & 30.675 & 30.794 & 0.000 & 0.940 & 0.950 &    & 29.958 & 30.006 & 0.000 & 0.657 & 1.063 &    & 30.675 & 30.794 & 0.000 & 0.940 & 0.950\\
\hspace{1mm}FW   & -0.038 & 2.354 & 93.519 & 0.811 & 0.944 &    & -0.044 & 1.618 & 93.056 & 0.570 & 0.965 &    & -0.038 & 2.354 & 93.519 & 0.811 & 0.944\\
\hline
\multicolumn{10}{l}{Model specification: QR--True, PM--True}   &      &      &    &  &   \\
\hline
\hspace{1mm}GPPP   & 0.054 & 2.473 & 96.759 & 0.935 & 1.035 &    & 0.014 & 2.171 & 95.370 & 0.862 & 1.087 &    & 0.003 & 2.039 & 95.370 & 0.750 & 1.007\\
\hspace{1mm}LWP   & 0.034 & 2.586 & 97.222 & 1.502 & 1.591 &    & -0.027 & 2.193 & 94.907 & 0.860 & 1.074 &    & -0.071 & 2.107 & 94.907 & 0.763 & 0.992\\
\hspace{1mm}AIPW    & -0.042 & 2.528 & 93.981 & 0.867 & 0.940 &    & 0.001 & 2.166 & 93.519 & 0.772 & 0.976 &    & -0.094 & 2.086 & 93.981 & 0.709 & 0.932\\
\hspace{1mm}PAPP   & 0.899 & 2.641 & 90.741 & 0.873 & 0.963 &    & 0.646 & 2.118 & 93.519 & 0.735 & 0.998 &    & 1.119 & 2.455 & 88.426 & 0.766 & 0.960\\
\hline
\multicolumn{10}{l}{Model specification: QR--True, PM--False}   &      &      &    &  &   \\
\hline
\hspace{1mm}GPPP   & 0.025 & 2.465 & 96.759 & 0.934 & 1.038 &    & 0.003 & 2.177 & 93.981 & 0.856 & 1.078 &    & -0.007 & 2.027 & 95.833 & 0.752 & 1.016\\
\hspace{1mm}LWP    & 0.022 & 2.462 & 96.759 & 0.935 & 1.041 &    & 0.007 & 2.161 & 94.907 & 0.858 & 1.087 &    & -0.028 & 2.036 & 95.833 & 0.749 & 1.007\\
\hspace{1mm}AIPW   & -0.002 & 2.452 & 92.593 & 0.844 & 0.943 &    & 0.003 & 2.150 & 93.981 & 0.757 & 0.964 &    & -0.068 & 2.019 & 93.519 & 0.697 & 0.947\\
\hline
\multicolumn{10}{l}{Model specification: QR--False, PM--True}   &      &      &    &  &   \\
\hline
\hspace{1mm}GPPP   & 0.945 & 2.855 & 95.370 & 0.983 & 0.999 &    & 0.798 & 2.453 & 96.759 & 0.898 & 1.061 &    & 0.990 & 2.461 & 91.204 & 0.792 & 0.963\\
\hspace{1mm}LWP    & 3.989 & 5.740 & 76.852 & 1.413 & 0.937 &    & 4.233 & 5.511 & 75.463 & 1.278 & 0.992 &    & 3.959 & 5.245 & 71.759 & 1.136 & 0.904\\
\hspace{1mm}AIPW   & 0.215 & 2.532 & 93.056 & 0.855 & 0.929 &    & 0.092 & 2.068 & 93.981 & 0.744 & 0.987 &    & 0.173 & 2.138 & 93.519 & 0.752 & 0.967\\
\hline
\multicolumn{10}{l}{Model specification: QR--False, PM--False}   &      &      &    &  &   \\
\hline
\hspace{1mm}GPPP   & 27.303 & 27.460 & 0.000 & 1.591 & 1.485 &    & 26.513 & 26.590 & 0.000 & 1.443 & 1.962 &    & 27.308 & 27.451 & 0.000 & 1.291 & 1.263\\
\hspace{1mm}LWP   & 27.132 & 27.295 & 0.000 & 1.600 & 1.470 &    & 26.437 & 26.514 & 0.000 & 1.441 & 1.955 &    & 27.194 & 27.341 & 0.000 & 1.307 & 1.262\\
\hspace{1mm}AIPW    & 27.162 & 27.322 & 0.000 & 0.986 & 0.914 &    & 26.453 & 26.531 & 0.000 & 0.725 & 0.978 &    & 27.110 & 27.252 & 0.000 & 0.947 & 0.934\\
\hspace{1mm}PAPP   & 27.946 & 28.097 & 0.000 & 0.976 & 0.918 &    & 26.996 & 27.070 & 0.000 & 0.715 & 0.979 &    & 28.166 & 28.303 & 0.000 & 0.954 & 0.941\\
\bottomrule
\end{tabular}}
 \begin{tablenotes}
 \tiny
 \item NOTE 1: GPPP: Gaussian Process of Propensity Prediction; LWP: Linear-in-weight Prediction; AIPW: Augmented Inverse Propensity Weighting; PAPP: Propensity-adjusted Probability Prediction.
 \item NOTE 2: The PAPP and AIPW methods have been implemented through a bootstrap method.
 \item NOTE 3: The PAPP method is non-robust, while the rest of the methods, i.e. GPPP, LWP, and AIPW, are doubly robust.
 \end{tablenotes}
\end{threeparttable}
\end{table}

AIPW and PAPP have slightly narrower CIs than the Bayesian methods, GPPP and LWP. The LWP performs poorly with respect to efficiency when the PM is incorrectly specified. Generally, the values of rSE suggest that variance estimation is unbiased across different model specification scenarios with a slight overestimation and underestimation in the Bayesian and bootstrap methods, respectively. Under the situations where the working model for QR is correct while that for PM is incorrect, LWP tends to underestimate the variance. The coverage rates of 95\% CIs are also close to the nominal value when at least one of the QR and PM models is correctly specified. However, we observe that 95\% CIs based on the frequentist methods tend to undercover the true population mean to some degrees, and the poorest result of crCI belongs to the LWP method when the PM is wrongly specified. These findings are generalizable to all other sample size combinations, and to the other extensions of the simulation for $\rho={0.3, 0.5}$, whose tables are displayed in Appendix~\ref{S:6.4.1}.\par 

\subsection{Simulation II}\label{S:3.2} 
\subsubsection{Design}\label{S:3.2.1}
In the previous simulation study, the ignorable assumption was violated to misspecify the working model by dropping a key auxiliary variable. Now, we focus on a situation where models misspecified with respect to the functional form of their conditional means. To this end, we consider (non-)linear associations and two-way interactions in construction of the outcome variables. In addition, to build a more realistic situation, two separate sets of auxiliary variables are generated, $D$ associated with the design of $S_A$, and $X$ associated with the design of $S_R$. However, we allow the two variables to be correlated through a bivariate Gaussian distribution as below:
\begin{equation}\label{eq:3.7}
\begin{pmatrix}
	d\\
	x
\end{pmatrix} \sim MVN \left(
\begin{pmatrix}
	0\\
	0
\end{pmatrix} , 
\begin{pmatrix}
	4   &  2\rho \\
	2\rho &  1
\end{pmatrix}
\right)
\end{equation}
Note that $\rho$ controls how strongly the sampling design of $S_R$ is associated with that of $S_A$. Primarily, we set $\rho=0.5$, but later we check other values ranging from $0$ to $0.9$ as well.\par

\begin{figure}[hbt!]
\centering\includegraphics[scale=0.20]{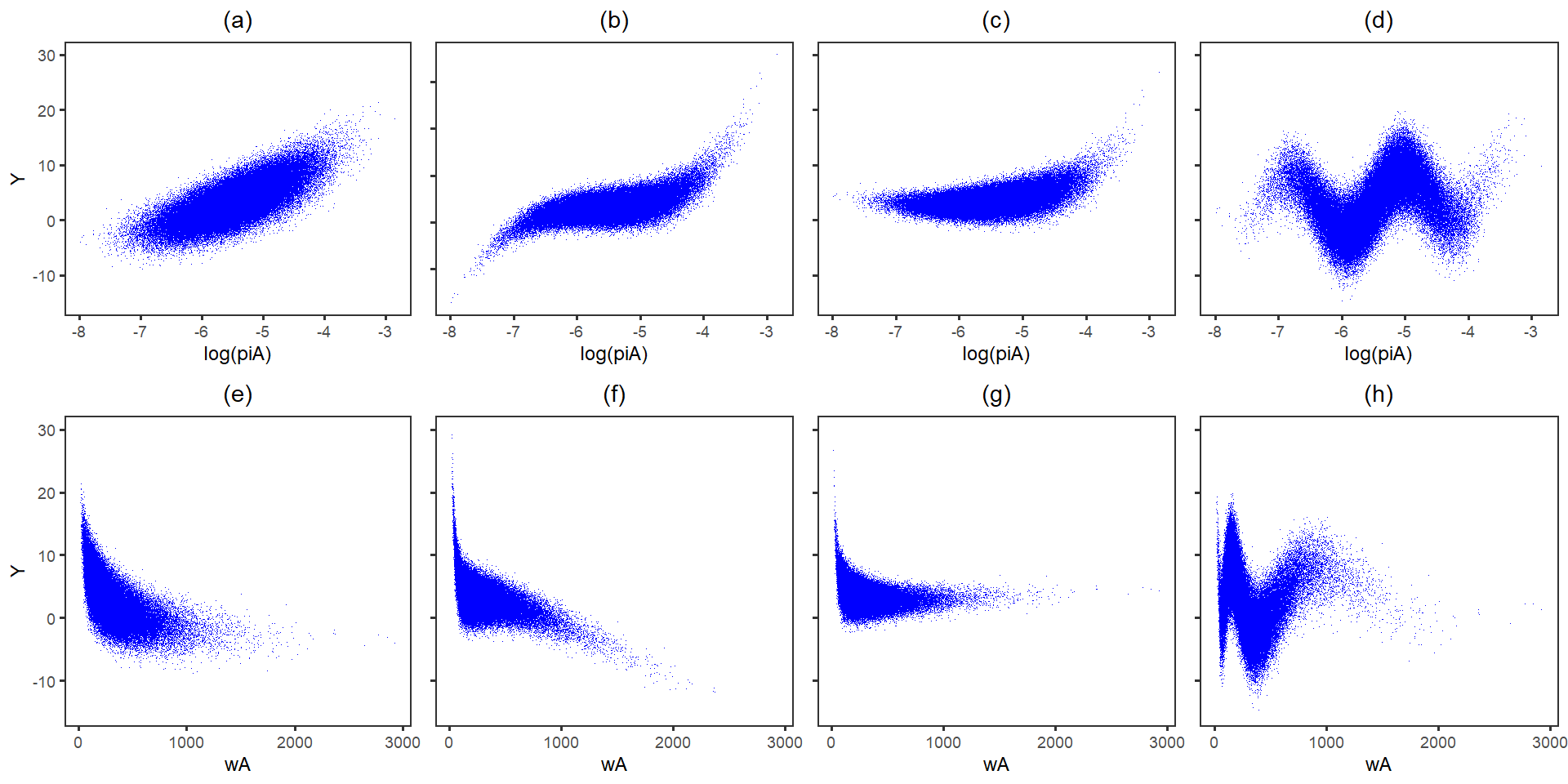}
\caption{The proposed relationships between the outcome variable $Y$ and $log(\pi^A)$ in $U$ for (a) $LIN$, (b) $CUB$, (c) $EXP$ and (d) $SIN$ scenarios, and between the outcome $Y$ and sampling weights $w^A$ for (e) $LIN$, (f) $CUB$, (g) $EXP$ and (h) $SIN$ scenarios.}\label{fig:1}
\end{figure}

We then generate a continuous outcome variable ($y_i^c$) and a binary outcome variable ($y_i^b$) for $i\in U$ as below:
\begin{equation}\label{eq:3.8}
\begin{aligned}
y^c_i &= 3 + f_k(x_i) + d_i + 0.2x_id_i + \sigma\epsilon_i\\
p(y^b_i=1|x_i, d_i) &= \frac{exp\{-1 + f_k(x_i) + d_i + 0.2x_id_i\}}{1+exp\{-1 + f_k(x_i) + d_i + 0.2x_id_i\}}
\end{aligned}
\end{equation}
where $\epsilon_i\sim N(0, 1)$, and $\sigma$ is determined such that the correlation between $y^c_i$ and $f_k(x_i) + d_i + 0.2x_id_i$ equals $0.8$ for $i\in U$. The function $f_k(.)$ is assumed to take one of the following forms:
\begin{equation}\label{eq:3.9}
\begin{aligned}
LIN: f_1(x)&=x \hspace{32mm} CUB:f_2(x)=(x/3)^3\\ EXP: f_3(x)&=exp(x/2)/5 \hspace{15mm} SIN: f_4(x)=5sin(\pi x/3)
\end{aligned}
\end{equation}
Figure~\ref{fig:1} depicts the relationships between $y^c$ and $\pi^A$, and between $y^c$ and $w^A=1/\pi^A$.

We then consider an informative sampling strategy with unequal probabilities of inclusion, where the selection mechanism of $S_A$ and $S_R$ depends on $x$ and $d$, respectively. Thus, each $i\in U$ is assigned two values within $(0, 1)$ corresponding to the probabilities of selection in $S_R$ and $S_A$ through a $logistic$ function as below:
\begin{align}\label{eq:3.10}
\pi^R(d_i)=p(\delta^R_i=1|d_i) & = \frac{exp\{\gamma_0-0.4d_i\}}{1+exp\{\gamma_0-0.4d_i\}}\\
\pi^A(x_i)=p(\delta^A_i=1|x_i) & = \frac{exp\{\gamma_1+\gamma_2x_i\}}{1+exp\{\gamma_1+\gamma_2x_i\}}
\end{align}
where $\delta^R_i$ and $\delta^A_i$ are the indicators of being selected in $S_R$ and $S_A$, respectively, for $i\in U$. we initially set $\gamma_2=0.3$, which yields PS with a normal range. To assess how the adjustments behave in presence of influential weights, later we set $\gamma_2=0.6$, which yields relatively extreme weights.\par

Associated with $S_R$ and $S_A$, independent samples of expected sizes $n_R=1,000$ and $n_A=500$ are selected randomly from $U$ with a Poisson sampling design. We choose $n_A<n_R$ as is the case in the two applications of this study. The model intercepts, $\gamma_0$ and $\gamma_1$ in~\ref{eq:3.10}, are obtained such that $\sum_{i=1}^{N}\pi^R_i=n_R$ and $\sum_{i=1}^{N}\pi^A_i=n_A$, respectively. The rest of the simulation design is similar to that defined in Simulation I, except for the way we specify a working model. A QR model is misspecified by replacing $x_i$ with $x^2_i$, and a PM model is misspecified by replacing $f_k(x_i)$ with $x^2_i$ and $d_i$ with $d^2_i$, and also by dropping the interaction term $x_id_i$.\par

\subsubsection{Results}\label{S:4.2.2}
\begin{figure}[hbt!]
\centering\includegraphics[scale=0.22]{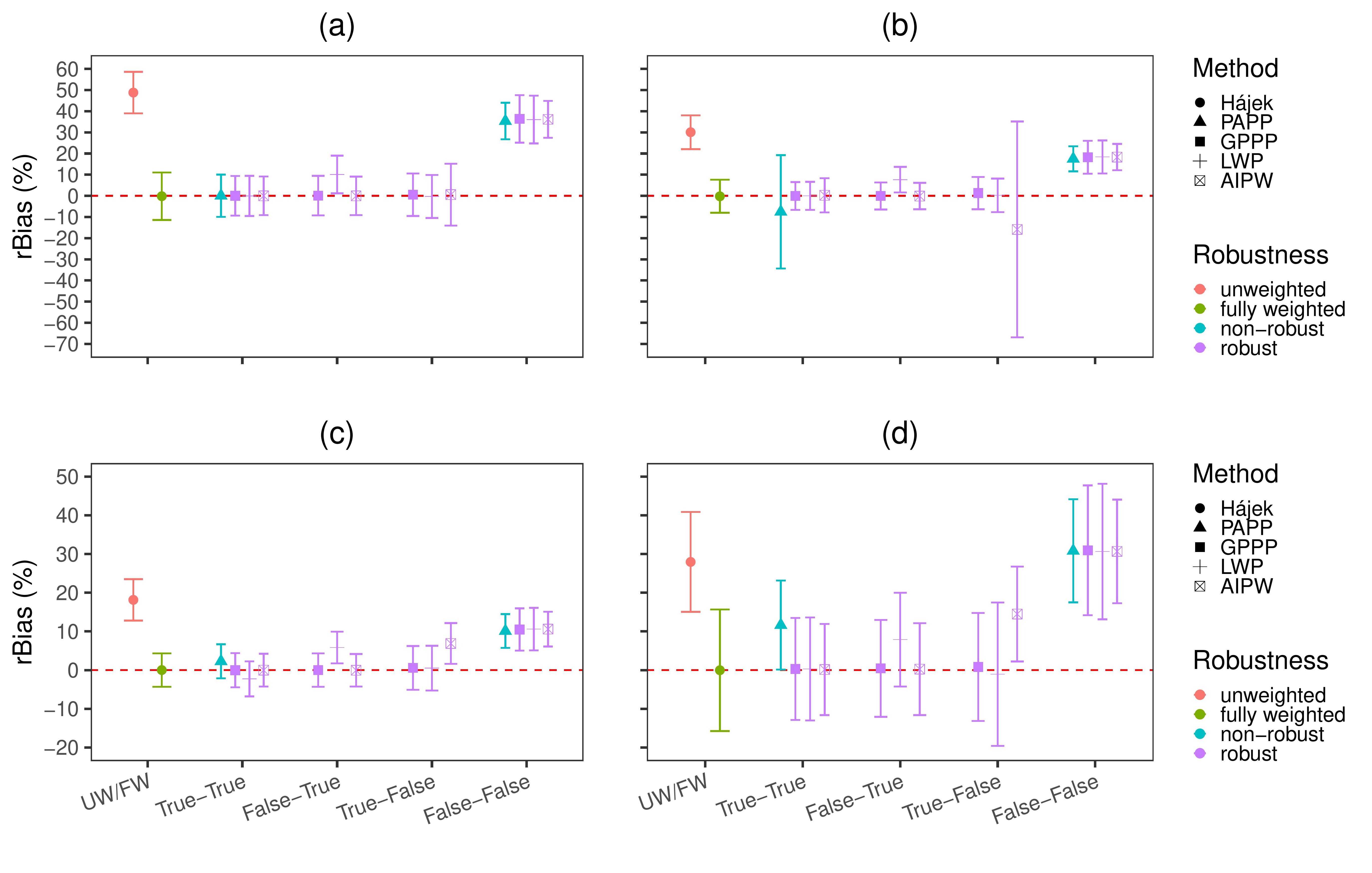}
\caption{Comparing the performance of the adjusted estimators under different model-specification scenarios for the \textit{continuous} outcome variable with $\gamma_2=0.3$ under (a) $LIN$, (b) $CUB$, (c) $EXP$, and (d) $SIN$ scenarios. The error bars have been drawn based on the 2.5\% and 97.5\% percentiles of the empirical distribution of bias over the simulation iterations. UW: unweighted; FW: Fully weighted; PAPP: Propensity-adjusted Probability Prediction; GPPP: Gaussian Processes of Propensity Prediction; LWP: Linear-in-weight Prediction; AIPW: Augmented Inverse Propensity Weighting}\label{fig:2}
\end{figure}

Figure~\ref{fig:2} compares the relative bias (rBias) magnitude and efficiency of the competing methods for the continuous outcome variable, $y^c$, across different scenarios of model specification while $\gamma_2=0.3$. Note that the error bars reflect the relative length of 95\% CIs (rlCI). As illustrated, point estimates from both $S_R$ and $S_A$ are biased if the sampling true weights are ignored. At the first glance, one can infer that for all $f_k$, $k=1, 2, 3, 4$, the magnitude of rBias is close to $zero$ as long as either QR or PM model is valid. However, in situations where $\pi^A$ is non-linearly associated with $y^c$, i.e. plots (b), (c), and (d), the AIPW and PAPP estimators are biased when the PM is misspecified, but the QR model is valid. In contrast, the LWP method yields slightly biased estimates in all plots when the QR model is misspecified, but the PM is correct. It turns out that the GPPP is the only method that leads to unbiased estimates in all the scenarios with respect to model specification and functional form of the PM. We did not observe consistent results across the adjustment methods with respect to efficiency. However, the GPPP method consistently shows high efficiency compared to the other methods across all the studied scenarios.\par  

\begin{figure}[hbt!]
\centering\includegraphics[scale=0.22]{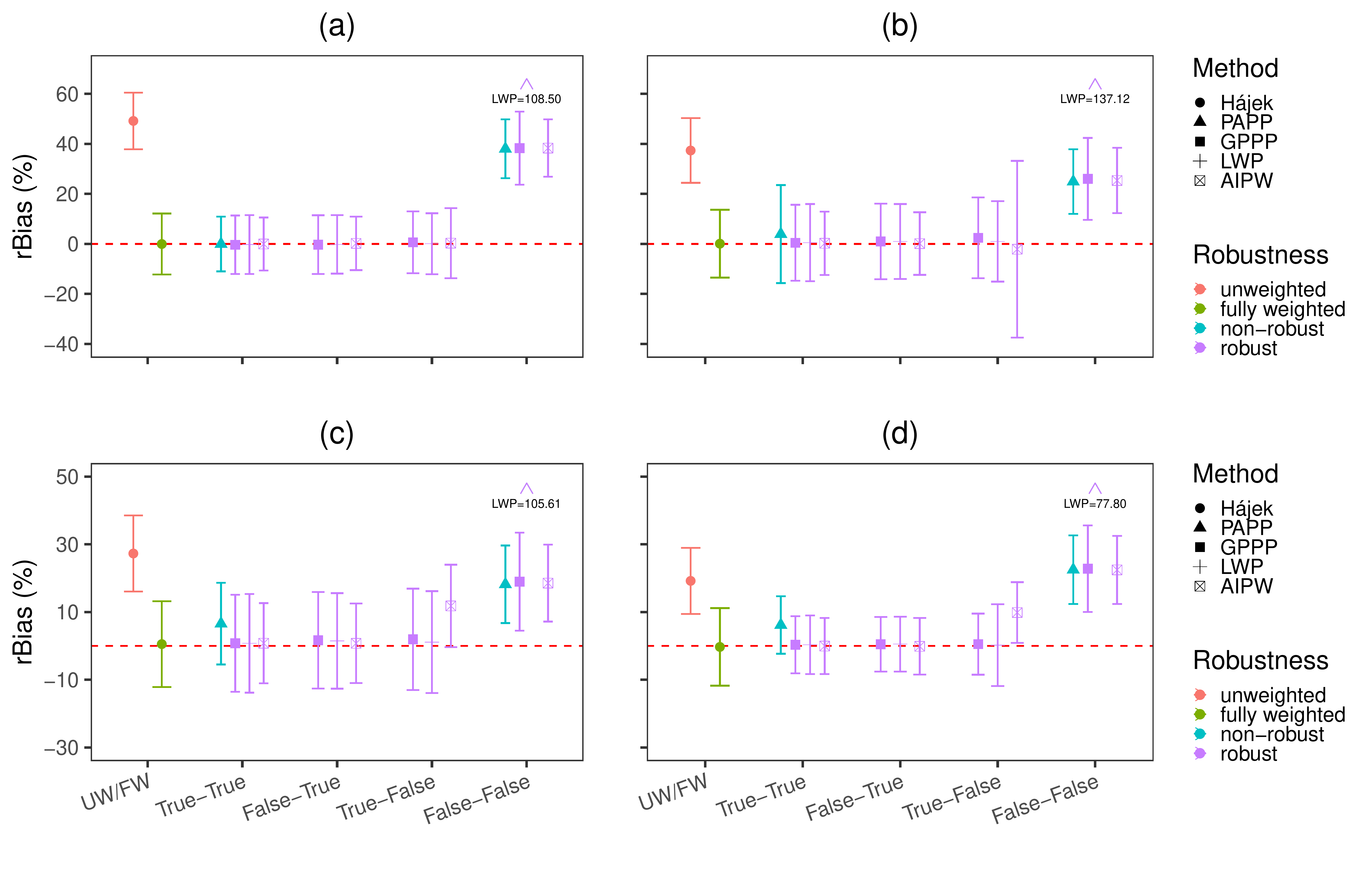}
\caption{Comparing the performance of the adjusted estimators under different model-specification scenarios for the \emph{binary} outcome variable with $\gamma_2=0.3$ under (a) $LIN$, (b) $CUB$, (c) $EXP$, and (d) $SIN$ scenarios. The error bars have been drawn based on the 2.5\% and 97.5\% percentiles of the empirical distribution of bias over the simulation iterations. UW: unweighted; FW: Fully weighted; PAPP: Propensity-adjusted Probability Prediction; GPPP: Gaussian Processes of Propensity Prediction; LWP: Linear-in-weight Prediction; AIPW: Augmented Inverse Propensity Weighting}\label{fig:3}
\end{figure}

We summarize the simulation results for the binary outcome, $y^b$, with $\gamma_2=0.3$ in Figure~\ref{fig:3}. Again, adjusted estimates are unbiased if the working model for either QR or PM holds. Exceptions are seen for the PAPP and AIPW methods with residual bias in the plots related to (c) EXP, and (d) SIN when the PM is incorrectly specified. Unlike the simulation results for the continuous variable, the LWP consistently produces unbiased estimates for the binary outcome when the working model for QR fails. However, the magnitude of bias seems to be much larger in the LWP method when both underlying models for QR and PM are misspecified. Again, as for the continuous outcome, the proposed GPPP method consistently gives unbiased and efficient estimates. The lowest efficiency is associated with the AIPW and PAPP methods in the EXP scenario when the PM is misspecified.\par

\begin{figure}[hbt!]
\centering\includegraphics[scale=0.20]{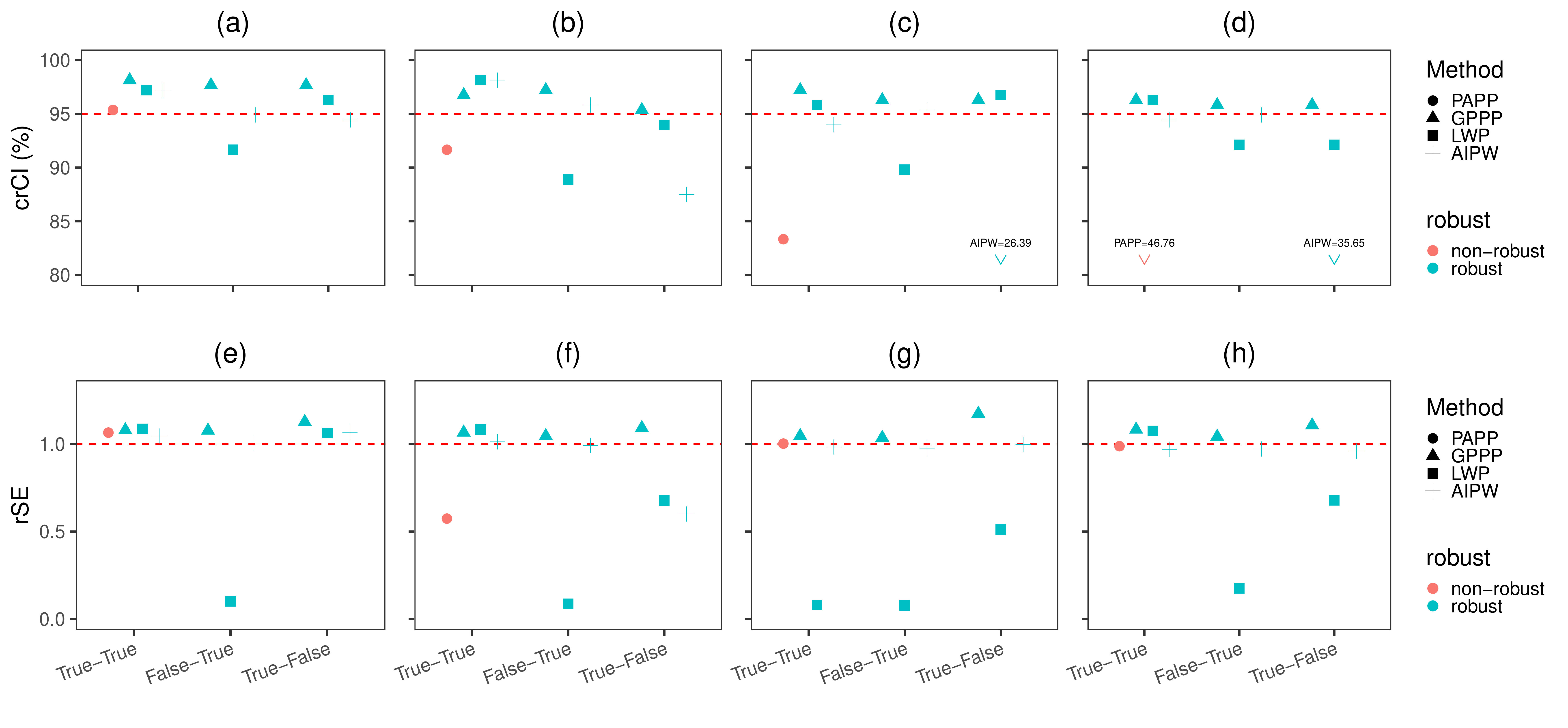}
\caption{Comparing the 95\% CI coverage rates (crCI) of the DR adjusted means for the \textit{continuous} outcome variable with $\gamma_2=0.3$ under (a) $LIN$, (b) $CUB$, (c) $EXP$, and (d) $SIN$ scenarios, and SE ratios (rSE) under (e) $LIN$, (f) $CUB$, (g) $EXP$, and (h) $SIN$ scenarios, across different DR methods under different model specification scenarios. UW: unweighted; FW: Fully weighted; PAPP: Propensity-adjusted Probability Prediction; GPPP: Gaussian Processes of Propensity Prediction; LWP: Linear-in-weight Prediction; AIPW: Augmented Inverse Propensity Weighting}\label{fig:4}
\end{figure}

\begin{figure}[hbt!]
\centering\includegraphics[scale=0.20]{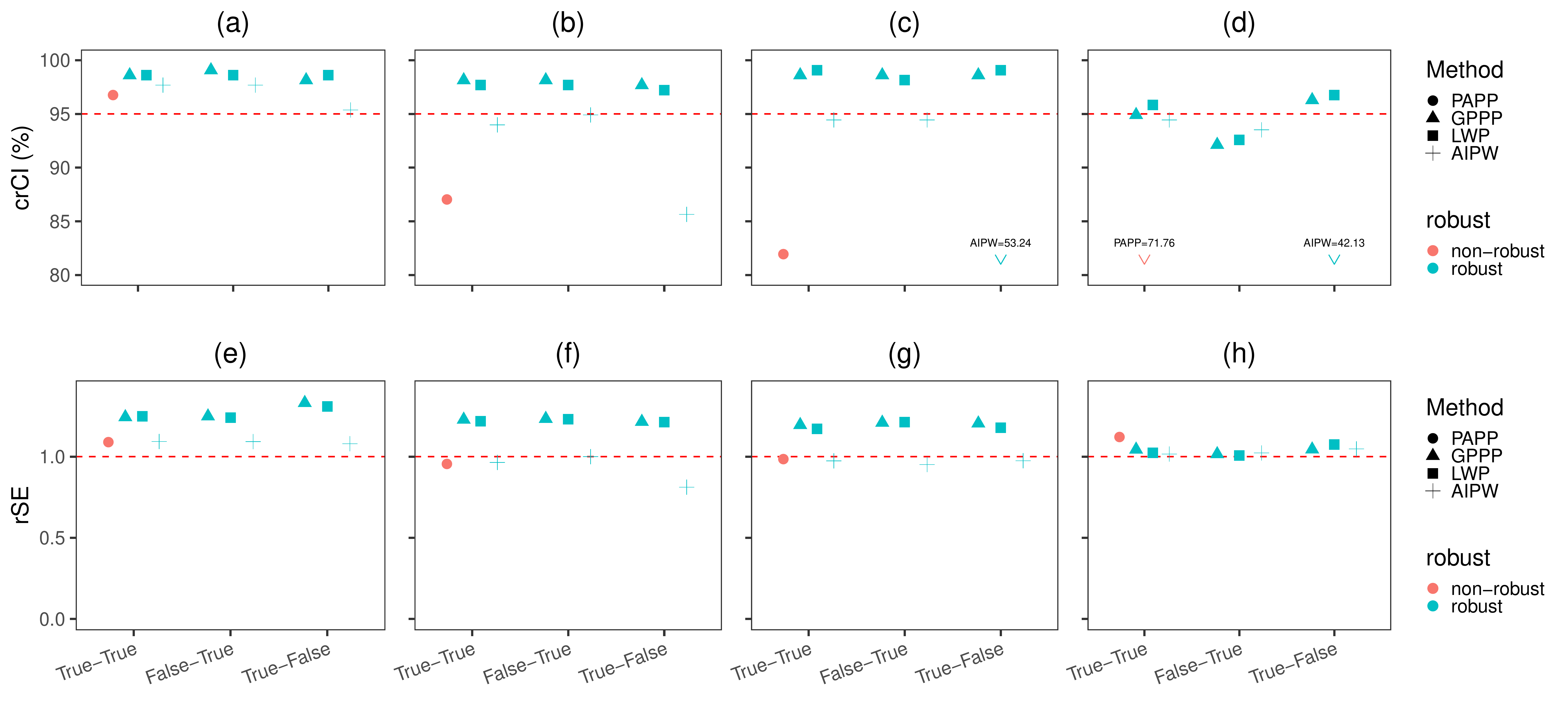}
\caption{Comparing the 95\% CI coverage rates (crCI) of the DR adjusted means for the \emph{binary} outcome variable with $\gamma_2=0.3$ under (a) $LIN$, (b) $CUB$, (c) $EXP$, and (d) $SIN$ scenarios, and SE ratios (rSE) under (e) $LIN$, (f) $CUB$, (g) $EXP$, and (h) $SIN$ scenarios, across different DR methods under different model specification scenarios. UW: unweighted; FW: Fully weighted; PAPP: Propensity-adjusted Probability Prediction; GPPP: Gaussian Processes of Propensity Prediction; LWP: Linear-in-weight Prediction; AIPW: Augmented Inverse Propensity Weighting}\label{fig:5}
\end{figure}

Figure~\ref{fig:4} displays the results of crCI and rSE for the continuous outcomes where $\gamma_2=0.3$. According to the rSE values, all methods perform well in variance estimation except for the LWP method which consistently underestimates the variance. A similar problem appears in the PAPP and AIPW methods for the EXP scenario when the outcome model is invalid. Generally, the Bayesian methods, i.e. GPPP and LWP, tend to slightly overestimate the variance. The values of crCI seem to be close to the nominal level for all the methods across almost all the scenarios, as long as at least one of the underlying models holds. For the non-linear associations, i.e. (b) CUB, (c) EXP and (d) SIN, the 95\% CIs associated with frequentist methods, i.e. AIPW and PAPP, tend to undercover the population mean when the outcome model is false. Figure~\ref{fig:5} depicts similar results for the binary outcome when $\gamma_2=0.3$. Overall, the results look analogous to those obtained for the continuous outcome. However, the degree of overestimation of variance by the Bayesian methods seems to be larger in the binary outcome than the continuous outcome.\par

Extensions of the simulation for other sample size combinations, i.e. $(n_A, n_R)=(500, 500)$ and $(n_A, n_R)=(1,000, 500)$ and also for $\gamma_2=0.6$, which creates extreme sampling weights in $S_A$, are included in Appendix~\ref{S:6.4}. While we observe no major discrepancy in the simulation results for other sample size scenarios than $(n_A, n_R)=(500, 1,000)$, having influential weights presented in $S_A$ leads to a larger magnitude of bias and lower efficiency in the estimates of PAPP, AIPW when the PM is incorrectly specified, but the QR model is valid. However, the GPPP method seems to be least affected by the presence of extreme weights.\par

In Figures~\ref{fig:4} and \ref{fig:5}, we depict the measures associated with the accuracy of the variance methods for GPPP/AIPW estimators. One can immediately infer that for both employed methods, the variance estimator is approximately unbiased when at least one of the underlying models holds. However, in situations where both models are invalid, according to the rSE values, the AIPW estimator tends to underestimate/overestimate the variance to a significant extent, while the variance estimator under GPPP shows more robustness across the model specification scenarios as well as outcome variables. Last but not least, the proximity of the crCI values to $95\%$ for the GPPP methods, especially when both underlying models are wrong, reflects the accuracy of both point and variance estimates under the GPPP method.\par 

\begin{figure}[hbt!]
\centering\includegraphics[scale=0.28]{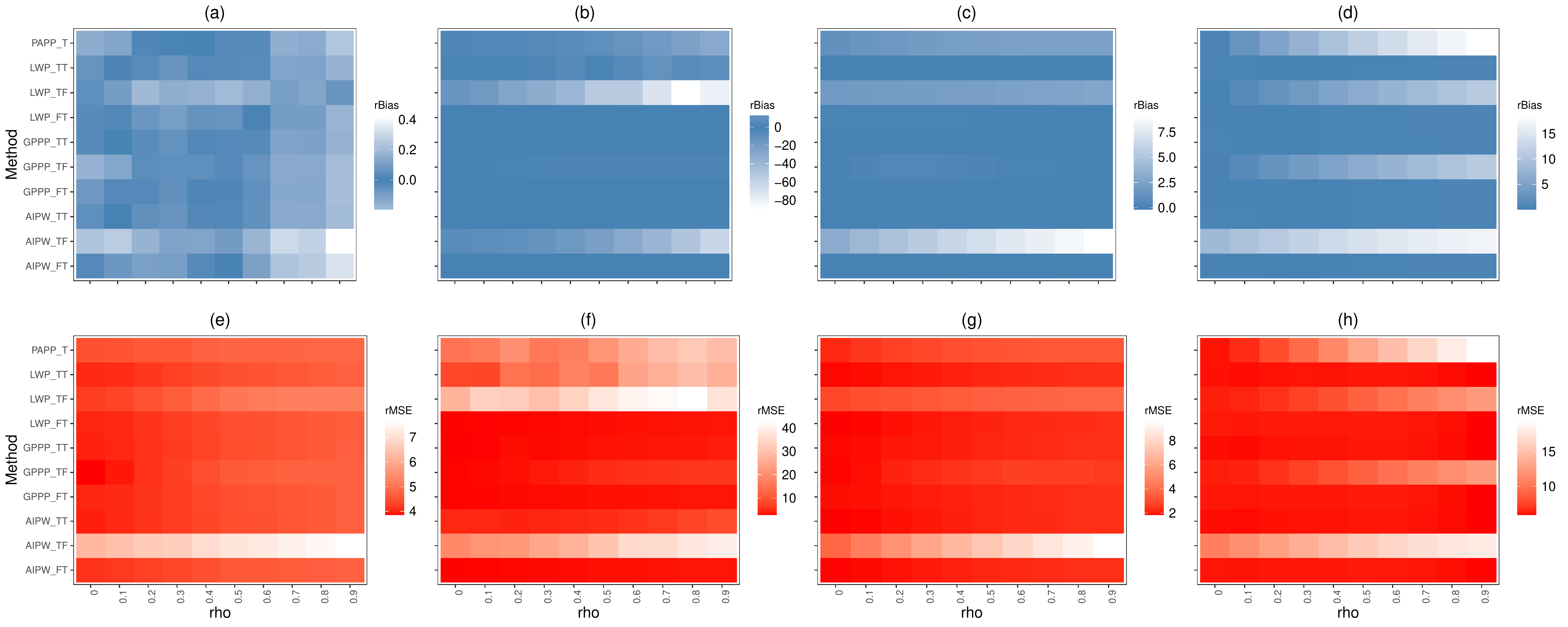}
\caption{Comparing the magnitude of rBias of the DR adjusted means for the \textit{continuous} outcome variable with $\gamma_2=0.3$ under (a) $LIN$, (b) $CUB$, (c) $EXP$, and (d) $SIN$, and rMSE under (e) $LIN$, (f) $CUB$, (g) $EXP$, and (h) SIN across different model specification scenarios and different values of $\rho$. UW: unweighted; FW: Fully weighted; PAPP: Propensity-adjusted Probability Prediction; GPPP: Gaussian Processes of Propensity Prediction; LWP: Linear-in-weight Prediction; AIPW: Augmented Inverse Propensity Weighting}\label{fig:6}
\end{figure}

\begin{figure}[hbt!]
\centering\includegraphics[scale=0.28]{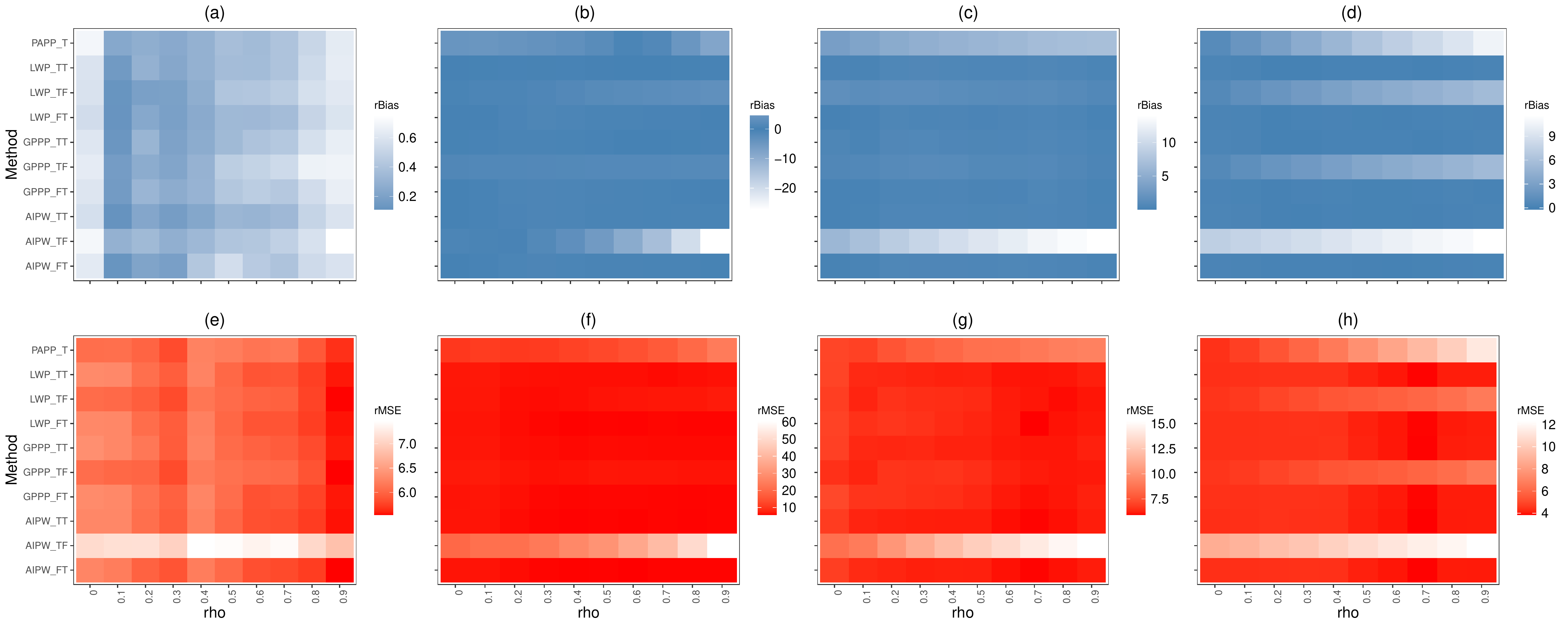}
\caption{Comparing the magnitude of rBias of the DR adjusted means for the \emph{binary} outcome variable with $\gamma_2=0.3$ under (a) $LIN$, (b) $CUB$, (c) $EXP$, and (d) $SIN$, and rMSE under (e) $LIN$, (f) $CUB$, (g) $EXP$, and (h) SIN across different model specification scenarios and different values of $\rho$. UW: unweighted; FW: Fully weighted; PAPP: Propensity-adjusted Probability Prediction; GPPP: Gaussian Processes of Propensity Prediction; LWP: Linear-in-weight Prediction; AIPW: Augmented Inverse Propensity Weighting}\label{fig:7}
\end{figure}

So far, the results we discussed were limited to a case where $\rho=0.5$. As the final step, we replicate the simulation for different values of $\rho$ ranging from $0$ to $0.9$ to show how stable the competing methods perform in terms of rbias and rMSE. Figure~\ref{fig:6} depicts changes in the values of rBias and rMSE in the continuous outcome, $y^c$, for different adjustment methods and across different model specification scenarios as the value of $\rho$ increases. Generally, it seems that the values of rBias and rMSE decline for all competing methods with an increase in $\rho$. In addition, for all values of $\rho$, it is evident that the GPPP method outperforms the PAPP, AIPW, LWP methods when the outcome model is wrong. This strength in GPPP is more evident when the association between the outcome and the PS is non-linear, i.e. in (b) CUB, (c) EXP, and (d) SIN. In Figure~\ref{fig:7}, we display corresponding comparisons for the binary outcome. The results are similar to those based on the continuous outcome, with a difference in that the values of rMSE increase with an increase in the value of $\rho$. Detailed numerical results of Simulation II is available in Appendix~\ref{S:6.4}.\par

\section{Applications}\label{S:4}
As the application part, we conduct an empirical study involving inference for a non-probability sample. Our goal is to estimate police-reportable crash rates per 100M miles driven using the sensor-based data from the second phase of the Strategic Highway Research Program (SHRP2). To this end, we consider the National Household Travel Survey (NHTS) 2017 as the reference survey to adjust for the potential selection bias in crash rates. The following two subsections briefly describe the design of these samples and discuss the findings in detail.

\subsection{Strategic Highway Research Program 2}\label{S:4.1}
SHRP2 is the largest naturalistic driving study (NDS) conducted to date, with the primary aim to assess how people interact with their vehicle and traffic conditions while driving \citep{SHRP2013}. About $A=3,140$ drivers aged $16-95$ years were recruited from six geographically dispersed sites across the United States, and over five million trips and $50$ million driven miles have been recorded during their participation time. The average follow-up time per person was $440$ days. A quasi-random approach was initially employed to select samples by random cold calling from a pool of $17,000$ pre-registered volunteers. However, because of the low success rate along with budgetary constraints, the investigators later chose to pursue voluntary recruitment. Sites were assigned one of three pre-determined sample sizes according to their population density \citep{campbell2012shrp}. The youngest and eldest age groups were oversampled in the sense that crash risk is expected to be higher among those subgroups. Thus, one can conclude that the selection mechanism in SHRP2 is a combination of convenience and quota sampling methods. Further description of the study design and recruitment process can be found in \cite{antin2015naturalistic}. \par

SHRP2 data are collected in multiple stages. Selected participants are initially asked to complete multiple assessment tests, including executive function and cognition, visual perception, visual-cognitive, physical and psychomotor capabilities, personality factors, sleep-related factors, general medical condition, driving knowledge, etc. In addition, demographic information such as age, gender, household income, education level, and marital status as well as vehicle characteristics such as vehicle type, model year, manufacturer, and annual mileage are gathered at the screening stage. A trip in SHRP2 is defined as the time interval during which the vehicle is operating. The in-vehicle sensors start recording kinematic information, the driver's behaviors, and traffic events continuously as soon as the vehicle is switched on. Encrypted data are stored in a removable hard drive, and participants are asked to provide access to the vehicle every four to six months, so that hard drives with accumulated data are removed and replaced. Then, trip-related information such as average speed, duration, distance, and GPS trajectory coordinates are obtained by aggregating the sensor records at the trip level \citep{antin2019second, campbell2012shrp}. 

\subsection{National Household Travel Survey 2017}\label{S:4.2}
In the present study, we use data from the eighth round of the NHTS conducted from March 2016 through May 2017 as the reference survey. The NHTS is a nationally representative survey, repeated cross-sectionally almost every seven years. It is aimed at characterizing personal travel behaviors among the civilian, non-institutionalized population of the United States. The 2017 NHTS was a mixed-mode survey, in which households were initially recruited by mailing through an address-based sampling (ABS) technique. Within the selected households, all eligible individuals aged $\geq5$ years were requested to report the trips they made on a randomly assigned weekday through a web-based travel log. Proxy interviews were requested for younger household members who were $\leq15$ years old. \par 

The overall sample size was $129,696$, of which roughly $20$\% was used for national representativity and the remaining $80$\% was regarded as add-ons for the state-level analysis. The recruitment response rate was $30.4$\%, of which $51.4$\% reported their trips via the travel logs \citep{santos2011summary}. In NHTS, a travel day is defined from $4:00$ AM of the assigned day to $3:59$ AM of the following day on a typical weekday. A trip is defined as that made by one person using any mode of transportation. While trip distance was measured by online geocoding, the rest of the trip-related information was based on self-reporting. A total of $264,234$ eligible individuals aged $\geq$5 took part in the study, for which $923,572$ trips were recorded \citep{mcguckin2018summary}.\par

\subsection{Auxiliary variables and analysis plan}\label{S:4.3}
Our focus here was on inference at the participant level, so both SHRP2 and NHTS data were aggregated. Considering this, we calculated the total distance driven and total number of police-reported crashes for each participant of SHRP2. The total of these quantities by all SHRP2 participants were $28$M miles and $210$, respectively. To make the two datasets more comparable, we also filtered out all the subjects in NHTS who were not drivers or were younger than $16$ years old or used public transportation or transportation modes other than cars, SUVs, vans, or light pickup trucks. In addition, we restricted the NHTS sample to those respondents who reside in the SHRP2-specific six states, so our inferences are only generalizable to those six states. The final sample sizes of the complete datasets were $n_A=2,862$ and $n_R=29,572$ in SHRP2 and NHTS, respectively.

To address the expressed objective of the present study, we set the outcome variable to be the frequency of police-reported crashes by SHRP2 participants throughout their follow-up time. In addition, we utilize the total miles driven by each SHRP2 participant as the model offset to obtain the rates by a driven mile. Particular attention was paid to identify as many relevant common auxiliary variables as possible in the combined sample that are expected to govern both selection mechanism and response surface in SHRP2. Two distinct sets of variables were considered: (i) demographic and socio-economic information of the drivers including sex, age groups, race, ethnicity, birth country, education level, household size, number of owned vehicles, and state of residence, and (ii) vehicle characteristics including vehicle age, vehicle manufacturer, vehicle type and fuel type.\par 

We chose to use a Bayesian negative binomial (NB) regression for modeling the response surface because the outcome variable was count data and effects of overdispersion were present. The \emph{log} of the total miles driven by SHRP2 participants is included as offset in the model such that crash rates can be predicted by unit of distance driven. conditional on $x_i$ and $\hat\pi^i_i$, the outcome model is given by
\begin{equation}
y_i|x_i, t_i, \theta, \phi, \gamma, \alpha,\tau, \sigma \sim NB\left(exp\big\{\theta_0 + x_i^T\theta_1 + f\left(x^T_A(\phi-\gamma), \alpha, \rho, \tau\right) + log(t_i)\big\}, 1/\sigma\right)
\end{equation}
where $t_i$ is the total distance driven by respondent $i$, and $\sigma\sim Cauchy^+(0, 3)$. Note that we also checked and found no evidence of zero-inflation in the distribution of the outcome by comparing the observed zeros with the expected number of zeros under the proposed NB model.\par

\subsubsection{Results}\label{S:4.4}
\begin{figure}[hbt!]
 \begin{adjustbox}{addcode={\begin{minipage}{\width}}{\caption{%
   Comparing the distribution of common auxiliary variables in SHRP2 with weighted NHTS
   }\label{fig:4.8}\end{minipage}},rotate=90,center}\vspace{-5mm}
   \includegraphics[width=1.4\linewidth]{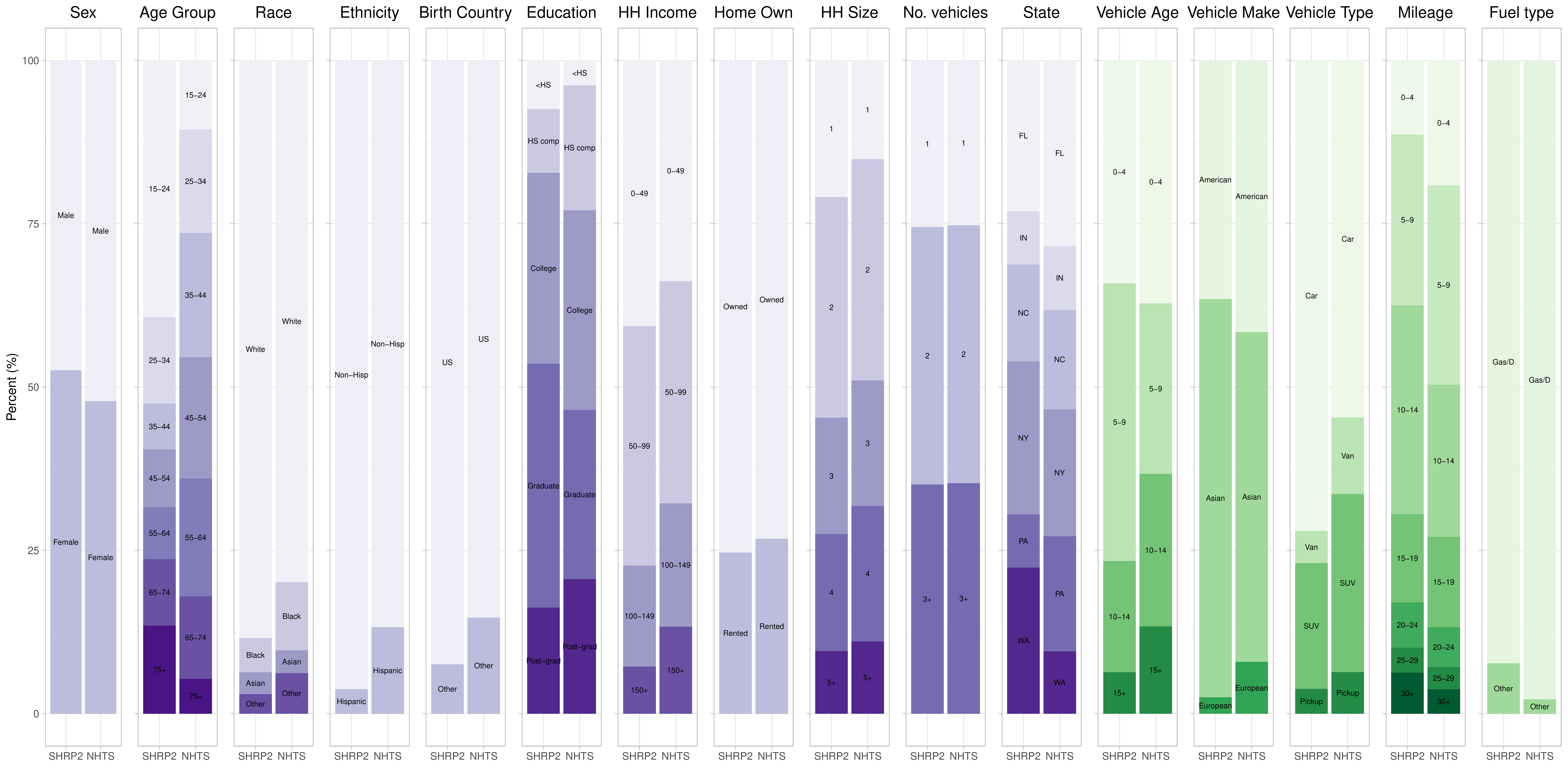}%
 \end{adjustbox}
 \vspace{-25mm}
\end{figure}

According to Figure~\ref{fig:4.8}, one can visually infer that the largest discrepancies between the sample distribution of auxiliary variables in SHRP2 and that in the population stem from participants' age, race, and population size of the residential area as well as vehicles' age and vehicles' type. The youngest and oldest age groups are overrepresented as are Whites and non-Hispanics. In addition, we found that the proportion of urban dwellers is higher in SHRP2 than that in the NHTS. In terms of vehicle characteristics, SHRP2 participants tend to own passenger cars more than the population average, whereas individuals with other vehicle types were underrepresented in SHRP2.\par

\begin{figure}[hbt!]
\centering\includegraphics[scale=0.32]{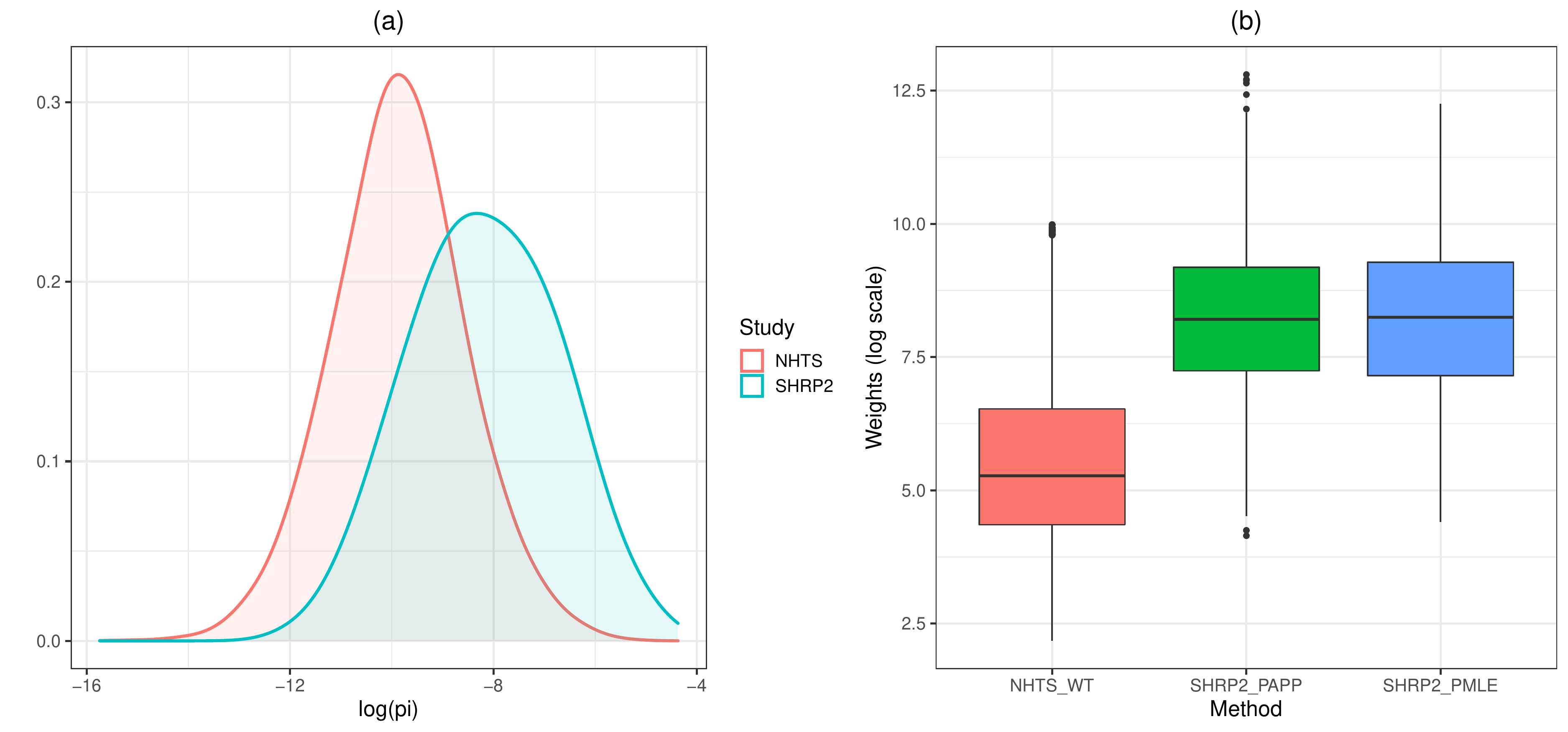}
\caption{Comparing the empirical density of (a) estimated propensity scores between SHRP2 and NHTS and (b) estimated pseudo-weights in SHRP2 across the applied quasi-randomization methods}\label{fig:4.9}
\end{figure}

\begin{figure}[hbt!]
 \begin{adjustbox}{addcode={\begin{minipage}{\width}}{\caption{%
   Comparing the distribution of common auxiliary variables in pseudo-weighted SHRP2 based on the PAPP method with weighted NHTS
   }\label{fig:4.10}\end{minipage}},rotate=90,center}\vspace{-5mm}
   \includegraphics[width=1.4\linewidth]{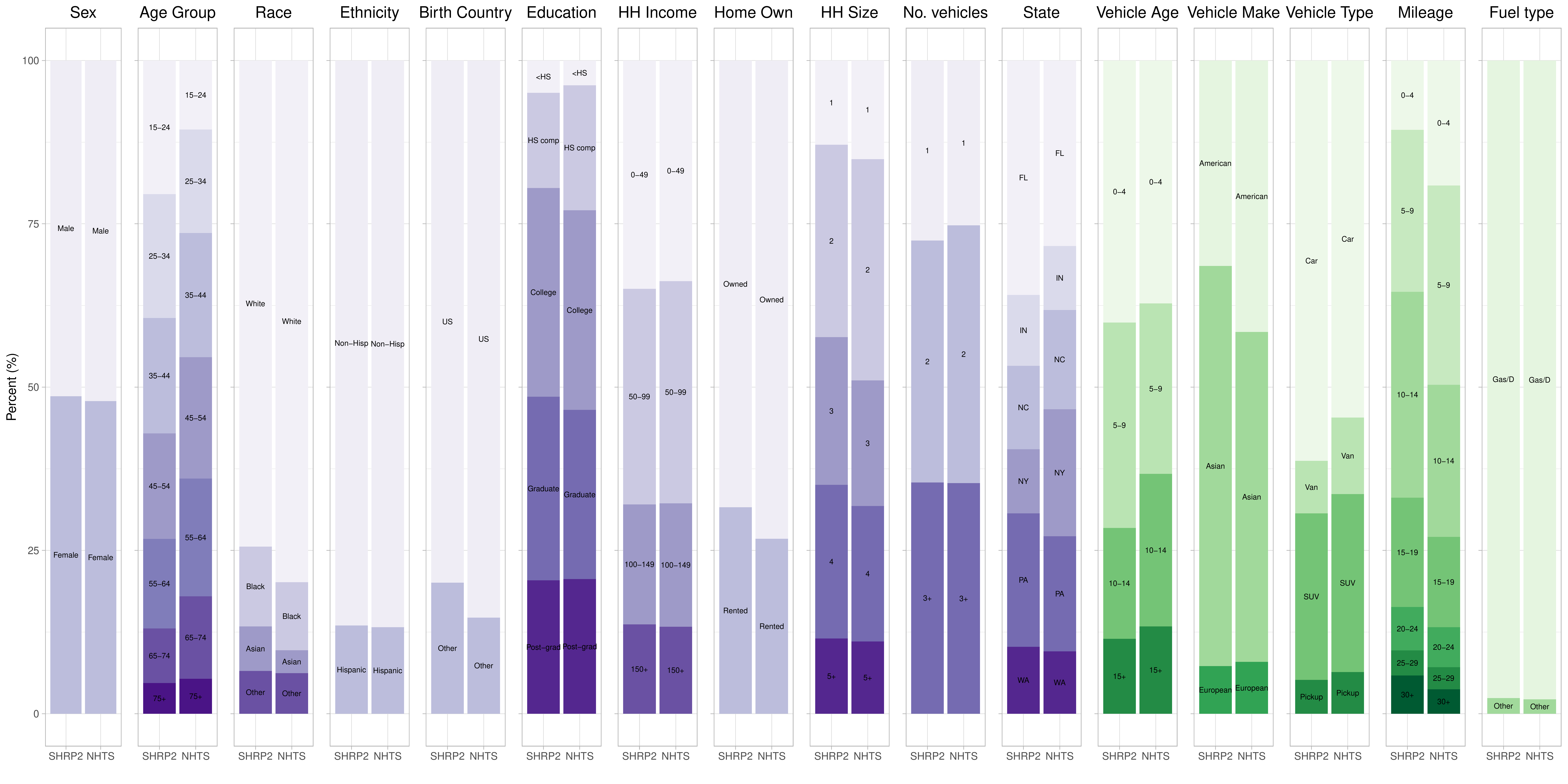}%
 \end{adjustbox}
 \vspace{-25mm}
\end{figure}

Before any attempt for bias adjustment, we check the positivity assumption as well as the existence of influential pseudo-weights. To this end, we estimate the pseudo-selection probabilities for the units of the SHRP2 sample using the PAPP method as well as the PMLE method by \cite{wang2020adjusted}. Figure~\ref{fig:4.9}a compares the distribution of estimated PS in log scale between the SHRP2 and NHTS samples. As illustrated, there is a slight lack of common support in the distribution of PS, which may lead to extreme weights. The box-plot on the right side (Figure~\ref{fig:4.9}b) confirms the presence of outlying pseudo-weights based on the PAPP method. However, it seems no outliers exist in the pseudo-weights based on the PMLE method. Figure~\ref{fig:4.10} compares the distribution of auxiliary variables between the two samples after (pseudo-)weighting. As illustrated, pseudo-weighting obviates most of the previously seen discrepancies in the distribution of common covariates.\par

\begin{figure}[hbt!]
\centering\includegraphics[scale=0.45]{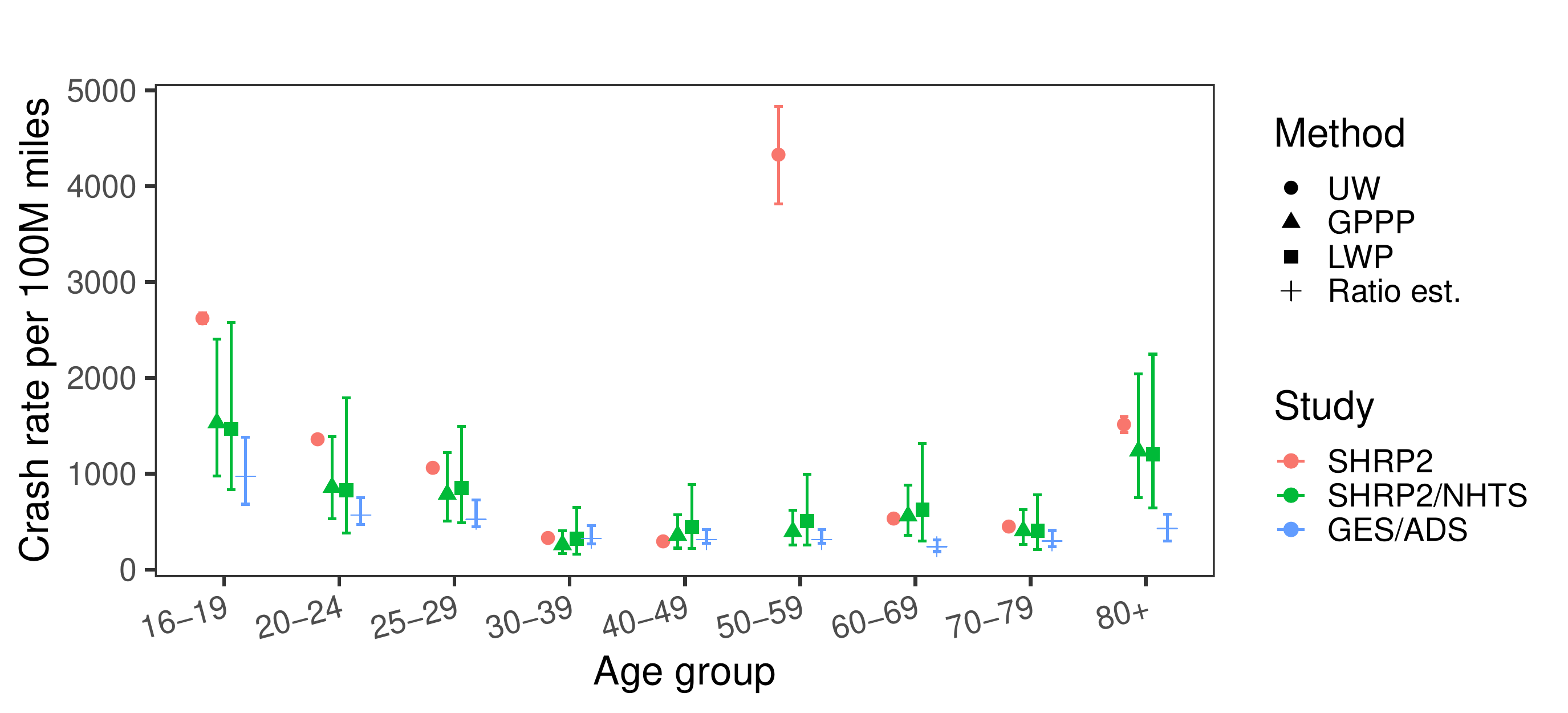}
\caption{Comparing the performance of adjustment methods for estimating crash rates per 100M miles and associated 95\% CIs in SHRP2/NHTS with native estimates and those based on CES/ADS as benchmark across age groups. UW: unweighted; FW: Fully weighted; PAPP: Propensity-adjusted Probability Prediction; GPPP: Gaussian Processes of Propensity Prediction; LWP: Linear-in-weight Prediction.}\label{fig:4.11}
\end{figure}

Figure~\ref{fig:4.11} displays the adjusted estimates of police-reported crash rates per 100M miles driven and associated 95\% CIs using the LWP and GPPP methods by age groups. The plot also compares the adjusted estimates in SHRP2/NHTS data with the naive estimate using SHRP2-only data and that based on the GES/ADS data, which is here considered as the benchmark \cite{tefft2017rates}. Note that the latter represents the entire population of American drivers while our adjusted estimates represent the SHRP2 target population. As illustrated, for most of the age groups, adjustments shift the unweighted crash rates to the true population value, and the associated 95\% CIs overlap, except for the last age group, i.e $80+$ years old. In particular, the unweighted crash rate for the age group 50-59 years seems to be severely biased while adjusted estimates are desirably close to the true population value. While we observe no significant differences in the performance of the GPPP and LWP methods, it is evident that GPPP offers more efficient estimates than the LWP method, as the length of 95\% CIs is consistently lower in GPPP than LWP. Finally, one can infer from Figure~\ref{fig:4.11} that the risk of traffic accidents is higher among young and elder people.\par

\begin{figure}[hbt!]
\centering\includegraphics[scale=0.28]{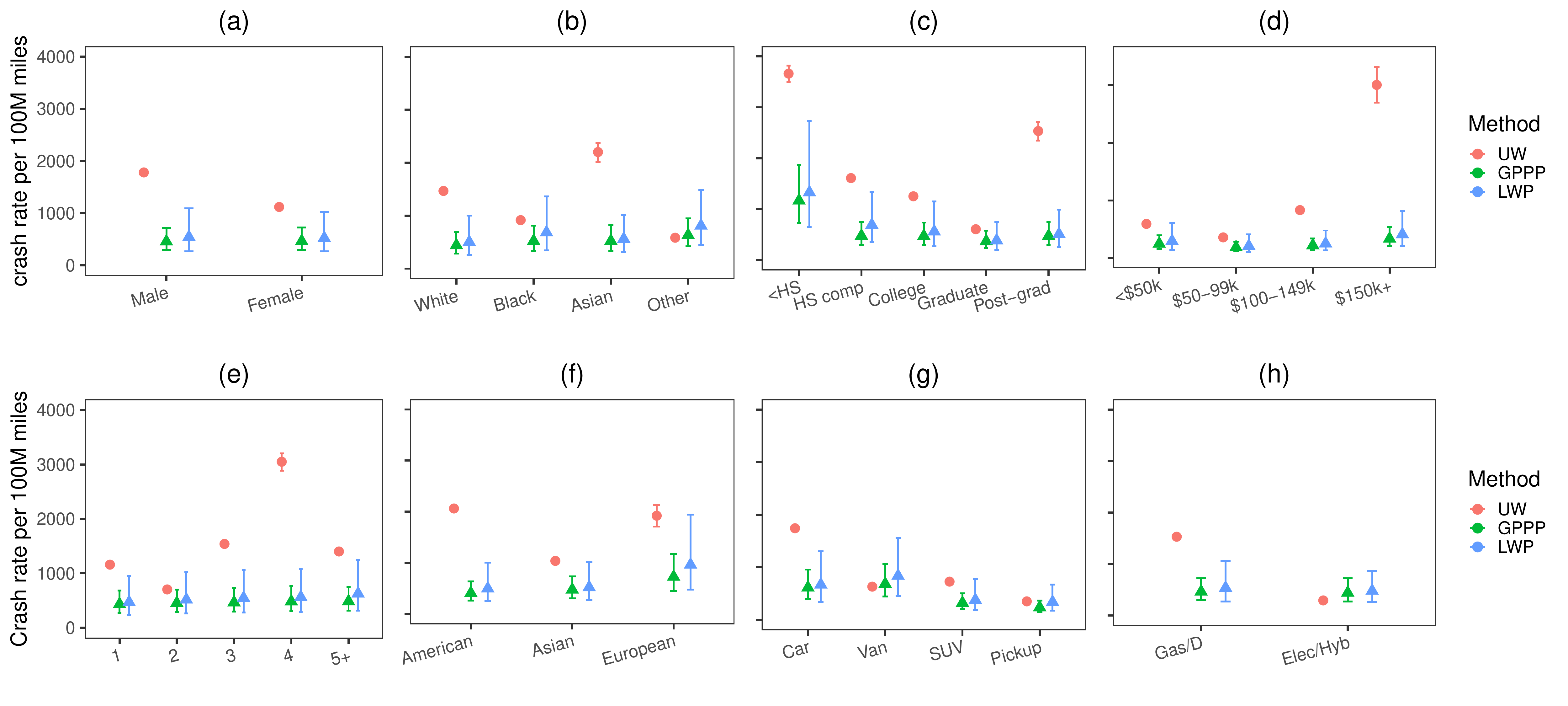}
\caption{Comparing the performance of adjustment methods for estimating crash rates per 100M miles and associated 95\% CIs in SHRP2/NHTS with native estimates across levels of (a) sex, (b) race, (c) education, (d) household income, (e) household size, (f) vehicle make, (g) vehicle type, and (d) fuel type. UW: unweighted; GPPP: Gaussian Process of Propensity Prediction; LWP: Linear-in-weight Prediction.}\label{fig:4.12}
\end{figure}

In Figure~\ref{fig:4.12}, we assess the adjusted rates of police-reportable crashes across levels of auxiliary variables. The major associations we observe are as follows: Whites, more educated drivers, and those in middle-income families are at lower risk of traffic accidents. In addition, there is a positive relationship between the crash risk and household size. There is also evidence of higher crash rates among Vans, European, and gas/diesel vehicles. Numerical values associated with this plot have been provided in Table~\ref{tab:6.20} of Appendix~\ref{S:6.5}.\par

\newpage
\clearpage


\section{Discussion}\label{S:5}
In the present article, we proposed alternative methods for inference based non-probability samples with the goal to fill some of the gaps in the current literature. To our knowledge, this was the first study proposing a fully Bayesian method in a non-probability sample setting that jointly estimates the PS and outcome, and allows one to directly simulate the posterior predictive distribution of the population quantity under a non-probability sample setting. Bayesian approaches provide a unified framework for deriving the variance of the point estimator by simulating the posterior predictive distribution of the population unknown parameters. A \emph{well-calibrated} Bayesian method can appropriately capture all sources of uncertainty, and therefore, yields desirable frequentist repeated sampling properties \citep{dawid1982well}. Unlike the Bayesian two-step methods \citep{kaplan2012two, rafei2021robust}, it is well-understood that joint modeling of the PS and the outcome, as was the case in this article, results in accurate variance estimation \citep{little2004model}.\par


The alternative design-based approaches, such as the AIPW estimator, are sensitive to the presence of influential pseudo-weights if the outcome model is invalid. In addition, the variance estimator proposed by \cite{chen2019doubly} relies on asymptotic theory, and there is no guarantee that simultaneously solving the estimating leads to a unique set of solutions. As another major limitation, such a method works only when the auxiliary variables are identical in the QR and PM models. According to the likelihood we factorized in Eq.~\ref{eq:2.1}, the dimension of the auxiliary variables may vary across the QR and PM methods in a non-probability sample setting ($\{X, D\}$ vs $X$), which makes it impossible to use Chen's AIPW method in practice. On the other hand, employing a fully model-based approach can be extremely expensive computationally, as one has to fit the propensity model on a synthesized population, and predict the outcome variable for its entire non-sampled units \cite{little2007bayesian, mercer2018selection}. However, the method we proposed requires fitting the model only on the combined sample, which makes it computationally more parsimonious, especially in a Big Data setting.\par

The results of our simulation studies reveal that the proposed GPPP method is doubly robust in terms of both point and variance estimates. Furthermore, estimates based on the GPPP method were more efficient than those based on the LWP method especially when outlying pseudo-weights are present and the PM is misspecified. We also showed that the use of GP with a stationary isotropic covariance structure provides a stronger rationale than Spline in the PSPP method as it is equivalent to a non-parametric matching technique based on the estimated PS \citep{huang2019gpmatch}. While Bayesian joint modeling demonstrates good frequentist properties, feedback occurs between the two models \citep{zigler2013model}. This can be controversial in the sense that PS estimates should not be informed by the outcome model \citep{rubin2007design}.\par

It is worth noting that although our proposed method limited computations to the combined samples, Bayesian joint modeling can still turn out computationally very expensive, even using the low-rank approximation techniques. To conduct the simulation studies of this research, we had to hire high-performance computing servers to be able to do parallel processing, but these resources are costly to hire and may not be available for every researcher. Finally, we want to pinpoint that our proposed estimator still contains a design-based term, and therefore, adjusted estimates can still be inefficient if there are extreme values in the sampling weights of $S_R$. However, this should not be of a big concern compared to the pseudo-weights of $S_A$ as $S_R$ is supposed to be a well-designed probability sample. As discussed earlier, a fully model-based estimator will require generating synthetic populations which cannot be fully implemented on the current Bayesian platforms.\par


\section{Acknowledgement}
The present study was part of the doctoral research of the first author of this article at the Michigan Program in Survey and Data Science. Therefore, we would like to thank the respected members of the dissertation committee, Professors Brady T. West, Roderick J. Little, and Philip S. Boonstra at the University of Michigan, who have continuously supported this research with their excellent comments and critical feedback. Our gratitude also goes to Professors Katharine Abraham, Stanley Presser, and Joseph Sedarski at the University of Maryland who have significantly contributed to the development of the main idea of this paper with their valuable comments and feedback over the doctoral seminar course. Last but not least, the authors would like to thank all the researchers and staff who have been involved in collecting the data of SHRP2 and NHTS.

\section{Conflict of Interest} The authors declare that there was no conflict of interest in the current research.


\bibliographystyle{chicago}
\bibliography{arXiv-Rafei-BA-paper}


\setcounter{page}{1}

\section{Appendix}\label{S:6}
\subsection{Gaussian Processes and kernel weighting}\label{S:6.1}
\noindent
Suppose $\hat\pi^A_i$ is the estimated PS for $i\in S_A$ based on a pseudo-weighting approach. Consider the following Gaussian Process (GP) regression model:
\begin{equation}\label{eq:6.1}
y_i = f(u_i)+\epsilon_i
\end{equation}
where $u_i=log(\hat\pi^A_i)$, and $f\sim Gp(0, K)$ with $K(u_i, u_j;\alpha, \rho)=Cov\left(f(u_i), f(u_j)\right)$. From a weight-space viewpoint, one can show that the model~\ref{eq:6.1} predicts $y_i$ for $i\in S_R$ using a weighted sum of observed $y_i$ in $S_A$ as below:
\begin{equation}\label{eq:6.2}
\hat y_i = \sum_{j=1}^{n_A} \tilde w_{ij}y_j
\end{equation}
where
\begin{equation}\label{eq:6.3}
\tilde w_{ij}=\frac{k_{ij}}{\sum_{j=1}^{n_A}k_{ij}} \hspace{5mm} and \hspace{5mm} k_{ij}=k^T(u_j)\Sigma^{-1}
\end{equation}
with $k(\hat\pi^A_j)=k(\hat\pi^A_j, \hat\pi^A_i)_{n_A\times 1}$. According to \cite{huang2019gpmatch}, $\hat y_i$ can be regarded as the Nadaraya-Watson estimator of the observed outcome and selection indicator in the population.\par

Considering an isotropic covariance structure, which is a function of $||u_j - u_i||$, $k_{ij}$ quite resembles the kernel weights \cite{wang2020improving}, with the bandwidth $h$ equivalent to the GP length-scale parameter $\rho$. Since the kernel weights obtained by GP is used in the PM estimator, it is clear that the final weights will be multiplied by $w^R$, i.e.
\begin{equation}\label{eq:6.4}
\hat w_j=\sum_{i=1}^{n_R}k_{ij}w^R_i
\end{equation}
One can show that the major kernel-related condition determined by \cite{wang2020improving} to obtain consistency in the kernel-weighted estimates holds for a Mat\'ern family covariance structure, i.e. $K(u)$, $\int K(u)du=1$, $Sup_u|K(u)|<\infty$, and $lim_{|u|\rightarrow\infty}|u||K(u)|=0$.


\subsection{Further extensions of the simulation study}\label{S:6.4}

\subsubsection{Simulation study I}\label{S:6.4.1}
\noindent
This subsection provides additional results associated with Simulation I. Table~\ref{tab:6.2} and Table~\ref{tab:6.3} summarize the findings of the simulation in Section~\ref{S:3.1} for $\rho=0.5$ and $\rho=0.3$, respectively.

\begin{sidewaystable}[hbt!]
\centering
\caption{Comparing the performance of the bias adjustment methods in the first simulation study for $\rho=0.5$}\label{tab:6.2}
\begin{threeparttable}
\scriptsize{\begin{tabular}{l l l l l l l l l l l l l l l l l l l}
\toprule
& \multicolumn{5}{c}{\textbf{$n_A=500, \hspace{2mm} n_R=500$}} & & \multicolumn{5}{c}{\textbf{$n_A=1,000, \hspace{2mm} n_R=500$}} & & \multicolumn{5}{c}{\textbf{$n_A=500, \hspace{2mm} n_R=1,000$}}\\\cline{2-6}\cline{8-12}\cline{14-18}
\textbf{Measure} & rBias & rMSE & crCI & lCI & rSE & & rBias & rMSE & crCI & lCI & rSE & & rBias & rMSE & crCI & lCI & rSE \\
\midrule
\multicolumn{10}{l}{\textbf{Probability sample ($S_R$)}}  &    &    &  &  \\
\hline
\hspace{2mm} UW & 8.866 & 9.445 & 21.759 & 1.149 & 0.965 & & 8.866 & 9.445 & 21.759 & 1.149 & 0.965 & & 8.810 & 9.115 & 2.778 & 0.816 & 0.954\\
\hspace{2mm} FW & 0.260 & 3.770 & 93.056 & 1.363 & 0.991 & & 0.260 & 3.770 & 93.056 & 1.363 & 0.991 & & 0.080 & 2.736 & 95.833 & 0.967 & 0.967\\
\hline
\multicolumn{10}{l}{\textbf{Non-probability sample ($S_A$)}} & & & & & \\
\hline
\hspace{2mm} UW & 30.708 & 30.926 & 0.000 & 1.304 & 0.974 & & 30.008 & 30.104 & 0.000 & 0.916 & 1.039 & & 30.708 & 30.926 & 0.000 & 1.304 & 0.974\\
\hspace{2mm} FW & -0.134 & 3.763 & 93.981 & 1.363 & 0.991 & & -0.146 & 2.659 & 95.833 & 0.957 & 0.986 & & -0.134 & 3.763 & 93.981 & 1.363 & 0.991\\
\hline
\multicolumn{10}{l}{Model specification: QR--True, PM--True} & & & & & \\
\hline
\hspace{2mm} GPPP & 0.129 & 3.876 & 99.537 & 1.912 & 1.350 & & 0.000 & 3.055 & 99.537 & 1.724 & 1.544 & & 0.092 & 3.567 & 98.148 & 1.576 & 1.209\\
\hspace{2mm} LWP & -0.063 & 4.200 & 99.537 & 2.516 & 1.638 & & -0.145 & 3.140 & 99.537 & 1.737 & 1.514 & & -0.135 & 3.742 & 97.685 & 1.613 & 1.179\\
\hspace{2mm} AIPW & -0.101 & 4.035 & 93.981 & 1.406 & 0.953 & & -0.098 & 3.139 & 94.907 & 1.107 & 0.965 & & -0.232 & 3.760 & 94.907 & 1.311 & 0.955\\
\hspace{2mm} PAPP & 0.832 & 4.012 & 93.056 & 1.406 & 0.980 & & 0.547 & 2.998 & 93.981 & 1.078 & 1.000 & & 0.984 & 3.844 & 95.370 & 1.339 & 0.985\\
\hline
\multicolumn{10}{l}{Model specification: QR--True, PM--False} & & & & & \\
\hline
\hspace{2mm} GPPP & 0.146 & 3.829 & 99.537 & 1.903 & 1.360 & & -0.017 & 3.068 & 99.537 & 1.720 & 1.533 & & 0.161 & 3.555 & 98.148 & 1.575 & 1.213\\
\hspace{2mm} LWP & -0.007 & 3.876 & 99.537 & 1.910 & 1.348 & & -0.124 & 3.084 & 99.537 & 1.730 & 1.535 & & -0.015 & 3.575 & 98.148 & 1.585 & 1.213\\
\hspace{2mm} AIPW & -0.056 & 3.916 & 94.907 & 1.358 & 0.948 & & -0.087 & 3.061 & 95.833 & 1.070 & 0.957 & & -0.180 & 3.590 & 93.981 & 1.272 & 0.970\\
\hline
\multicolumn{10}{l}{Model specification: QR--False, PM--True} & & & & & \\
\hline
\hspace{2mm} GPPP & 1.317 & 4.321 & 99.074 & 1.934 & 1.285 & & 1.146 & 3.477 & 99.537 & 1.748 & 1.456 & & 1.314 & 4.001 & 96.296 & 1.607 & 1.163\\
\hspace{2mm} LWP & 4.052 & 6.752 & 89.815 & 2.236 & 1.132 & & 4.147 & 6.026 & 92.130 & 2.004 & 1.253 & & 4.090 & 6.299 & 87.037 & 1.850 & 1.056\\
\hspace{2mm} AIPW & 0.040 & 3.987 & 95.833 & 1.392 & 0.954 & & -0.010 & 3.025 & 94.444 & 1.074 & 0.971 & & 0.073 & 3.716 & 93.519 & 1.338 & 0.984\\
\hline
\multicolumn{10}{l}{Model specification: QR--False, PM--False} & & & & & \\
\hline
\hspace{2mm} GPPP & 27.400 & 27.678 & 0.000 & 2.239 & 1.564 & & 26.595 & 26.733 & 0.000 & 2.042 & 2.055 & & 27.430 & 27.692 & 0.000 & 1.814 & 1.306\\
\hspace{2mm} LWP & 27.075 & 27.361 & 0.000 & 2.247 & 1.558 & & 26.434 & 26.575 & 0.000 & 2.056 & 2.053 & & 27.127 & 27.395 & 0.000 & 1.837 & 1.314\\
\hspace{2mm} AIPW & 27.115 & 27.402 & 0.000 & 1.366 & 0.944 & & 26.432 & 26.571 & 0.000 & 0.981 & 0.987 & & 27.056 & 27.328 & 0.000 & 1.334 & 0.949\\
\hspace{2mm} PAPP & 27.912 & 28.191 & 0.000 & 1.354 & 0.936 & & 27.024 & 27.158 & 0.000 & 0.968 & 0.983 & & 28.121 & 28.381 & 0.000 & 1.329 & 0.950\\
\bottomrule
\end{tabular}}
 \begin{tablenotes}
 \footnotesize
 \item GPPP: Gaussian Process of Propensity Prediction; LWP: Linear-in-weight prediction; AIPW: Augmented Inverse Propensity Weighting; PAPP: Propensity-adjusted Probability Prediction\\
 NOTE: The PAPP and AIPW methods have been implemented through a bootstrap method. 
 \end{tablenotes}
\end{threeparttable}
\end{sidewaystable}

\begin{sidewaystable}[hbt!]
\centering
\caption{Comparing the performance of the bias adjustment methods in the first simulation study for $\rho=0.3$}\label{tab:6.3}
\begin{threeparttable}
\scriptsize{\begin{tabular}{l l l l l l l l l l l l l l l l l l l}
\toprule
& \multicolumn{5}{c}{\textbf{$n_A=500, \hspace{2mm} n_R=500$}} & & \multicolumn{5}{c}{\textbf{$n_A=1,000, \hspace{2mm} n_R=500$}} & & \multicolumn{5}{c}{\textbf{$n_A=500, \hspace{2mm} n_R=1,000$}}\\\cline{2-6}\cline{8-12}\cline{14-18}
\textbf{Measure} & rBias & rMSE & crCI & lCI & rSE & & rBias & rMSE & crCI & lCI & rSE & & rBias & rMSE & crCI & lCI & rSE \\
\midrule
\multicolumn{10}{l}{\textbf{Probability sample ($S_R$)}}  &    &    &  &  \\
\hline
\hspace{2mm} UW & 8.867 & 10.406 & 58.796 & 1.927 & 0.965 & & 8.867 & 10.406 & 58.796 & 1.927 & 0.965 & & 8.797 & 9.612 & 38.889 & 1.369 & 0.964\\
\hspace{2mm} FW & 0.424 & 6.361 & 92.593 & 2.303 & 0.989 & & 0.424 & 6.361 & 92.593 & 2.303 & 0.989 & & 0.154 & 4.580 & 95.370 & 1.633 & 0.973\\
\hline
\multicolumn{10}{l}{\textbf{Non-probability sample ($S_A$)}} & & & & & \\
\hline
\hspace{2mm} UW & 30.759 & 31.251 & 0.000 & 2.037 & 1.006 & & 30.082 & 30.322 & 0.000 & 1.435 & 1.028 & & 30.759 & 31.251 & 0.000 & 2.037 & 1.006\\
\hspace{2mm} FW & -0.277 & 6.284 & 95.833 & 2.335 & 1.014 & & -0.298 & 4.529 & 93.519 & 1.639 & 0.989 & & -0.277 & 6.284 & 95.833 & 2.335 & 1.014\\
\hline
\multicolumn{10}{l}{Model specification: QR--True, PM--True} & & & & & \\
\hline
\hspace{2mm} GPPP & 0.438 & 6.297 & 100.000 & 3.413 & 1.482 & & 0.061 & 4.754 & 99.537 & 3.094 & 1.775 & & 0.489 & 6.052 & 99.074 & 2.821 & 1.275\\
\hspace{2mm} LWP & -0.342 & 6.811 & 100.000 & 4.061 & 1.628 & & -0.266 & 4.917 & 99.537 & 3.131 & 1.739 & & -0.232 & 6.429 & 98.148 & 2.932 & 1.244\\
\hspace{2mm} AIPW & -0.390 & 6.666 & 95.370 & 2.378 & 0.975 & & -0.296 & 4.911 & 94.444 & 1.737 & 0.966 & & -0.395 & 6.337 & 95.370 & 2.328 & 1.004\\
\hspace{2mm} PAPP & 0.574 & 6.537 & 95.370 & 2.366 & 0.991 & & 0.360 & 4.696 & 95.370 & 1.709 & 0.995 & & 0.873 & 6.291 & 94.907 & 2.334 & 1.022\\
\hline
\multicolumn{10}{l}{Model specification: QR--True, PM--False} & & & & & \\
\hline
\hspace{2mm} GPPP & 0.553 & 6.241 & 100.000 & 3.391 & 1.488 & & 0.077 & 4.733 & 99.537 & 3.080 & 1.775 & & 0.585 & 6.033 & 98.611 & 2.817 & 1.280\\
\hspace{2mm} LWP & -0.115 & 6.302 & 100.000 & 3.464 & 1.499 & & -0.255 & 4.748 & 99.537 & 3.105 & 1.786 & & -0.020 & 6.114 & 99.074 & 2.879 & 1.284\\
\hspace{2mm} AIPW & -0.271 & 6.301 & 95.370 & 2.312 & 1.002 & & -0.275 & 4.789 & 94.444 & 1.688 & 0.963 & & -0.287 & 6.042 & 94.907 & 2.233 & 1.009\\
\hline
\multicolumn{10}{l}{Model specification: QR--False, PM--True} & & & & & \\
\hline
\hspace{2mm} GPPP & 1.891 & 6.750 & 99.537 & 3.409 & 1.435 & & 1.372 & 5.092 & 99.537 & 3.097 & 1.722 & & 1.981 & 6.536 & 97.685 & 2.837 & 1.242\\
\hspace{2mm} LWP & 4.179 & 8.638 & 98.611 & 3.692 & 1.332 & & 4.373 & 7.326 & 99.537 & 3.309 & 1.535 & & 4.277 & 8.280 & 95.833 & 3.098 & 1.192\\
\hspace{2mm} AIPW & -0.146 & 6.594 & 94.907 & 2.377 & 0.983 & & -0.187 & 4.794 & 94.444 & 1.708 & 0.973 & & -0.125 & 6.319 & 95.370 & 2.333 & 1.007\\
\hline
\multicolumn{10}{l}{Model specification: QR--False, PM--False} & & & & & \\
\hline
\hspace{2mm} GPPP & 27.642 & 28.234 & 5.093 & 3.483 & 1.652 & & 26.734 & 27.051 & 2.315 & 3.212 & 2.121 & & 27.792 & 28.364 & 0.463 & 2.792 & 1.344\\
\hspace{2mm} LWP & 26.915 & 27.545 & 11.111 & 3.542 & 1.649 & & 26.278 & 26.601 & 2.315 & 3.242 & 2.142 & & 27.022 & 27.636 & 1.852 & 2.900 & 1.366\\
\hspace{2mm} AIPW & 26.987 & 27.603 & 0.000 & 2.120 & 0.997 & & 26.405 & 26.734 & 0.000 & 1.528 & 0.996 & & 27.022 & 27.630 & 0.000 & 2.076 & 0.982\\
\hspace{2mm} PAPP & 27.881 & 28.473 & 0.000 & 2.122 & 1.002 & & 26.994 & 27.300 & 0.000 & 1.501 & 1.004 & & 28.146 & 28.716 & 0.000 & 2.104 & 1.008\\
\bottomrule
\end{tabular}}
 \begin{tablenotes}
 \footnotesize
 \item GPPP: Gaussian Process of Propensity Prediction; LWP: Linear-in-weight prediction; AIPW: Augmented Inverse Propensity Weighting; PAPP: Propensity-adjusted Probability Prediction\\
 NOTE: The PAPP and AIPW methods have been implemented through a bootstrap method. 
 \end{tablenotes}
\end{threeparttable}
\end{sidewaystable}

\newpage
\clearpage

\newpage

\begin{sidewaystable}[hbt!]
\subsubsection{Simulation study II}\label{S:6.4.2}

Table~\ref{tab:6.4} exhibits the numerical results associated with Figure~\ref{fig:2} and Figure~\ref{fig:3}.
\caption{Comparing the performance of the bias adjustment methods in the second simulation study for the \emph{continuous} outcome with $(n_A, n_R)=(500, 1,000)$ and $\gamma_1=0.3$}\label{tab:6.4}
\begin{threeparttable}
\scriptsize{\begin{tabular}{l l l l l l l l l l l l l l l l l l l l l l l l l}
\toprule
& \multicolumn{5}{c}{\textbf{$LIN$}} & & \multicolumn{5}{c}{\textbf{$CUB$}} & & \multicolumn{5}{c}{\textbf{$EXP$}} & & \multicolumn{5}{c}{\textbf{$SIN$}}\\\cline{2-6}\cline{8-12}\cline{14-18}\cline{20-24}
\textbf{Measure} & rBias & rMSE & crCI & lCI & rSE & & rBias & rMSE & crCI & lCI & rSE & & rBias & rMSE & crCI & lCI & rSE & & rBias & rMSE & crCI & lCI & rSE \\
\midrule
\multicolumn{24}{l}{\textbf{Probability sample ($S_R$)}}\\ 
\hline 
\hspace{2mm} UW & -23.870 & 24.078 & 0.000 & 12.411 & 1.001 & & -17.120 & 17.243 & 0.000 & 8.442 & 1.045 & & -12.097 & 12.179 & 0.000 & 5.427 & 0.976 & & -18.475 & 19.159 & 3.241 & 18.535 & 0.93\\
\hspace{2mm} FW & -0.180 & 3.921 & 93.519 & 15.087 & 0.980 & & -0.156 & 2.749 & 94.907 & 10.799 & 1.001 & & -0.091 & 1.947 & 94.444 & 7.392 & 0.967 & & 0.015 & 5.549 & 93.519 & 20.503 & 0.94\\
\hline 
\multicolumn{24}{l}{\textbf{Non-probability sample ($S_A$)}}\\ 
\hline 
\hspace{2mm} UW & 48.795 & 49.086 & 0.000 & 19.644 & 0.936 & & 30.073 & 30.366 & 0.000 & 15.989 & 0.966 & & 18.146 & 18.363 & 0.000 & 10.710 & 0.969 & & 27.948 & 28.716 & 0.926 & 25.807 & 0.995\\
\hspace{2mm} FW & -0.178 & 6.184 & 91.204 & 22.458 & 0.925 & & -0.218 & 4.249 & 92.593 & 15.612 & 0.936 & & 0.004 & 2.282 & 93.981 & 8.639 & 0.964 & & -0.034 & 8.276 & 93.519 & 31.425 & 0.966\\
\hline 
\multicolumn{24}{l}{\textbf{Non-robust method}}\\ 
\hline 
\multicolumn{24}{l}{Model specification: QR--True}\\ 
\hline 
\hspace{2mm} PAPP & 0.066 & 4.771 & 95.370 & 19.982 & 1.066 & & -7.545 & 24.919 & 91.667 & 53.569 & 0.574 & & 2.269 & 3.182 & 83.333 & 8.794 & 1.003 & & 11.642 & 13.063 & 46.759 & 23.010 & 0.988\\
\hline 
\multicolumn{24}{l}{Model specification: QR--False}\\ 
\hline 
\hspace{2mm} PAPP & 35.346 & 35.660 & 0.000 & 17.238 & 0.929 & & 17.490 & 17.760 & 0.000 & 11.878 & 0.980 & & 10.111 & 10.362 & 0.000 & 8.707 & 0.977 & & 30.831 & 31.610 & 0.463 & 26.668 & 0.973\\
\hline 
\multicolumn{24}{l}{\textbf{Doubly robust methods}}\\ 
\hline 
\multicolumn{24}{l}{Model specification: QR--True, PM--True}\\ 
\hline 
\hspace{2mm} GPPP & 0.025 & 4.439 & 98.148 & 18.703 & 1.081 & & -0.060 & 3.150 & 96.759 & 13.124 & 1.067 & & -0.027 & 2.164 & 97.222 & 8.823 & 1.048 & & 0.301 & 6.250 & 96.296 & 26.369 & 1.084\\
\hspace{2mm} LWP & -0.038 & 4.478 & 97.222 & 18.960 & 1.088 & & -0.005 & 3.133 & 98.148 & 13.227 & 1.084 & & -2.254 & 4.458 & 95.833 & 9.065 & 0.080 & & 0.299 & 6.335 & 96.296 & 26.621 & 1.076\\
\hspace{2mm} AIPW & 0.000 & 4.439 & 97.222 & 18.266 & 1.047 & & 0.250 & 4.060 & 98.148 & 16.137 & 1.014 & & -0.022 & 2.194 & 93.981 & 8.482 & 0.984 & & 0.157 & 6.179 & 94.444 & 23.568 & 0.971\\
\hline 
\multicolumn{24}{l}{Model specification: QR--True, PM--False}\\ 
\hline 
\hspace{2mm} GPPP & 0.099 & 4.467 & 97.685 & 18.732 & 1.078 & & -0.061 & 3.136 & 97.222 & 12.796 & 1.047 & & 0.005 & 2.140 & 96.296 & 8.654 & 1.037 & & 0.439 & 6.189 & 95.833 & 25.041 & 1.043\\
\hspace{2mm} LWP & 10.099 & 4.776 & 91.667 & 17.809 & 0.100 & & 7.636 & 3.382 & 88.889 & 12.078 & 0.087 & & 5.845 & 17.130 & 89.815 & 8.171 & 0.078 & & 7.886 & 14.740 & 92.130 & 24.236 & 0.175\\
\hspace{2mm} AIPW & -0.033 & 4.608 & 94.907 & 18.233 & 1.007 & & -0.109 & 3.198 & 95.833 & 12.461 & 0.992 & & -0.048 & 2.188 & 95.370 & 8.400 & 0.977 & & 0.245 & 6.221 & 94.907 & 23.752 & 0.972\\
\hline 
\multicolumn{24}{l}{Model specification: QR--False, PM--True}\\ 
\hline 
\hspace{2mm} GPPP & 0.523 & 4.628 & 97.685 & 20.143 & 1.128 & & 1.310 & 3.829 & 95.370 & 15.280 & 1.093 & & 0.575 & 2.531 & 96.296 & 11.294 & 1.175 & & 0.819 & 6.523 & 95.833 & 27.890 & 1.108\\
\hspace{2mm} LWP & -0.294 & 4.923 & 96.296 & 20.356 & 1.064 & & 0.231 & 6.026 & 93.981 & 15.853 & 0.677 & & 0.509 & 5.817 & 96.759 & 11.564 & 0.512 & & -1.060 & 14.048 & 92.130 & 37.060 & 0.679\\
\hspace{2mm} AIPW & 0.546 & 6.979 & 94.444 & 29.210 & 1.069 & & -15.872 & 46.090 & 87.500 & 102.060 & 0.600 & & 6.885 & 7.391 & 26.389 & 10.549 & 0.999 & & 14.490 & 15.884 & 35.648 & 24.548 & 0.96\\
\hline 
\multicolumn{24}{l}{Model specification: QR--False, PM--False}\\ 
\hline 
\hspace{2mm} GPPP & 36.369 & 36.676 & 0.000 & 22.484 & 1.217 & & 18.222 & 18.524 & 0.000 & 15.525 & 1.195 & & 10.492 & 10.758 & 0.926 & 10.941 & 1.182 & & 30.956 & 31.750 & 3.241 & 33.560 & 1.222\\
\hspace{2mm} LWP & 36.049 & 36.361 & 0.000 & 22.546 & 1.219 & & 18.396 & 18.690 & 0.000 & 15.611 & 1.213 & & 10.590 & 10.856 & 0.926 & 11.005 & 1.185 & & 30.650 & 31.450 & 2.778 & 35.040 & 1.276\\
\hspace{2mm} AIPW & 36.146 & 36.468 & 0.000 & 17.448 & 0.919 & & 18.293 & 18.583 & 0.000 & 12.436 & 0.967 & & 10.587 & 10.848 & 0.000 & 8.968 & 0.965 & & 30.664 & 31.483 & 0.000 & 26.807 & 0.956\\
\bottomrule
\end{tabular}}
 \begin{tablenotes}
 \footnotesize
 \item UW: Unweighted; FW: Fully weighted; GPPP: Gaussian Process of Propensity Prediction; LWP: Linear-in-weight prediction; AIPW: Augmented Inverse Propensity Weighting.\\
 NOTE: The PAPP and AIPW methods have been implemented through a bootstrap method. 
 \end{tablenotes}
\end{threeparttable}
\end{sidewaystable}

\begin{sidewaystable}[hbt!]
\centering
\caption{Comparing the performance of the bias adjustment methods in the second simulation study for the \emph{continuous} outcome with $(n_A, n_R)=(1,000, 500)$ and $\gamma_1=0.3$}\label{tab:6.5}
\begin{threeparttable}
\scriptsize{\begin{tabular}{l l l l l l l l l l l l l l l l l l l l l l l l l}
\toprule
& \multicolumn{5}{c}{\textbf{$LIN$}} & & \multicolumn{5}{c}{\textbf{$CUB$}} & & \multicolumn{5}{c}{\textbf{$EXP$}} & & \multicolumn{5}{c}{\textbf{$SIN$}}\\\cline{2-6}\cline{8-12}\cline{14-18}\cline{20-24}
\textbf{Measure} & rBias & rMSE & crCI & lCI & rSE & & rBias & rMSE & crCI & lCI & rSE & & rBias & rMSE & crCI & lCI & rSE & & rBias & rMSE & crCI & lCI & rSE \\
\midrule
\multicolumn{24}{l}{\textbf{Probability sample ($S_R$)}}\\ 
\hline 
\hspace{2mm} UW & -23.868 & 24.297 & 0.000 & 17.587 & 0.985 & & -17.158 & 17.440 & 0.000 & 11.959 & 0.974 & & -12.107 & 12.274 & 0.000 & 7.683 & 0.967 & & -18.706 & 19.966 & 22.222 & 26.219 & 0.956\\
\hspace{2mm} FW & 0.118 & 5.477 & 93.519 & 21.395 & 0.994 & & 0.016 & 3.998 & 94.444 & 15.257 & 0.971 & & 0.015 & 2.673 & 95.833 & 10.454 & 0.995 & & -0.081 & 7.162 & 96.296 & 29.064 & 1.033\\
\hline 
\multicolumn{24}{l}{\textbf{Non-probability sample ($S_A$)}}\\ 
\hline 
\hspace{2mm} UW & 48.581 & 48.727 & 0.000 & 13.878 & 0.935 & & 29.971 & 30.129 & 0.000 & 11.296 & 0.933 & & 18.042 & 18.159 & 0.000 & 7.569 & 0.935 & & 27.647 & 28.037 & 0.000 & 18.285 & 0.999\\
\hspace{2mm} FW & 0.056 & 4.211 & 95.833 & 15.872 & 0.959 & & 0.107 & 2.862 & 94.907 & 10.960 & 0.975 & & 0.055 & 1.586 & 96.296 & 6.127 & 0.984 & & -0.297 & 5.417 & 95.833 & 22.208 & 1.045\\
\hline 
\multicolumn{24}{l}{\textbf{Non-robust method}}\\ 
\hline 
\multicolumn{24}{l}{Model specification: QR--True}\\ 
\hline 
\hspace{2mm} PAPP & 0.504 & 4.495 & 96.296 & 18.667 & 1.064 & & -16.241 & 32.851 & 98.611 & 74.882 & 0.667 & & 0.403 & 1.995 & 95.833 & 7.859 & 1.024 & & 11.569 & 13.086 & 53.241 & 24.302 & 1.011\\
\hline 
\multicolumn{24}{l}{Model specification: QR--False}\\ 
\hline 
\hspace{2mm} PAPP & 35.317 & 35.521 & 0.000 & 14.654 & 0.982 & & 17.616 & 17.838 & 0.000 & 10.819 & 0.982 & & 10.136 & 10.324 & 0.000 & 7.535 & 0.977 & & 30.431 & 30.802 & 0.000 & 18.895 & 1.008\\
\hline 
\multicolumn{24}{l}{\textbf{Doubly robust methods}}\\ 
\hline 
\multicolumn{24}{l}{Model specification: QR--True, PM--True}\\ 
\hline 
\hspace{2mm} GPPP & 0.326 & 4.758 & 98.611 & 21.123 & 1.142 & & 0.169 & 3.505 & 96.759 & 14.875 & 1.091 & & 0.120 & 2.338 & 96.759 & 10.083 & 1.108 & & 0.291 & 6.396 & 96.759 & 29.612 & 1.186\\
\hspace{2mm} LWP & 0.285 & 4.746 & 98.611 & 21.279 & 1.152 & & -1.313 & 3.504 & 96.759 & 16.475 & 0.196 & & -0.655 & 8.930 & 96.296 & 10.952 & 0.316 & & 0.246 & 6.403 & 97.685 & 29.663 & 1.19\\
\hspace{2mm} AIPW & 0.287 & 4.808 & 97.222 & 20.161 & 1.069 & & 0.767 & 5.589 & 97.222 & 20.610 & 0.947 & & 0.140 & 2.361 & 96.296 & 9.838 & 1.062 & & 0.292 & 6.293 & 95.370 & 26.295 & 1.065\\
\hline 
\multicolumn{24}{l}{Model specification: QR--True, PM--False}\\ 
\hline 
\hspace{2mm} GPPP & 0.397 & 4.731 & 97.685 & 21.125 & 1.151 & & 0.123 & 3.530 & 97.222 & 14.813 & 1.072 & & 0.096 & 2.337 & 98.148 & 10.056 & 1.097 & & 0.340 & 6.358 & 97.685 & 29.112 & 1.175\\
\hspace{2mm} LWP & 21.964 & 5.322 & 84.259 & 19.456 & 0.077 & & 16.925 & 4.571 & 81.019 & 13.776 & 0.071 & & 12.837 & 31.570 & 78.241 & 9.029 & 0.061 & & 17.658 & 14.123 & 85.648 & 26.438 & 0.131\\
\hspace{2mm} AIPW & 0.430 & 4.877 & 97.685 & 20.009 & 1.048 & & 0.153 & 3.531 & 95.833 & 14.218 & 1.026 & & 0.108 & 2.359 & 96.759 & 9.660 & 1.043 & & 0.357 & 6.324 & 95.370 & 26.422 & 1.065\\
\hline 
\multicolumn{24}{l}{Model specification: QR--False, PM--True}\\ 
\hline 
\hspace{2mm} GPPP & 0.679 & 4.644 & 98.611 & 22.199 & 1.242 & & 1.028 & 3.562 & 98.611 & 16.651 & 1.251 & & 0.538 & 2.275 & 99.074 & 12.101 & 1.401 & & 0.734 & 6.435 & 98.148 & 30.583 & 1.228\\
\hspace{2mm} LWP & 0.107 & 4.712 & 99.074 & 22.396 & 1.223 & & -1.191 & 3.593 & 97.222 & 17.681 & 0.335 & & 0.118 & 2.437 & 99.537 & 12.407 & 1.310 & & -2.162 & 8.254 & 96.759 & 38.983 & 1.248\\
\hspace{2mm} AIPW & 1.102 & 6.334 & 95.833 & 25.788 & 1.052 & & -33.598 & 67.410 & 95.833 & 153.677 & 0.669 & & 6.226 & 6.774 & 39.815 & 10.775 & 1.028 & & 14.128 & 15.372 & 36.111 & 24.900 & 1.046\\
\hline 
\multicolumn{24}{l}{Model specification: QR--False, PM--False}\\ 
\hline 
\hspace{2mm} GPPP & 36.360 & 36.574 & 0.000 & 24.865 & 1.615 & & 18.538 & 18.801 & 0.000 & 17.376 & 1.427 & & 10.648 & 10.868 & 0.463 & 12.232 & 1.443 & & 30.417 & 30.823 & 1.852 & 36.965 & 1.893\\
\hspace{2mm} LWP & 36.137 & 36.354 & 0.000 & 24.980 & 1.612 & & 18.636 & 18.902 & 0.000 & 17.596 & 1.426 & & 10.702 & 10.920 & 0.926 & 12.309 & 1.452 & & 30.259 & 30.656 & 0.926 & 38.086 & 1.988\\
\hspace{2mm} AIPW & 36.248 & 36.471 & 0.000 & 15.439 & 0.976 & & 18.546 & 18.800 & 0.000 & 12.121 & 1.003 & & 10.730 & 10.951 & 0.000 & 8.629 & 1.005 & & 30.357 & 30.749 & 0.000 & 19.215 & 0.999\\
\bottomrule
\end{tabular}}
 \begin{tablenotes}
 \footnotesize
 \item UW: Unweighted; FW: Fully weighted; GPPP: Gaussian Process of Propensity Prediction; LWP: Linear-in-weight Prediction; AIPW: Augmented Inverse Propensity Weighting.\\
 NOTE: The PAPP and AIPW methods have been implemented through a bootstrap method. 
 \end{tablenotes}
\end{threeparttable}
\end{sidewaystable}

\begin{sidewaystable}[hbt!]
\centering
\caption{Comparing the performance of the bias adjustment methods in the second simulation study for the \emph{continuous} outcome with $(n_A, n_R)=(500, 500)$ and $\gamma_1=0.3$}\label{tab:6.6}
\begin{threeparttable}
\scriptsize{\begin{tabular}{l l l l l l l l l l l l l l l l l l l l l l l l l}
\toprule
& \multicolumn{5}{c}{\textbf{$LIN$}} & & \multicolumn{5}{c}{\textbf{$CUB$}} & & \multicolumn{5}{c}{\textbf{$EXP$}} & & \multicolumn{5}{c}{\textbf{$SIN$}}\\\cline{2-6}\cline{8-12}\cline{14-18}\cline{20-24}
\textbf{Measure} & rBias & rMSE & crCI & lCI & rSE & & rBias & rMSE & crCI & lCI & rSE & & rBias & rMSE & crCI & lCI & rSE & & rBias & rMSE & crCI & lCI & rSE \\
\midrule
\multicolumn{24}{l}{\textbf{Probability sample ($S_R$)}}\\ 
\hline 
\hspace{2mm} UW & -23.868 & 24.297 & 0.000 & 17.587 & 0.985 & & -17.158 & 17.440 & 0.000 & 11.959 & 0.974 & & -12.107 & 12.274 & 0.000 & 7.683 & 0.967 & & -18.706 & 19.966 & 22.222 & 26.219 & 0.956\\
\hspace{2mm} FW & 0.118 & 5.477 & 93.519 & 21.395 & 0.994 & & 0.016 & 3.998 & 94.444 & 15.257 & 0.971 & & 0.015 & 2.673 & 95.833 & 10.454 & 0.995 & & -0.081 & 7.162 & 96.296 & 29.064 & 1.033\\
\hline 
\multicolumn{24}{l}{\textbf{Non-probability sample ($S_A$)}}\\ 
\hline 
\hspace{2mm} UW & 48.795 & 49.086 & 0.000 & 19.644 & 0.936 & & 30.073 & 30.366 & 0.000 & 15.989 & 0.966 & & 18.146 & 18.363 & 0.000 & 10.710 & 0.969 & & 27.948 & 28.716 & 0.926 & 25.807 & 0.995\\
\hspace{2mm} FW & -0.178 & 6.184 & 91.204 & 22.458 & 0.925 & & -0.218 & 4.249 & 92.593 & 15.612 & 0.936 & & 0.004 & 2.282 & 93.981 & 8.639 & 0.964 & & -0.034 & 8.276 & 93.519 & 31.425 & 0.966\\
\hline 
\multicolumn{24}{l}{\textbf{Non-robust method}}\\ 
\hline 
\multicolumn{24}{l}{Model specification: QR--True}\\ 
\hline 
\hspace{2mm} PAPP & 0.487 & 5.489 & 94.907 & 22.317 & 1.039 & & -10.305 & 26.492 & 94.444 & 62.285 & 0.650 & & 1.287 & 2.842 & 91.667 & 9.682 & 0.973 & & 11.663 & 13.627 & 60.648 & 27.601 & 0.997\\
\hline 
\multicolumn{24}{l}{Model specification: QR--False}\\ 
\hline 
\hspace{2mm} PAPP & 35.408 & 35.754 & 0.000 & 18.412 & 0.945 & & 17.598 & 17.930 & 0.000 & 13.143 & 0.974 & & 10.135 & 10.422 & 0.000 & 9.452 & 0.991 & & 30.805 & 31.598 & 0.463 & 26.762 & 0.968\\
\hline 
\multicolumn{24}{l}{\textbf{Doubly robust methods}}\\ 
\hline 
\multicolumn{24}{l}{Model specification: QR--True, PM--True}\\ 
\hline 
\hspace{2mm} GPPP & 0.359 & 5.406 & 97.222 & 23.039 & 1.098 & & 0.147 & 3.855 & 96.296 & 16.208 & 1.080 & & 0.112 & 2.546 & 97.685 & 10.919 & 1.105 & & 0.235 & 7.323 & 96.296 & 32.247 & 1.129\\
\hspace{2mm} LWP & 0.293 & 5.384 & 97.222 & 23.243 & 1.108 & & 0.130 & 3.860 & 96.759 & 16.270 & 1.083 & & 0.060 & 2.595 & 97.222 & 11.233 & 1.111 & & 0.306 & 7.389 & 96.296 & 32.653 & 1.133\\
\hspace{2mm} AIPW & 0.292 & 5.410 & 96.296 & 22.264 & 1.049 & & 0.570 & 4.891 & 97.222 & 20.056 & 1.051 & & 0.100 & 2.610 & 96.759 & 10.631 & 1.037 & & 0.311 & 7.290 & 96.296 & 29.106 & 1.017\\
\hline 
\multicolumn{24}{l}{Model specification: QR--True, PM--False}\\ 
\hline 
\hspace{2mm} GPPP & 0.402 & 5.359 & 96.759 & 23.118 & 1.110 & & 0.149 & 3.926 & 95.370 & 15.959 & 1.039 & & 0.137 & 2.580 & 96.759 & 10.813 & 1.077 & & 0.389 & 7.230 & 97.685 & 31.411 & 1.112\\
\hspace{2mm} LWP & 10.150 & 5.659 & 91.204 & 22.008 & 0.126 & & 7.552 & 4.184 & 88.426 & 14.928 & 0.110 & & 5.872 & 23.999 & 88.426 & 10.056 & 0.098 & & 7.821 & 17.007 & 91.204 & 29.704 & 0.214\\
\hspace{2mm} AIPW & 0.291 & 5.475 & 95.370 & 22.112 & 1.029 & & 0.136 & 3.982 & 95.833 & 15.500 & 0.991 & & 0.086 & 2.662 & 95.370 & 10.677 & 1.022 & & 0.156 & 7.197 & 94.907 & 28.904 & 1.022\\
\hline 
\multicolumn{24}{l}{Model specification: QR--False, PM--True}\\ 
\hline 
\hspace{2mm} GPPP & 0.879 & 5.451 & 97.685 & 24.471 & 1.169 & & 1.589 & 4.338 & 95.833 & 18.450 & 1.172 & & 0.715 & 2.775 & 96.759 & 13.472 & 1.290 & & 0.827 & 7.547 & 96.296 & 33.826 & 1.156\\
\hspace{2mm} LWP & -1.094 & 5.826 & 97.685 & 24.830 & 0.530 & & -1.064 & 4.293 & 96.759 & 19.139 & 0.535 & & 0.132 & 2.775 & 98.148 & 13.657 & 1.262 & & -1.868 & 11.807 & 94.444 & 43.580 & 0.96\\
\hspace{2mm} AIPW & 1.226 & 7.554 & 94.444 & 31.434 & 1.073 & & -20.691 & 52.415 & 92.130 & 121.585 & 0.643 & & 6.562 & 7.243 & 43.981 & 12.248 & 1.017 & & 14.329 & 16.096 & 50.000 & 28.830 & 1.001\\
\hline 
\multicolumn{24}{l}{Model specification: QR--False, PM--False}\\ 
\hline 
\hspace{2mm} GPPP & 36.537 & 36.882 & 0.000 & 27.271 & 1.392 & & 18.359 & 18.731 & 0.000 & 19.077 & 1.320 & & 10.586 & 10.905 & 8.333 & 13.442 & 1.317 & & 30.765 & 31.605 & 9.259 & 40.674 & 1.443\\
\hspace{2mm} LWP & 36.191 & 36.544 & 0.000 & 27.447 & 1.391 & & 18.536 & 18.894 & 0.000 & 19.235 & 1.350 & & 10.681 & 10.998 & 7.870 & 13.431 & 1.314 & & 30.488 & 31.305 & 9.259 & 42.309 & 1.529\\
\hspace{2mm} AIPW & 36.291 & 36.644 & 0.000 & 19.219 & 0.964 & & 18.489 & 18.836 & 0.000 & 14.274 & 1.010 & & 10.726 & 11.039 & 0.463 & 10.142 & 0.989 & & 30.643 & 31.452 & 0.000 & 26.975 & 0.968\\
\bottomrule
\end{tabular}}
 \begin{tablenotes}
 \footnotesize
 \item UW: Unweighted; FW: Fully weighted; GPPP: Gaussian Process of Propensity Prediction; LWP: Linear-in-weight Prediction; AIPW: Augmented Inverse Propensity Weighting.\\
 NOTE: The PAPP and AIPW methods have been implemented through a bootstrap method. 
 \end{tablenotes}
\end{threeparttable}
\end{sidewaystable}

\begin{sidewaystable}[hbt!]
\centering
\caption{Comparing the performance of the bias adjustment methods in the second simulation study for the \emph{continuous} outcome with $(n_A, n_R)=(500, 1,000)$ and $\gamma_1=0.6$}\label{tab:6.7}
\begin{threeparttable}
\scriptsize{\begin{tabular}{l l l l l l l l l l l l l l l l l l l l l l l l l}
\toprule
& \multicolumn{5}{c}{\textbf{$LIN$}} & & \multicolumn{5}{c}{\textbf{$CUB$}} & & \multicolumn{5}{c}{\textbf{$EXP$}} & & \multicolumn{5}{c}{\textbf{$SIN$}}\\\cline{2-6}\cline{8-12}\cline{14-18}\cline{20-24}
\textbf{Measure} & rBias & rMSE & crCI & lCI & rSE & & rBias & rMSE & crCI & lCI & rSE & & rBias & rMSE & crCI & lCI & rSE & & rBias & rMSE & crCI & lCI & rSE \\
\midrule
\multicolumn{24}{l}{\textbf{Probability sample ($S_R$)}}\\ 
\hline 
\hspace{2mm} UW & -23.870 & 24.078 & 0.000 & 12.411 & 1.001 & & -17.120 & 17.243 & 0.000 & 8.442 & 1.045 & & -12.097 & 12.179 & 0.000 & 5.427 & 0.976 & & -18.475 & 19.159 & 3.241 & 18.535 & 0.93\\
\hspace{2mm} FW & -0.180 & 3.921 & 93.519 & 15.087 & 0.980 & & -0.156 & 2.749 & 94.907 & 10.799 & 1.001 & & -0.091 & 1.947 & 94.444 & 7.392 & 0.967 & & 0.015 & 5.549 & 93.519 & 20.503 & 0.94\\
\hline 
\multicolumn{24}{l}{\textbf{Non-probability sample ($S_A$)}}\\ 
\hline 
\hspace{2mm} UW & 100.237 & 100.399 & 0.000 & 20.820 & 0.929 & & 73.900 & 74.114 & 0.000 & 21.935 & 0.992 & & 44.946 & 45.091 & 0.000 & 13.875 & 0.978 & & 36.869 & 37.411 & 0.000 & 25.372 & 1.018\\
\hspace{2mm} FW & -0.428 & 10.721 & 92.130 & 40.191 & 0.955 & & -0.515 & 8.917 & 91.667 & 27.339 & 0.782 & & -0.070 & 3.570 & 92.593 & 13.283 & 0.947 & & -0.451 & 12.399 & 94.907 & 49.982 & 1.027\\
\hline 
\multicolumn{24}{l}{\textbf{Non-robust method}}\\ 
\hline 
\multicolumn{24}{l}{Model specification: QR--True}\\ 
\hline 
\hspace{2mm} PAPP & 0.267 & 8.677 & 94.444 & 36.302 & 1.065 & & 1.511 & 27.294 & 48.148 & 46.558 & 0.435 & & 5.295 & 5.987 & 49.537 & 10.631 & 0.969 & & 27.459 & 28.131 & 0.926 & 22.936 & 0.955\\
\hline 
\multicolumn{24}{l}{Model specification: QR--False}\\ 
\hline 
\hspace{2mm} PAPP & 52.850 & 53.104 & 0.000 & 18.790 & 0.921 & & 24.244 & 24.516 & 0.000 & 13.272 & 0.928 & & 15.205 & 15.440 & 0.000 & 9.782 & 0.928 & & 54.554 & 55.056 & 0.000 & 28.544 & 0.979\\
\hline 
\multicolumn{24}{l}{\textbf{Doubly robust methods}}\\ 
\hline 
\multicolumn{24}{l}{Model specification: QR--True, PM--True}\\ 
\hline 
\hspace{2mm} GPPP & 0.023 & 6.062 & 95.833 & 23.490 & 0.995 & & -0.014 & 4.219 & 97.685 & 16.700 & 1.012 & & 0.017 & 2.871 & 95.833 & 11.306 & 1.011 & & 0.083 & 8.500 & 93.519 & 33.426 & 1.007\\
\hspace{2mm} LWP & -0.400 & 6.959 & 96.759 & 28.647 & 1.061 & & -0.168 & 4.808 & 96.759 & 19.394 & 1.035 & & -0.076 & 3.301 & 96.296 & 12.962 & 1.008 & & -0.139 & 9.842 & 94.444 & 39.758 & 1.038\\
\hspace{2mm} AIPW & -0.244 & 6.410 & 93.981 & 24.890 & 0.989 & & -0.214 & 5.745 & 93.056 & 19.238 & 0.853 & & -0.055 & 2.696 & 93.981 & 9.824 & 0.927 & & -0.070 & 6.698 & 95.370 & 25.116 & 0.954\\
\hline 
\multicolumn{24}{l}{Model specification: QR--True, PM--False}\\ 
\hline 
\hspace{2mm} GPPP & 0.578 & 6.177 & 95.370 & 24.644 & 1.027 & & -0.227 & 3.652 & 94.444 & 14.732 & 1.038 & & -0.118 & 2.529 & 95.370 & 9.677 & 0.985 & & 0.252 & 7.465 & 93.519 & 28.027 & 0.964\\
\hspace{2mm} LWP & 102.755 & 6.228 & 61.111 & 36.944 & 0.065 & & 77.157 & 4.030 & 58.796 & 26.171 & 0.062 & & 52.013 & 28.166 & 60.185 & 21.390 & 0.074 & & 68.011 & 7.895 & 61.111 & 27.053 & 0.074\\
\hspace{2mm} AIPW & 0.108 & 6.499 & 92.593 & 22.724 & 0.890 & & -0.148 & 3.903 & 92.593 & 13.978 & 0.912 & & -0.124 & 2.678 & 95.833 & 9.389 & 0.893 & & 0.111 & 7.136 & 92.593 & 25.796 & 0.92\\
\hline 
\multicolumn{24}{l}{Model specification: QR--False, PM--True}\\ 
\hline 
\hspace{2mm} GPPP & 1.435 & 6.880 & 94.907 & 27.203 & 1.039 & & 5.206 & 8.273 & 81.944 & 21.919 & 0.874 & & 2.127 & 4.765 & 90.278 & 16.145 & 0.971 & & -0.452 & 9.473 & 94.444 & 39.581 & 1.071\\
\hspace{2mm} LWP & 38.447 & 24.018 & 62.037 & 66.505 & 0.166 & & 45.933 & 22.233 & 78.704 & 68.192 & 0.145 & & 32.100 & 19.000 & 82.407 & 52.561 & 0.202 & & 36.851 & 46.251 & 56.481 & 92.284 & 0.236\\
\hspace{2mm} AIPW & 1.156 & 14.115 & 90.278 & 53.458 & 0.967 & & -5.904 & 64.920 & 48.611 & 92.613 & 0.365 & & 11.560 & 11.978 & 3.704 & 11.967 & 0.972 & & 34.579 & 35.388 & 0.000 & 28.031 & 0.948\\
\hline 
\multicolumn{24}{l}{Model specification: QR--False, PM--False}\\ 
\hline 
\hspace{2mm} GPPP & 51.343 & 51.589 & 0.000 & 20.925 & 1.069 & & 22.943 & 23.242 & 0.000 & 15.244 & 1.053 & & 14.179 & 14.437 & 0.000 & 11.227 & 1.060 & & 55.161 & 55.704 & 0.000 & 31.475 & 1.04\\
\hspace{2mm} LWP & 195.557 & 96.444 & 0.000 & 28.177 & 0.255 & & 177.674 & 118.013 & 0.463 & 28.728 & 0.250 & & 146.864 & 113.140 & 0.000 & 25.490 & 0.260 & & 166.531 & 125.351 & 0.000 & 28.693 & 0.288\\
\hspace{2mm} AIPW & 51.698 & 51.947 & 0.000 & 18.450 & 0.924 & & 22.823 & 23.115 & 0.000 & 13.586 & 0.944 & & 14.321 & 14.575 & 0.000 & 9.978 & 0.936 & & 54.681 & 55.217 & 0.000 & 28.899 & 0.958\\
\bottomrule
\end{tabular}}
 \begin{tablenotes}
 \footnotesize
 \item UW: Unweighted; FW: Fully weighted; GPPP: Gaussian Process of Propensity Prediction; LWP: Linear-in-weight Prediction; AIPW: Augmented Inverse Propensity Weighting.\\
 NOTE: The PAPP and AIPW methods have been implemented through a bootstrap method. 
 \end{tablenotes}
\end{threeparttable}
\end{sidewaystable}

\begin{sidewaystable}[hbt!]
\centering
\caption{Comparing the performance of the bias adjustment methods in the second simulation study for the \emph{continuous} outcome with $(n_A, n_R)=(1,000, 500)$ and $\gamma_1=0.6$}\label{tab:6.8}
\begin{threeparttable}
\scriptsize{\begin{tabular}{l l l l l l l l l l l l l l l l l l l l l l l l l}
\toprule
& \multicolumn{5}{c}{\textbf{$LIN$}} & & \multicolumn{5}{c}{\textbf{$CUB$}} & & \multicolumn{5}{c}{\textbf{$EXP$}} & & \multicolumn{5}{c}{\textbf{$SIN$}}\\\cline{2-6}\cline{8-12}\cline{14-18}\cline{20-24}
\textbf{Measure} & rBias & rMSE & crCI & lCI & rSE & & rBias & rMSE & crCI & lCI & rSE & & rBias & rMSE & crCI & lCI & rSE & & rBias & rMSE & crCI & lCI & rSE \\
\midrule
\multicolumn{24}{l}{\textbf{Probability sample ($S_R$)}}\\ 
\hline 
\hspace{2mm} UW & -23.868 & 24.297 & 0.000 & 17.587 & 0.985 & & -17.158 & 17.440 & 0.000 & 11.959 & 0.974 & & -12.107 & 12.274 & 0.000 & 7.683 & 0.967 & & -18.706 & 19.966 & 22.222 & 26.219 & 0.956\\
\hspace{2mm} FW & 0.118 & 5.477 & 93.519 & 21.395 & 0.994 & & 0.016 & 3.998 & 94.444 & 15.257 & 0.971 & & 0.015 & 2.673 & 95.833 & 10.454 & 0.995 & & -0.081 & 7.162 & 96.296 & 29.064 & 1.033\\
\hline 
\multicolumn{24}{l}{\textbf{Non-probability sample ($S_A$)}}\\ 
\hline 
\hspace{2mm} UW & 98.629 & 98.709 & 0.000 & 14.606 & 0.936 & & 71.503 & 71.609 & 0.000 & 15.102 & 0.988 & & 43.622 & 43.693 & 0.000 & 9.597 & 0.980 & & 37.338 & 37.581 & 0.000 & 17.907 & 1.067\\
\hspace{2mm} FW & -0.355 & 8.567 & 90.278 & 29.835 & 0.887 & & -0.483 & 7.679 & 88.889 & 21.564 & 0.716 & & -0.008 & 2.547 & 92.130 & 9.577 & 0.957 & & -0.590 & 9.229 & 93.519 & 36.092 & 0.997\\
\hline 
\multicolumn{24}{l}{\textbf{Non-robust method}}\\ 
\hline 
\multicolumn{24}{l}{Model specification: QR--True}\\ 
\hline 
\hspace{2mm} PAPP & 0.837 & 6.971 & 93.981 & 28.496 & 1.048 & & -15.335 & 47.984 & 81.944 & 80.081 & 0.448 & & 2.217 & 3.140 & 86.574 & 9.040 & 1.035 & & 27.350 & 27.937 & 0.463 & 23.160 & 1.035\\
\hline 
\multicolumn{24}{l}{Model specification: QR--False}\\ 
\hline 
\hspace{2mm} PAPP & 51.938 & 52.102 & 0.000 & 16.106 & 0.990 & & 23.574 & 23.768 & 0.000 & 11.833 & 0.993 & & 14.805 & 14.960 & 0.000 & 8.330 & 0.987 & & 55.512 & 55.750 & 0.000 & 20.689 & 1.022\\
\hline 
\multicolumn{24}{l}{\textbf{Doubly robust methods}}\\ 
\hline 
\multicolumn{24}{l}{Model specification: QR--True, PM--True}\\ 
\hline 
\hspace{2mm} GPPP & 0.467 & 5.136 & 97.685 & 23.457 & 1.177 & & 0.310 & 3.766 & 97.222 & 16.493 & 1.126 & & 0.237 & 2.519 & 96.759 & 11.192 & 1.145 & & 0.247 & 6.833 & 98.148 & 32.759 & 1.234\\
\hspace{2mm} LWP & 0.696 & 9.836 & 97.222 & 25.617 & 0.670 & & 0.169 & 3.997 & 96.759 & 17.628 & 1.134 & & 0.129 & 2.695 & 98.611 & 12.199 & 1.161 & & 0.240 & 7.532 & 97.685 & 35.788 & 1.216\\
\hspace{2mm} AIPW & 0.328 & 5.304 & 96.296 & 23.712 & 1.140 & & 0.323 & 9.295 & 96.759 & 26.822 & 0.735 & & 0.236 & 2.504 & 96.296 & 10.389 & 1.061 & & 0.292 & 6.326 & 95.833 & 27.065 & 1.09\\
\hline 
\multicolumn{24}{l}{Model specification: QR--True, PM--False}\\ 
\hline 
\hspace{2mm} GPPP & 0.847 & 5.217 & 96.296 & 23.740 & 1.191 & & -0.091 & 3.657 & 97.222 & 15.703 & 1.102 & & -0.035 & 2.398 & 97.222 & 10.482 & 1.121 & & 0.301 & 6.461 & 97.685 & 30.374 & 1.209\\
\hspace{2mm} LWP & 160.667 & 5.868 & 44.444 & 57.995 & 0.103 & & 121.455 & 15.991 & 38.889 & 50.917 & 0.120 & & 79.393 & 20.870 & 42.130 & 40.288 & 0.139 & & 99.141 & 6.863 & 45.370 & 34.496 & 0.092\\
\hspace{2mm} AIPW & 0.764 & 5.471 & 96.759 & 22.254 & 1.046 & & 0.188 & 3.787 & 95.370 & 14.917 & 1.004 & & 0.103 & 2.524 & 97.222 & 10.169 & 1.026 & & 0.301 & 6.422 & 94.907 & 27.431 & 1.088\\
\hline 
\multicolumn{24}{l}{Model specification: QR--False, PM--True}\\ 
\hline 
\hspace{2mm} GPPP & 1.462 & 5.582 & 97.685 & 25.713 & 1.227 & & 4.015 & 6.286 & 89.352 & 20.044 & 1.062 & & 1.752 & 3.622 & 97.685 & 14.822 & 1.197 & & 0.278 & 7.406 & 97.222 & 36.447 & 1.265\\
\hspace{2mm} LWP & -3.714 & 21.454 & 78.704 & 31.849 & 0.335 & & 1.510 & 13.601 & 91.667 & 23.485 & 0.307 & & 1.474 & 13.319 & 92.593 & 19.695 & 0.383 & & -7.084 & 31.041 & 68.981 & 51.814 & 0.44\\
\hspace{2mm} AIPW & 1.481 & 12.221 & 88.889 & 44.394 & 0.931 & & -35.926 & 106.302 & 75.926 & 175.824 & 0.447 & & 10.829 & 11.221 & 5.093 & 12.150 & 1.051 & & 35.268 & 35.750 & 0.000 & 26.074 & 1.134\\
\hline 
\multicolumn{24}{l}{Model specification: QR--False, PM--False}\\ 
\hline 
\hspace{2mm} GPPP & 51.234 & 51.404 & 0.000 & 22.464 & 1.382 & & 23.626 & 23.875 & 0.000 & 16.807 & 1.249 & & 14.541 & 14.741 & 0.000 & 12.211 & 1.293 & & 54.802 & 55.140 & 0.000 & 33.025 & 1.393\\
\hspace{2mm} LWP & 186.628 & 124.670 & 0.000 & 36.048 & 0.281 & & 168.280 & 123.557 & 0.000 & 36.124 & 0.283 & & 138.694 & 108.095 & 0.000 & 32.502 & 0.287 & & 158.745 & 122.172 & 0.000 & 34.886 & 0.292\\
\hspace{2mm} AIPW & 51.843 & 52.021 & 0.000 & 16.918 & 1.001 & & 23.384 & 23.626 & 0.000 & 13.378 & 1.008 & & 14.678 & 14.877 & 0.000 & 9.732 & 1.020 & & 55.491 & 55.756 & 0.000 & 21.258 & 0.997\\
\bottomrule
\end{tabular}}
 \begin{tablenotes}
 \footnotesize
 \item UW: Unweighted; FW: Fully weighted; GPPP: Gaussian Process of Propensity Prediction; LWP: Linear-in-weight Prediction; AIPW: Augmented Inverse Propensity Weighting.\\
 NOTE: The PAPP and AIPW methods have been implemented through a bootstrap method. 
 \end{tablenotes}
\end{threeparttable}
\end{sidewaystable}

\begin{sidewaystable}[hbt!]
\centering
\caption{Comparing the performance of the bias adjustment methods in the second simulation study for the \emph{continuous} outcome with $(n_A, n_R)=(500, 500)$ and $\gamma_1=0.6$}\label{tab:6.9}
\begin{threeparttable}
\scriptsize{\begin{tabular}{l l l l l l l l l l l l l l l l l l l l l l l l l}
\toprule
& \multicolumn{5}{c}{\textbf{$LIN$}} & & \multicolumn{5}{c}{\textbf{$CUB$}} & & \multicolumn{5}{c}{\textbf{$EXP$}} & & \multicolumn{5}{c}{\textbf{$SIN$}}\\\cline{2-6}\cline{8-12}\cline{14-18}\cline{20-24}
\textbf{Measure} & rBias & rMSE & crCI & lCI & rSE & & rBias & rMSE & crCI & lCI & rSE & & rBias & rMSE & crCI & lCI & rSE & & rBias & rMSE & crCI & lCI & rSE \\
\midrule
\multicolumn{24}{l}{\textbf{Probability sample ($S_R$)}}\\ 
\hline 
\hspace{2mm} UW & -23.868 & 24.297 & 0.000 & 17.587 & 0.985 & & -17.158 & 17.440 & 0.000 & 11.959 & 0.974 & & -12.107 & 12.274 & 0.000 & 7.683 & 0.967 & & -18.706 & 19.966 & 22.222 & 26.219 & 0.956\\
\hspace{2mm} FW & 0.118 & 5.477 & 93.519 & 21.395 & 0.994 & & 0.016 & 3.998 & 94.444 & 15.257 & 0.971 & & 0.015 & 2.673 & 95.833 & 10.454 & 0.995 & & -0.081 & 7.162 & 96.296 & 29.064 & 1.033\\
\hline 
\multicolumn{24}{l}{\textbf{Non-probability sample ($S_A$)}}\\ 
\hline 
\hspace{2mm} UW & 100.237 & 100.399 & 0.000 & 20.820 & 0.929 & & 73.900 & 74.114 & 0.000 & 21.935 & 0.992 & & 44.946 & 45.091 & 0.000 & 13.875 & 0.978 & & 36.869 & 37.411 & 0.000 & 25.372 & 1.018\\
\hspace{2mm} FW & -0.428 & 10.721 & 92.130 & 40.191 & 0.955 & & -0.515 & 8.917 & 91.667 & 27.339 & 0.782 & & -0.070 & 3.570 & 92.593 & 13.283 & 0.947 & & -0.451 & 12.399 & 94.907 & 49.982 & 1.027\\
\hline 
\multicolumn{24}{l}{\textbf{Non-robust method}}\\ 
\hline 
\multicolumn{24}{l}{Model specification: QR--True}\\ 
\hline 
\hspace{2mm} PAPP & 0.893 & 9.134 & 95.833 & 37.207 & 1.042 & & -3.022 & 30.594 & 66.667 & 56.091 & 0.469 & & 3.650 & 4.758 & 76.852 & 11.529 & 0.961 & & 27.447 & 28.332 & 1.852 & 26.590 & 0.963\\
\hline 
\multicolumn{24}{l}{Model specification: QR--False}\\ 
\hline 
\hspace{2mm} PAPP & 52.215 & 52.495 & 0.000 & 20.075 & 0.943 & & 23.753 & 24.077 & 0.000 & 14.584 & 0.942 & & 14.865 & 15.131 & 0.000 & 10.533 & 0.949 & & 54.727 & 55.253 & 0.000 & 28.872 & 0.965\\
\hline 
\multicolumn{24}{l}{\textbf{Doubly robust methods}}\\ 
\hline 
\multicolumn{24}{l}{Model specification: QR--True, PM--True}\\ 
\hline 
\hspace{2mm} GPPP & 0.319 & 6.749 & 94.444 & 27.157 & 1.035 & & 0.154 & 4.755 & 95.370 & 19.240 & 1.037 & & 0.134 & 3.192 & 96.759 & 12.998 & 1.044 & & 0.017 & 9.290 & 96.759 & 38.202 & 1.055\\
\hspace{2mm} LWP & 0.090 & 7.864 & 95.833 & 32.102 & 1.049 & & 0.034 & 5.204 & 95.370 & 21.564 & 1.063 & & 0.019 & 3.837 & 94.907 & 14.712 & 0.987 & & -2.658 & 10.462 & 97.222 & 43.943 & 0.343\\
\hspace{2mm} AIPW & 0.180 & 6.969 & 95.370 & 28.099 & 1.027 & & -0.103 & 7.074 & 94.444 & 24.196 & 0.871 & & 0.098 & 3.079 & 94.444 & 11.765 & 0.973 & & -0.011 & 7.659 & 94.907 & 30.253 & 1.005\\
\hline 
\multicolumn{24}{l}{Model specification: QR--True, PM--False}\\ 
\hline 
\hspace{2mm} GPPP & 0.876 & 6.819 & 96.759 & 27.893 & 1.060 & & -0.061 & 4.357 & 95.833 & 17.596 & 1.036 & & -0.015 & 2.943 & 95.833 & 11.661 & 1.017 & & 0.150 & 8.338 & 95.370 & 33.641 & 1.037\\
\hspace{2mm} LWP & 103.450 & 6.510 & 62.963 & 44.391 & 0.078 & & 77.847 & 4.802 & 59.722 & 36.575 & 0.085 & & 51.800 & 18.153 & 58.333 & 26.690 & 0.093 & & 67.796 & 8.286 & 62.037 & 33.876 & 0.091\\
\hspace{2mm} AIPW & 0.490 & 7.184 & 91.667 & 25.904 & 0.920 & & -0.024 & 4.629 & 93.981 & 16.579 & 0.912 & & -0.042 & 3.065 & 94.444 & 11.341 & 0.942 & & -0.031 & 7.974 & 94.907 & 30.870 & 0.985\\
\hline 
\multicolumn{24}{l}{Model specification: QR--False, PM--True}\\ 
\hline 
\hspace{2mm} GPPP & 1.843 & 7.386 & 94.444 & 30.532 & 1.097 & & 5.574 & 8.660 & 86.111 & 24.260 & 0.939 & & 2.304 & 4.919 & 94.907 & 17.944 & 1.057 & & -0.476 & 10.124 & 95.370 & 43.960 & 1.117\\
\hspace{2mm} LWP & 8.759 & 20.713 & 74.074 & 46.601 & 0.157 & & 10.563 & 16.870 & 87.963 & 38.325 & 0.187 & & 8.852 & 17.395 & 88.426 & 30.616 & 0.222 & & -4.843 & 41.854 & 68.519 & 74.878 & 0.382\\
\hspace{2mm} AIPW & 1.835 & 14.943 & 92.593 & 54.945 & 0.943 & & -11.650 & 71.384 & 60.185 & 116.080 & 0.420 & & 11.277 & 11.832 & 12.500 & 13.911 & 0.988 & & 34.536 & 35.418 & 0.926 & 31.247 & 1.012\\
\hline 
\multicolumn{24}{l}{Model specification: QR--False, PM--False}\\ 
\hline 
\hspace{2mm} GPPP & 51.474 & 51.753 & 0.000 & 25.043 & 1.197 & & 23.106 & 23.478 & 0.000 & 18.473 & 1.139 & & 14.274 & 14.585 & 0.926 & 13.478 & 1.153 & & 54.945 & 55.521 & 0.000 & 37.361 & 1.199\\
\hspace{2mm} LWP & 191.826 & 108.237 & 0.000 & 35.914 & 0.226 & & 173.883 & 112.649 & 0.000 & 35.981 & 0.224 & & 143.469 & 104.426 & 0.000 & 32.503 & 0.231 & & 163.575 & 116.824 & 0.000 & 36.164 & 0.257\\
\hspace{2mm} AIPW & 51.656 & 51.943 & 0.000 & 20.474 & 0.957 & & 23.094 & 23.448 & 0.000 & 15.597 & 0.978 & & 14.519 & 14.814 & 0.000 & 11.295 & 0.978 & & 54.858 & 55.414 & 0.000 & 29.376 & 0.955\\
\bottomrule
\end{tabular}}
 \begin{tablenotes}
 \footnotesize
 \item UW: Unweighted; FW: Fully weighted; GPPP: Gaussian Process of Propensity Prediction; LWP: Linear-in-weight Prediction; AIPW: Augmented Inverse Propensity Weighting.\\
 NOTE: The PAPP and AIPW methods have been implemented through a bootstrap method. 
 \end{tablenotes}
\end{threeparttable}
\end{sidewaystable}


\begin{sidewaystable}[hbt!]
\centering
\caption{Comparing the performance of the bias adjustment methods in the second simulation study for the \emph{binary} outcome with $(n_A, n_R)=(500, 1,000)$ and $\gamma_1=0.3$}\label{tab:6.10}
\begin{threeparttable}
\scriptsize{\begin{tabular}{l l l l l l l l l l l l l l l l l l l l l l l l l}
\toprule
& \multicolumn{5}{c}{\textbf{$LIN$}} & & \multicolumn{5}{c}{\textbf{$CUB$}} & & \multicolumn{5}{c}{\textbf{$EXP$}} & & \multicolumn{5}{c}{\textbf{$SIN$}}\\\cline{2-6}\cline{8-12}\cline{14-18}\cline{20-24}
\textbf{Measure} & rBias & rMSE & crCI & lCI & rSE & & rBias & rMSE & crCI & lCI & rSE & & rBias & rMSE & crCI & lCI & rSE & & rBias & rMSE & crCI & lCI & rSE \\
\midrule
\multicolumn{24}{l}{\textbf{Probability sample ($S_R$)}}\\ 
\hline 
\hspace{2mm} UW & -24.160 & 24.480 & 0.000 & 14.690 & 0.948 & & -23.859 & 24.179 & 0.000 & 16.035 & 1.040 & & -20.146 & 20.452 & 0.000 & 14.709 & 1.061 & & -11.189 & 11.695 & 10.185 & 13.552 & 1.013\\
\hspace{2mm} FW & -0.042 & 4.761 & 94.907 & 17.832 & 0.953 & & -0.400 & 4.855 & 96.759 & 19.943 & 1.049 & & -0.706 & 4.262 & 94.907 & 17.407 & 1.054 & & 0.169 & 3.866 & 94.444 & 15.066 & 0.993\\
\hline 
\multicolumn{24}{l}{\textbf{Non-probability sample ($S_A$)}}\\ 
\hline 
\hspace{2mm} UW & 49.180 & 49.553 & 0.000 & 22.648 & 0.950 & & 37.359 & 38.019 & 0.000 & 25.871 & 0.934 & & 27.292 & 27.979 & 0.000 & 22.450 & 0.928 & & 19.208 & 19.788 & 1.852 & 19.537 & 1.046\\
\hspace{2mm} FW & -0.043 & 6.460 & 93.056 & 24.387 & 0.961 & & 0.091 & 6.891 & 94.444 & 27.157 & 1.003 & & 0.532 & 6.855 & 94.907 & 25.353 & 0.944 & & -0.300 & 5.720 & 94.444 & 22.909 & 1.021\\
\hline 
\multicolumn{24}{l}{\textbf{Non-robust method}}\\ 
\hline 
\multicolumn{24}{l}{Model specification: QR--True}\\ 
\hline 
\hspace{2mm} PAPP & -0.038 & 5.120 & 96.759 & 21.921 & 1.090 & & 3.926 & 11.150 & 87.037 & 39.172 & 0.955 & & 6.575 & 9.060 & 81.944 & 24.125 & 0.985 & & 6.183 & 7.285 & 71.759 & 16.975 & 1.121\\
\hline 
\multicolumn{24}{l}{Model specification: QR--False}\\ 
\hline 
\hspace{2mm} PAPP & 38.049 & 38.533 & 0.000 & 23.509 & 0.983 & & 24.934 & 25.843 & 3.704 & 25.873 & 0.969 & & 18.198 & 19.176 & 14.352 & 22.897 & 0.964 & & 22.508 & 23.055 & 0.000 & 20.223 & 1.031\\
\hline 
\multicolumn{24}{l}{\textbf{Doubly robust methods}}\\ 
\hline 
\multicolumn{24}{l}{Model specification: QR--True, PM--True}\\ 
\hline 
\hspace{2mm} GPPP & -0.345 & 4.843 & 98.611 & 23.428 & 1.244 & & 0.454 & 6.387 & 98.148 & 30.398 & 1.229 & & 0.770 & 6.202 & 98.611 & 28.665 & 1.196 & & 0.345 & 4.177 & 94.907 & 16.881 & 1.044\\
\hspace{2mm} LWP & -0.243 & 4.852 & 98.611 & 23.554 & 1.249 & & 0.511 & 6.541 & 97.685 & 30.906 & 1.218 & & 0.764 & 6.407 & 99.074 & 29.130 & 1.172 & & 0.327 & 4.356 & 95.833 & 17.307 & 1.024\\
\hspace{2mm} AIPW & -0.028 & 4.938 & 97.685 & 21.216 & 1.094 & & 0.248 & 6.688 & 93.981 & 25.328 & 0.964 & & 0.792 & 6.236 & 94.444 & 23.674 & 0.974 & & -0.032 & 4.149 & 94.444 & 16.568 & 1.016\\
\hline 
\multicolumn{24}{l}{Model specification: QR--True, PM--False}\\ 
\hline 
\hspace{2mm} GPPP & -0.286 & 4.831 & 99.074 & 23.492 & 1.249 & & 0.996 & 6.370 & 98.148 & 30.231 & 1.234 & & 1.659 & 6.255 & 98.611 & 28.479 & 1.210 & & 0.485 & 4.095 & 92.130 & 16.104 & 1.016\\
\hspace{2mm} LWP & -0.193 & 4.850 & 98.611 & 23.389 & 1.241 & & 0.968 & 6.336 & 97.685 & 29.999 & 1.231 & & 1.480 & 6.150 & 98.148 & 28.234 & 1.213 & & 0.543 & 4.156 & 92.593 & 16.206 & 1.007\\
\hspace{2mm} AIPW & 0.216 & 4.997 & 97.685 & 21.435 & 1.093 & & 0.137 & 6.371 & 94.907 & 25.029 & 1.000 & & 0.772 & 6.335 & 94.444 & 23.511 & 0.952 & & -0.115 & 4.158 & 93.519 & 16.704 & 1.023\\
\hline 
\multicolumn{24}{l}{Model specification: QR--False, PM--True}\\ 
\hline 
\hspace{2mm} GPPP & 0.623 & 4.813 & 98.148 & 24.745 & 1.331 & & 2.439 & 7.249 & 97.685 & 32.279 & 1.215 & & 1.942 & 6.670 & 98.611 & 29.934 & 1.205 & & 0.517 & 4.457 & 96.296 & 18.057 & 1.045\\
\hspace{2mm} LWP & 0.089 & 4.790 & 98.611 & 24.379 & 1.310 & & 1.007 & 6.884 & 97.222 & 32.125 & 1.213 & & 1.128 & 6.633 & 99.074 & 30.109 & 1.179 & & 0.225 & 5.787 & 96.759 & 24.201 & 1.075\\
\hspace{2mm} AIPW & 0.295 & 6.609 & 95.370 & 28.014 & 1.080 & & -2.126 & 22.250 & 85.648 & 70.680 & 0.812 & & 11.808 & 13.407 & 53.241 & 24.334 & 0.975 & & 9.860 & 10.779 & 42.130 & 17.942 & 1.048\\
\hline 
\multicolumn{24}{l}{Model specification: QR--False, PM--False}\\ 
\hline 
\hspace{2mm} GPPP & 38.289 & 38.703 & 0.000 & 29.233 & 1.333 & & 26.017 & 26.829 & 6.481 & 32.772 & 1.284 & & 18.980 & 19.864 & 24.074 & 28.967 & 1.270 & & 22.805 & 23.317 & 2.315 & 25.525 & 1.345\\
\hspace{2mm} LWP & 38.289 & 38.703 & 0.000 & 29.233 & 1.333 & & 26.017 & 26.829 & 6.481 & 32.772 & 1.284 & & 18.980 & 19.864 & 24.074 & 28.967 & 1.270 & & 22.805 & 23.317 & 2.315 & 25.525 & 1.345\\
\hspace{2mm} AIPW & 38.321 & 38.770 & 0.000 & 22.934 & 0.992 & & 25.373 & 26.267 & 4.167 & 26.143 & 0.979 & & 18.557 & 19.503 & 11.111 & 22.737 & 0.964 & & 22.449 & 23.014 & 0.463 & 20.115 & 1.01\\
\bottomrule
\end{tabular}}
 \begin{tablenotes}
 \footnotesize
 \item UW: Unweighted; FW: Fully weighted; GPPP: Gaussian Process of Propensity Prediction; LWP: Linear-in-weight Prediction; AIPW: Augmented Inverse Propensity Weighting.\\
 NOTE: The PAPP and AIPW methods have been implemented through a bootstrap method. 
 \end{tablenotes}
\end{threeparttable}
\end{sidewaystable}

\begin{sidewaystable}[hbt!]
\centering
\caption{Comparing the performance of the bias adjustment methods in the second simulation study for the \emph{binary} outcome with $(n_A, n_R)=(1,000, 500)$ and $\gamma_1=0.3$}\label{tab:6.11}
\begin{threeparttable}
\scriptsize{\begin{tabular}{l l l l l l l l l l l l l l l l l l l l l l l l l}
\toprule
& \multicolumn{5}{c}{\textbf{$LIN$}} & & \multicolumn{5}{c}{\textbf{$CUB$}} & & \multicolumn{5}{c}{\textbf{$EXP$}} & & \multicolumn{5}{c}{\textbf{$SIN$}}\\\cline{2-6}\cline{8-12}\cline{14-18}\cline{20-24}
\textbf{Measure} & rBias & rMSE & crCI & lCI & rSE & & rBias & rMSE & crCI & lCI & rSE & & rBias & rMSE & crCI & lCI & rSE & & rBias & rMSE & crCI & lCI & rSE \\
\midrule
\multicolumn{24}{l}{\textbf{Probability sample ($S_R$)}}\\ 
\hline 
\hspace{2mm} UW & -24.046 & 24.606 & 0.926 & 20.785 & 1.014 & & -23.445 & 24.144 & 0.926 & 22.713 & 1.003 & & -19.875 & 20.549 & 4.167 & 20.818 & 1.015 & & -11.430 & 12.367 & 35.648 & 19.157 & 1.032\\
\hspace{2mm} FW & 0.329 & 6.211 & 95.833 & 25.283 & 1.038 & & 0.153 & 7.020 & 97.222 & 28.280 & 1.026 & & -0.219 & 6.035 & 96.759 & 24.676 & 1.041 & & -0.014 & 4.947 & 96.759 & 21.354 & 1.099\\
\hline 
\multicolumn{24}{l}{\textbf{Non-probability sample ($S_A$)}}\\ 
\hline 
\hspace{2mm} UW & 48.655 & 48.836 & 0.000 & 16.025 & 0.970 & & 36.988 & 37.324 & 0.000 & 18.292 & 0.932 & & 26.828 & 27.149 & 0.000 & 15.876 & 0.970 & & 18.929 & 19.164 & 0.000 & 13.818 & 1.176\\
\hspace{2mm} FW & -0.155 & 4.474 & 94.907 & 17.199 & 0.979 & & -0.060 & 4.840 & 94.444 & 19.145 & 1.007 & & 0.380 & 4.728 & 94.444 & 17.877 & 0.965 & & -0.574 & 3.624 & 96.759 & 16.151 & 1.149\\
\hline 
\multicolumn{24}{l}{\textbf{Non-robust method}}\\ 
\hline 
\multicolumn{24}{l}{Model specification: QR--True}\\ 
\hline 
\hspace{2mm} PAPP & 0.236 & 4.935 & 95.370 & 20.326 & 1.049 & & -2.054 & 12.151 & 95.370 & 42.145 & 0.896 & & 3.062 & 5.794 & 91.667 & 19.074 & 0.987 & & 5.980 & 7.462 & 78.241 & 18.605 & 1.061\\
\hline 
\multicolumn{24}{l}{Model specification: QR--False}\\ 
\hline 
\hspace{2mm} PAPP & 37.580 & 37.873 & 0.000 & 18.098 & 0.979 & & 24.751 & 25.263 & 0.000 & 19.860 & 0.999 & & 17.868 & 18.407 & 1.852 & 17.047 & 0.981 & & 22.204 & 22.420 & 0.000 & 14.329 & 1.175\\
\hline 
\multicolumn{24}{l}{\textbf{Doubly robust methods}}\\ 
\hline 
\multicolumn{24}{l}{Model specification: QR--True, PM--True}\\ 
\hline 
\hspace{2mm} GPPP & -0.350 & 5.172 & 98.148 & 26.705 & 1.330 & & 1.049 & 5.831 & 99.074 & 33.701 & 1.509 & & 1.324 & 5.279 & 99.074 & 30.631 & 1.539 & & 0.701 & 4.796 & 96.296 & 19.836 & 1.072\\
\hspace{2mm} LWP & -0.260 & 5.178 & 98.148 & 26.714 & 1.324 & & 1.057 & 5.860 & 99.074 & 33.705 & 1.501 & & 1.427 & 5.421 & 99.537 & 30.879 & 1.517 & & 0.952 & 5.585 & 94.907 & 20.030 & 0.936\\
\hspace{2mm} AIPW & 0.121 & 5.365 & 96.759 & 22.128 & 1.050 & & 0.199 & 5.875 & 93.981 & 23.158 & 1.004 & & 0.594 & 5.128 & 95.833 & 20.331 & 1.016 & & -0.109 & 4.639 & 95.370 & 19.630 & 1.077\\
\hline 
\multicolumn{24}{l}{Model specification: QR--True, PM--False}\\ 
\hline 
\hspace{2mm} GPPP & -0.304 & 5.215 & 98.611 & 26.629 & 1.313 & & 1.361 & 5.859 & 99.537 & 33.405 & 1.505 & & 1.777 & 5.363 & 99.537 & 30.723 & 1.560 & & 0.915 & 4.710 & 94.907 & 19.554 & 1.087\\
\hspace{2mm} LWP & -0.221 & 5.209 & 98.148 & 26.602 & 1.311 & & 1.348 & 5.811 & 99.537 & 33.435 & 1.520 & & 1.679 & 5.205 & 100.000 & 30.677 & 1.596 & & 1.008 & 4.850 & 95.833 & 19.504 & 1.053\\
\hspace{2mm} AIPW & 0.251 & 5.271 & 95.833 & 21.900 & 1.059 & & 0.163 & 5.703 & 95.370 & 22.900 & 1.022 & & 0.565 & 5.074 & 95.370 & 20.202 & 1.020 & & -0.091 & 4.651 & 95.833 & 19.795 & 1.083\\
\hline 
\multicolumn{24}{l}{Model specification: QR--False, PM--True}\\ 
\hline 
\hspace{2mm} GPPP & 0.703 & 4.852 & 99.537 & 27.464 & 1.472 & & 2.832 & 6.303 & 100.000 & 34.562 & 1.578 & & 2.438 & 5.370 & 100.000 & 31.759 & 1.706 & & 1.026 & 5.008 & 95.833 & 20.459 & 1.075\\
\hspace{2mm} LWP & 0.519 & 4.836 & 100.000 & 27.407 & 1.466 & & 2.032 & 6.031 & 100.000 & 34.399 & 1.559 & & 1.868 & 5.186 & 100.000 & 31.708 & 1.685 & & 0.445 & 4.765 & 99.074 & 26.385 & 1.43\\
\hspace{2mm} AIPW & 0.619 & 6.290 & 93.519 & 24.931 & 1.014 & & -13.115 & 30.760 & 94.444 & 90.189 & 0.825 & & 9.924 & 11.185 & 53.241 & 20.331 & 1.003 & & 9.440 & 10.465 & 50.000 & 19.173 & 1.08\\
\hline 
\multicolumn{24}{l}{Model specification: QR--False, PM--False}\\ 
\hline 
\hspace{2mm} GPPP & 38.279 & 38.545 & 0.000 & 31.845 & 1.813 & & 26.986 & 27.438 & 3.704 & 35.974 & 1.863 & & 19.504 & 19.977 & 22.222 & 31.709 & 1.886 & & 22.778 & 23.046 & 0.463 & 27.540 & 2.019\\
\hspace{2mm} LWP & 38.279 & 38.545 & 0.000 & 31.845 & 1.813 & & 26.986 & 27.438 & 3.704 & 35.974 & 1.863 & & 19.504 & 19.977 & 22.222 & 31.709 & 1.886 & & 22.778 & 23.046 & 0.463 & 27.540 & 2.019\\
\hspace{2mm} AIPW & 38.013 & 38.302 & 0.000 & 18.348 & 0.995 & & 25.475 & 26.004 & 0.000 & 20.612 & 1.005 & & 18.409 & 18.940 & 2.315 & 17.500 & 1.001 & & 22.053 & 22.284 & 0.000 & 14.343 & 1.141\\
\bottomrule
\end{tabular}}
 \begin{tablenotes}
 \footnotesize
 \item UW: Unweighted; FW: Fully weighted; GPPP: Gaussian Process of Propensity Prediction; LWP: Linear-in-weight Prediction; AIPW: Augmented Inverse Propensity Weighting.\\
 NOTE: The PAPP and AIPW methods have been implemented through a bootstrap method. 
 \end{tablenotes}
\end{threeparttable}
\end{sidewaystable}

\begin{sidewaystable}[hbt!]
\centering
\caption{Comparing the performance of the bias adjustment methods in the second simulation study for the \emph{binary} outcome with $(n_A, n_R)=(500, 500)$ and $\gamma_1=0.3$}\label{tab:4.12}
\begin{threeparttable}
\scriptsize{\begin{tabular}{l l l l l l l l l l l l l l l l l l l l l l l l l}
\toprule
& \multicolumn{5}{c}{\textbf{$LIN$}} & & \multicolumn{5}{c}{\textbf{$CUB$}} & & \multicolumn{5}{c}{\textbf{$EXP$}} & & \multicolumn{5}{c}{\textbf{$SIN$}}\\\cline{2-6}\cline{8-12}\cline{14-18}\cline{20-24}
\textbf{Measure} & rBias & rMSE & crCI & lCI & rSE & & rBias & rMSE & crCI & lCI & rSE & & rBias & rMSE & crCI & lCI & rSE & & rBias & rMSE & crCI & lCI & rSE \\
\midrule
\multicolumn{24}{l}{\textbf{Probability sample ($S_R$)}}\\ 
\hline 
\hspace{2mm} UW & -24.046 & 24.606 & 0.926 & 20.785 & 1.014 & & -23.445 & 24.144 & 0.926 & 22.713 & 1.003 & & -19.875 & 20.549 & 4.167 & 20.818 & 1.015 & & -11.430 & 12.367 & 35.648 & 19.157 & 1.032\\
\hspace{2mm} FW & 0.329 & 6.211 & 95.833 & 25.283 & 1.038 & & 0.153 & 7.020 & 97.222 & 28.280 & 1.026 & & -0.219 & 6.035 & 96.759 & 24.676 & 1.041 & & -0.014 & 4.947 & 96.759 & 21.354 & 1.099\\
\hline 
\multicolumn{24}{l}{\textbf{Non-probability sample ($S_A$)}}\\ 
\hline 
\hspace{2mm} UW & 49.180 & 49.553 & 0.000 & 22.648 & 0.950 & & 37.359 & 38.019 & 0.000 & 25.871 & 0.934 & & 27.292 & 27.979 & 0.000 & 22.450 & 0.928 & & 19.208 & 19.788 & 1.852 & 19.537 & 1.046\\
\hspace{2mm} FW & -0.043 & 6.460 & 93.056 & 24.387 & 0.961 & & 0.091 & 6.891 & 94.444 & 27.157 & 1.003 & & 0.532 & 6.855 & 94.907 & 25.353 & 0.944 & & -0.300 & 5.720 & 94.444 & 22.909 & 1.021\\
\hline 
\multicolumn{24}{l}{\textbf{Non-robust method}}\\ 
\hline 
\multicolumn{24}{l}{Model specification: QR--True}\\ 
\hline 
\hspace{2mm} PAPP & 0.448 & 6.024 & 93.981 & 24.516 & 1.039 & & 1.301 & 12.124 & 93.981 & 44.579 & 0.941 & & 4.807 & 8.095 & 88.426 & 25.182 & 0.984 & & 6.097 & 7.752 & 78.241 & 20.696 & 1.1\\
\hline 
\multicolumn{24}{l}{Model specification: QR--False}\\ 
\hline 
\hspace{2mm} PAPP & 38.110 & 38.636 & 0.000 & 24.026 & 0.963 & & 24.997 & 25.965 & 6.944 & 26.728 & 0.968 & & 18.257 & 19.247 & 15.741 & 23.253 & 0.971 & & 22.514 & 23.073 & 0.000 & 20.041 & 1.01\\
\hline 
\multicolumn{24}{l}{\textbf{Doubly robust methods}}\\ 
\hline 
\multicolumn{24}{l}{Model specification: QR--True, PM--True}\\ 
\hline 
\hspace{2mm} GPPP & 0.041 & 5.778 & 98.148 & 29.040 & 1.290 & & 0.907 & 7.182 & 98.611 & 37.212 & 1.342 & & 1.092 & 6.455 & 99.537 & 34.268 & 1.382 & & 0.340 & 5.187 & 95.833 & 21.357 & 1.06\\
\hspace{2mm} LWP & 0.119 & 5.848 & 97.685 & 29.079 & 1.279 & & 0.923 & 7.407 & 98.148 & 37.189 & 1.304 & & 1.167 & 6.632 & 100.000 & 34.385 & 1.356 & & 0.372 & 5.302 & 97.685 & 21.684 & 1.054\\
\hspace{2mm} AIPW & 0.329 & 5.941 & 94.444 & 25.256 & 1.084 & & 0.566 & 7.494 & 93.056 & 28.540 & 0.972 & & 0.795 & 6.444 & 94.444 & 25.518 & 1.016 & & -0.091 & 5.004 & 95.833 & 21.217 & 1.079\\
\hline 
\multicolumn{24}{l}{Model specification: QR--True, PM--False}\\ 
\hline 
\hspace{2mm} GPPP & 0.133 & 5.778 & 98.611 & 29.108 & 1.294 & & 1.387 & 7.130 & 99.537 & 36.942 & 1.358 & & 1.909 & 6.569 & 99.537 & 34.123 & 1.395 & & 0.740 & 6.331 & 95.370 & 20.815 & 0.85\\
\hspace{2mm} LWP & 0.255 & 5.782 & 98.148 & 28.810 & 1.285 & & 1.365 & 7.102 & 98.611 & 36.637 & 1.356 & & 1.755 & 6.472 & 100.000 & 33.880 & 1.396 & & 0.512 & 4.994 & 94.907 & 20.754 & 1.075\\
\hspace{2mm} AIPW & 0.509 & 5.932 & 96.296 & 25.084 & 1.080 & & 0.439 & 7.218 & 93.056 & 27.731 & 0.980 & & 0.910 & 6.627 & 95.370 & 25.364 & 0.983 & & -0.118 & 5.034 & 95.370 & 21.062 & 1.065\\
\hline 
\multicolumn{24}{l}{Model specification: QR--False, PM--True}\\ 
\hline 
\hspace{2mm} GPPP & 1.066 & 5.654 & 99.537 & 30.395 & 1.408 & & 2.959 & 7.837 & 99.537 & 38.926 & 1.378 & & 2.388 & 6.928 & 100.000 & 35.583 & 1.406 & & 0.495 & 5.498 & 95.370 & 22.425 & 1.057\\
\hspace{2mm} LWP & 0.551 & 5.573 & 99.537 & 30.016 & 1.392 & & 1.474 & 7.532 & 99.537 & 38.708 & 1.346 & & 1.462 & 6.800 & 100.000 & 35.656 & 1.377 & & 0.175 & 6.160 & 98.611 & 29.437 & 1.226\\
\hspace{2mm} AIPW & 0.939 & 7.448 & 97.222 & 30.643 & 1.056 & & -6.206 & 26.540 & 92.593 & 84.822 & 0.837 & & 10.963 & 12.774 & 60.648 & 26.141 & 1.015 & & 9.697 & 10.937 & 57.407 & 21.505 & 1.082\\
\hline 
\multicolumn{24}{l}{Model specification: QR--False, PM--False}\\ 
\hline 
\hspace{2mm} GPPP & 38.601 & 39.056 & 0.463 & 35.225 & 1.524 & & 26.372 & 27.263 & 16.667 & 39.480 & 1.465 & & 19.248 & 20.128 & 36.574 & 35.137 & 1.535 & & 22.752 & 23.322 & 8.333 & 30.490 & 1.528\\
\hspace{2mm} LWP & 38.601 & 39.056 & 0.463 & 35.225 & 1.524 & & 26.372 & 27.263 & 16.667 & 39.480 & 1.465 & & 19.248 & 20.128 & 36.574 & 35.137 & 1.535 & & 22.752 & 23.322 & 8.333 & 30.490 & 1.528\\
\hspace{2mm} AIPW & 38.532 & 39.009 & 0.000 & 24.171 & 1.011 & & 25.585 & 26.540 & 4.630 & 26.995 & 0.974 & & 18.697 & 19.645 & 11.111 & 23.391 & 0.987 & & 22.342 & 22.938 & 0.926 & 20.299 & 0.995\\
\bottomrule
\end{tabular}}
 \begin{tablenotes}
 \footnotesize
 \item UW: Unweighted; FW: Fully weighted; GPPP: Gaussian Process of Propensity Prediction; LWP: Linear-in-weight Prediction; AIPW: Augmented Inverse Propensity Weighting.\\
 NOTE: The PAPP and AIPW methods have been implemented through a bootstrap method. 
 \end{tablenotes}
\end{threeparttable}
\end{sidewaystable}

\begin{sidewaystable}[hbt!]
\centering
\caption{Comparing the performance of the bias adjustment methods in the second simulation study for the \emph{continuous} outcome with $(n_A, n_R)=(500, 1,000)$ and $\gamma_1=0.6$}\label{tab:6.13}
\begin{threeparttable}
\scriptsize{\begin{tabular}{l l l l l l l l l l l l l l l l l l l l l l l l l}
\toprule
& \multicolumn{5}{c}{\textbf{$LIN$}} & & \multicolumn{5}{c}{\textbf{$CUB$}} & & \multicolumn{5}{c}{\textbf{$EXP$}} & & \multicolumn{5}{c}{\textbf{$SIN$}}\\\cline{2-6}\cline{8-12}\cline{14-18}\cline{20-24}
\textbf{Measure} & rBias & rMSE & crCI & lCI & rSE & & rBias & rMSE & crCI & lCI & rSE & & rBias & rMSE & crCI & lCI & rSE & & rBias & rMSE & crCI & lCI & rSE \\
\midrule
\multicolumn{24}{l}{\textbf{Probability sample ($S_R$)}}\\ 
\hline 
\hspace{2mm} UW & -24.160 & 24.480 & 0.000 & 14.690 & 0.948 & & -23.859 & 24.179 & 0.000 & 16.035 & 1.040 & & -20.146 & 20.452 & 0.000 & 14.709 & 1.061 & & -11.189 & 11.695 & 10.185 & 13.552 & 1.013\\
\hspace{2mm} FW & -0.042 & 4.761 & 94.907 & 17.832 & 0.953 & & -0.400 & 4.855 & 96.759 & 19.943 & 1.049 & & -0.706 & 4.262 & 94.907 & 17.407 & 1.054 & & 0.169 & 3.866 & 94.444 & 15.066 & 0.993\\
\hline 
\multicolumn{24}{l}{\textbf{Non-probability sample ($S_A$)}}\\ 
\hline 
\hspace{2mm} UW & 93.235 & 93.380 & 0.000 & 20.067 & 0.982 & & 79.514 & 79.811 & 0.000 & 25.341 & 0.937 & & 58.010 & 58.307 & 0.000 & 21.829 & 0.945 & & 22.488 & 22.962 & 0.000 & 19.491 & 1.069\\
\hspace{2mm} FW & -0.792 & 10.053 & 91.667 & 37.625 & 0.956 & & -1.068 & 10.904 & 93.981 & 41.622 & 0.976 & & -0.125 & 11.375 & 89.352 & 40.607 & 0.909 & & -0.444 & 9.445 & 94.444 & 37.557 & 1.013\\
\hline 
\multicolumn{24}{l}{\textbf{Non-robust method}}\\ 
\hline 
\multicolumn{24}{l}{Model specification: QR--True}\\ 
\hline 
\hspace{2mm} PAPP & -0.227 & 8.039 & 95.833 & 33.477 & 1.060 & & 13.742 & 22.143 & 56.019 & 50.263 & 0.737 & & 12.275 & 14.469 & 66.667 & 29.468 & 0.979 & & 14.626 & 15.234 & 6.944 & 17.300 & 1.033\\
\hline 
\multicolumn{24}{l}{Model specification: QR--False}\\ 
\hline 
\hspace{2mm} PAPP & 59.708 & 60.092 & 0.000 & 26.975 & 1.012 & & 37.229 & 38.028 & 0.000 & 29.931 & 0.982 & & 27.691 & 28.525 & 1.852 & 26.075 & 0.969 & & 41.013 & 41.367 & 0.000 & 21.652 & 1.02\\
\hline 
\multicolumn{24}{l}{\textbf{Doubly robust methods}}\\ 
\hline 
\multicolumn{24}{l}{Model specification: QR--True, PM--True}\\ 
\hline 
\hspace{2mm} GPPP & -2.388 & 6.355 & 96.296 & 27.586 & 1.201 & & -2.760 & 8.776 & 96.296 & 37.554 & 1.155 & & -1.181 & 8.926 & 93.056 & 37.261 & 1.087 & & -0.179 & 5.066 & 95.370 & 21.096 & 1.069\\
\hspace{2mm} LWP & -1.908 & 8.425 & 94.907 & 30.108 & 0.941 & & -2.168 & 10.318 & 93.981 & 39.769 & 1.011 & & 0.601 & 12.671 & 89.815 & 39.277 & 0.803 & & -0.367 & 7.087 & 91.667 & 23.998 & 0.879\\
\hspace{2mm} AIPW & -0.429 & 6.143 & 95.833 & 25.367 & 1.054 & & -0.311 & 7.571 & 95.370 & 29.165 & 0.981 & & 0.596 & 7.700 & 92.130 & 29.438 & 0.976 & & 0.211 & 4.192 & 94.444 & 17.162 & 1.043\\
\hline 
\multicolumn{24}{l}{Model specification: QR--True, PM--False}\\ 
\hline 
\hspace{2mm} GPPP & -1.367 & 9.372 & 96.759 & 28.818 & 0.802 & & 0.141 & 7.570 & 98.611 & 36.271 & 1.230 & & 2.018 & 7.965 & 96.759 & 35.445 & 1.183 & & 0.338 & 4.147 & 93.981 & 17.033 & 1.061\\
\hspace{2mm} LWP & -12.060 & 133.689 & 95.833 & 47.857 & 0.098 & & 0.028 & 14.769 & 98.148 & 33.546 & 0.584 & & --- & --- & 96.759 & --- & 0.024 & & 0.526 & 4.355 & 93.519 & 17.391 & 1.033\\
\hspace{2mm} AIPW & 0.315 & 6.036 & 95.370 & 25.426 & 1.074 & & 0.007 & 7.330 & 94.444 & 29.316 & 1.018 & & 0.783 & 7.440 & 94.444 & 28.708 & 0.988 & & 0.051 & 4.317 & 93.981 & 17.622 & 1.039\\
\hline 
\multicolumn{24}{l}{Model specification: QR--False, PM--True}\\ 
\hline 
\hspace{2mm} GPPP & -0.218 & 6.315 & 97.685 & 31.009 & 1.257 & & 3.623 & 10.786 & 96.296 & 43.539 & 1.105 & & 2.203 & 10.529 & 93.056 & 39.155 & 0.986 & & -0.932 & 6.565 & 92.593 & 23.803 & 0.941\\
\hspace{2mm} LWP & -2.276 & 6.534 & 96.759 & 27.687 & 1.160 & & -0.705 & 10.418 & 93.519 & 41.871 & 1.032 & & 0.000 & 11.326 & 90.278 & 38.376 & 0.871 & & -0.156 & 10.385 & 83.796 & 29.606 & 0.731\\
\hspace{2mm} AIPW & 0.857 & 12.269 & 94.907 & 50.214 & 1.044 & & 8.524 & 37.241 & 53.241 & 81.350 & 0.571 & & 19.411 & 20.900 & 26.389 & 28.560 & 0.938 & & 26.569 & 27.089 & 0.000 & 21.309 & 1.027\\
\hline 
\multicolumn{24}{l}{Model specification: QR--False, PM--False}\\ 
\hline 
\hspace{2mm} GPPP & 53.302 & 53.677 & 0.000 & 31.071 & 1.263 & & 32.374 & 33.276 & 2.778 & 34.849 & 1.163 & & 24.967 & 25.879 & 7.407 & 31.078 & 1.171 & & 40.756 & 41.104 & 0.000 & 25.637 & 1.24\\
\hspace{2mm} LWP & 53.302 & 53.677 & 0.000 & 31.071 & 1.263 & & 32.374 & 33.276 & 2.778 & 34.849 & 1.163 & & 24.967 & 25.879 & 7.407 & 31.078 & 1.171 & & 40.756 & 41.104 & 0.000 & 25.637 & 1.24\\
\hspace{2mm} AIPW & 55.682 & 56.049 & 0.000 & 26.339 & 1.047 & & 33.889 & 34.796 & 0.463 & 29.653 & 0.956 & & 25.863 & 26.799 & 3.704 & 26.079 & 0.945 & & 41.401 & 41.769 & 0.000 & 22.047 & 1.015\\
\bottomrule
\end{tabular}}
 \begin{tablenotes}
 \footnotesize
 \item UW: Unweighted; FW: Fully weighted; GPPP: Gaussian Process of Propensity Prediction; LWP: Linear-in-weight Prediction; AIPW: Augmented Inverse Propensity Weighting.\\
 NOTE: The PAPP and AIPW methods have been implemented through a bootstrap method. 
 \end{tablenotes}
\end{threeparttable}
\end{sidewaystable}

\begin{sidewaystable}
\centering
\caption{Comparing the performance of the bias adjustment methods in the second simulation study for the \emph{binary} outcome with $(n_A, n_R)=(1,000, 500)$ and $\gamma_1=0.6$}\label{tab:6.14}
\begin{threeparttable}
\scriptsize{\begin{tabular}{l l l l l l l l l l l l l l l l l l l l l l l l l}
\toprule
& \multicolumn{5}{c}{\textbf{$LIN$}} & & \multicolumn{5}{c}{\textbf{$CUB$}} & & \multicolumn{5}{c}{\textbf{$EXP$}} & & \multicolumn{5}{c}{\textbf{$SIN$}}\\\cline{2-6}\cline{8-12}\cline{14-18}\cline{20-24}
\textbf{Measure} & rBias & rMSE & crCI & lCI & rSE & & rBias & rMSE & crCI & lCI & rSE & & rBias & rMSE & crCI & lCI & rSE & & rBias & rMSE & crCI & lCI & rSE \\
\midrule
\multicolumn{24}{l}{\textbf{Probability sample ($S_R$)}}\\ 
\hline 
\hspace{2mm} UW & -24.046 & 24.606 & 0.926 & 20.785 & 1.014 & & -23.445 & 24.144 & 0.926 & 22.713 & 1.003 & & -19.875 & 20.549 & 4.167 & 20.818 & 1.015 & & -11.430 & 12.367 & 35.648 & 19.157 & 1.032\\
\hspace{2mm} FW & 0.329 & 6.211 & 95.833 & 25.283 & 1.038 & & 0.153 & 7.020 & 97.222 & 28.280 & 1.026 & & -0.219 & 6.035 & 96.759 & 24.676 & 1.041 & & -0.014 & 4.947 & 96.759 & 21.354 & 1.099\\
\hline 
\multicolumn{24}{l}{\textbf{Non-probability sample ($S_A$)}}\\ 
\hline 
\hspace{2mm} UW & 92.334 & 92.413 & 0.000 & 14.252 & 0.948 & & 78.359 & 78.506 & 0.000 & 17.952 & 0.951 & & 56.892 & 57.036 & 0.000 & 15.470 & 0.973 & & 22.907 & 23.094 & 0.000 & 13.777 & 1.195\\
\hspace{2mm} FW & -0.489 & 7.243 & 92.593 & 26.827 & 0.945 & & 0.076 & 7.212 & 93.519 & 29.956 & 1.057 & & 0.340 & 7.973 & 93.519 & 29.311 & 0.936 & & -1.071 & 6.592 & 96.296 & 26.831 & 1.05\\
\hline 
\multicolumn{24}{l}{\textbf{Non-robust method}}\\ 
\hline 
\multicolumn{24}{l}{Model specification: QR--True}\\ 
\hline 
\hspace{2mm} PAPP & 0.544 & 6.512 & 94.907 & 26.660 & 1.046 & & -0.222 & 21.896 & 81.019 & 62.841 & 0.731 & & 6.897 & 9.340 & 76.852 & 23.256 & 0.940 & & 14.493 & 15.104 & 11.574 & 18.440 & 1.104\\
\hline 
\multicolumn{24}{l}{Model specification: QR--False}\\ 
\hline 
\hspace{2mm} PAPP & 58.629 & 58.883 & 0.000 & 20.647 & 0.962 & & 36.297 & 36.745 & 0.000 & 23.009 & 1.025 & & 26.921 & 27.371 & 0.000 & 19.501 & 1.005 & & 41.494 & 41.660 & 0.000 & 15.831 & 1.084\\
\hline 
\multicolumn{24}{l}{\textbf{Doubly robust methods}}\\ 
\hline 
\multicolumn{24}{l}{Model specification: QR--True, PM--True}\\ 
\hline 
\hspace{2mm} GPPP & -3.903 & 6.864 & 95.833 & 28.565 & 1.295 & & -2.421 & 7.354 & 98.611 & 37.021 & 1.372 & & -0.993 & 7.089 & 99.537 & 35.230 & 1.293 & & -0.285 & 5.107 & 96.759 & 21.765 & 1.097\\
\hspace{2mm} LWP & -3.647 & 6.913 & 96.296 & 29.888 & 1.304 & & -2.105 & 7.801 & 97.222 & 38.341 & 1.312 & & -0.166 & 8.573 & 97.685 & 36.597 & 1.097 & & 0.242 & 8.279 & 93.056 & 23.532 & 0.734\\
\hspace{2mm} AIPW & -0.017 & 5.864 & 94.444 & 24.464 & 1.062 & & 0.362 & 6.715 & 95.370 & 26.837 & 1.019 & & 0.738 & 6.379 & 96.759 & 24.283 & 0.975 & & 0.158 & 4.599 & 96.759 & 19.998 & 1.107\\
\hline 
\multicolumn{24}{l}{Model specification: QR--True, PM--False}\\ 
\hline 
\hspace{2mm} GPPP & -3.729 & 6.809 & 96.759 & 28.835 & 1.297 & & -0.979 & 6.280 & 99.074 & 36.689 & 1.521 & & 0.846 & 6.314 & 99.537 & 34.586 & 1.425 & & 0.275 & 4.658 & 96.296 & 19.899 & 1.1\\
\hspace{2mm} LWP & -4.364 & 7.228 & 94.444 & 28.283 & 1.261 & & -1.685 & 6.469 & 99.074 & 35.185 & 1.447 & & -0.398 & 5.941 & 99.074 & 32.773 & 1.415 & & -128.960 & 1104.353 & 92.593 & 307.225 & 0.071\\
\hspace{2mm} AIPW & 0.392 & 5.823 & 94.907 & 24.088 & 1.055 & & 0.428 & 6.352 & 95.370 & 25.181 & 1.011 & & 0.772 & 6.207 & 96.296 & 23.403 & 0.967 & & 0.005 & 4.710 & 95.833 & 20.270 & 1.095\\
\hline 
\multicolumn{24}{l}{Model specification: QR--False, PM--True}\\ 
\hline 
\hspace{2mm} GPPP & -1.790 & 5.718 & 100.000 & 30.256 & 1.428 & & 3.429 & 8.509 & 98.611 & 40.538 & 1.339 & & 1.870 & 7.991 & 99.074 & 36.386 & 1.207 & & -0.683 & 5.880 & 94.444 & 23.425 & 1.029\\
\hspace{2mm} LWP & -2.976 & 6.166 & 99.074 & 29.028 & 1.380 & & 0.070 & 7.947 & 96.296 & 39.401 & 1.271 & & --- & --- & 98.148 & --- & 0.074 & & 0.056 & 9.039 & 89.352 & 28.784 & 0.816\\
\hspace{2mm} AIPW & 1.373 & 9.986 & 94.444 & 39.167 & 1.008 & & -13.701 & 54.515 & 73.611 & 128.401 & 0.619 & & 17.160 & 18.219 & 19.444 & 23.805 & 0.989 & & 26.793 & 27.156 & 0.000 & 19.895 & 1.145\\
\hline 
\multicolumn{24}{l}{Model specification: QR--False, PM--False}\\ 
\hline 
\hspace{2mm} GPPP & 51.913 & 52.162 & 0.000 & 32.167 & 1.621 & & 32.802 & 33.285 & 0.926 & 36.718 & 1.668 & & 25.110 & 25.585 & 3.241 & 32.362 & 1.697 & & 40.375 & 40.592 & 0.000 & 26.482 & 1.623\\
\hspace{2mm} LWP & 51.913 & 52.162 & 0.000 & 32.167 & 1.621 & & 32.802 & 33.285 & 0.926 & 36.718 & 1.668 & & 25.110 & 25.585 & 3.241 & 32.362 & 1.697 & & 40.375 & 40.592 & 0.000 & 26.482 & 1.623\\
\hspace{2mm} AIPW & 55.890 & 56.136 & 0.000 & 20.741 & 1.006 & & 34.495 & 34.980 & 0.000 & 23.541 & 1.033 & & 25.807 & 26.291 & 0.000 & 20.174 & 1.023 & & 41.565 & 41.743 & 0.000 & 16.317 & 1.077\\
\bottomrule
\end{tabular}}
 \begin{tablenotes}
 \footnotesize
 \item UW: Unweighted; FW: Fully weighted; GPPP: Gaussian Process of Propensity Prediction; LWP: Linear-in-weight Prediction; AIPW: Augmented Inverse Propensity Weighting.\\
 NOTE: The PAPP and AIPW methods have been implemented through a bootstrap method. 
 \end{tablenotes}
\end{threeparttable}
\end{sidewaystable}

\begin{sidewaystable}[hbt!]
\centering
\caption{Comparing the performance of the bias adjustment methods in the second simulation study for the \emph{binary} outcome with $(n_A, n_R)=(500, 500)$ and $\gamma_1=0.6$}\label{tab:6.15}
\begin{threeparttable}
\scriptsize{\begin{tabular}{l l l l l l l l l l l l l l l l l l l l l l l l l}
\toprule
& \multicolumn{5}{c}{\textbf{$LIN$}} & & \multicolumn{5}{c}{\textbf{$CUB$}} & & \multicolumn{5}{c}{\textbf{$EXP$}} & & \multicolumn{5}{c}{\textbf{$SIN$}}\\\cline{2-6}\cline{8-12}\cline{14-18}\cline{20-24}
\textbf{Measure} & rBias & rMSE & crCI & lCI & rSE & & rBias & rMSE & crCI & lCI & rSE & & rBias & rMSE & crCI & lCI & rSE & & rBias & rMSE & crCI & lCI & rSE \\
\midrule
\multicolumn{24}{l}{\textbf{Probability sample ($S_R$)}}\\ 
\hline 
\hspace{2mm} UW & -24.046 & 24.606 & 0.926 & 20.785 & 1.014 & & -23.445 & 24.144 & 0.926 & 22.713 & 1.003 & & -19.875 & 20.549 & 4.167 & 20.818 & 1.015 & & -11.430 & 12.367 & 35.648 & 19.157 & 1.032\\
\hspace{2mm} FW & 0.329 & 6.211 & 95.833 & 25.283 & 1.038 & & 0.153 & 7.020 & 97.222 & 28.280 & 1.026 & & -0.219 & 6.035 & 96.759 & 24.676 & 1.041 & & -0.014 & 4.947 & 96.759 & 21.354 & 1.099\\
\hline 
\multicolumn{24}{l}{\textbf{Non-probability sample ($S_A$)}}\\ 
\hline 
\hspace{2mm} UW & 93.235 & 93.380 & 0.000 & 20.067 & 0.982 & & 79.514 & 79.811 & 0.000 & 25.341 & 0.937 & & 58.010 & 58.307 & 0.000 & 21.829 & 0.945 & & 22.488 & 22.962 & 0.000 & 19.491 & 1.069\\
\hspace{2mm} FW & -0.792 & 10.053 & 91.667 & 37.625 & 0.956 & & -1.068 & 10.904 & 93.981 & 41.622 & 0.976 & & -0.125 & 11.375 & 89.352 & 40.607 & 0.909 & & -0.444 & 9.445 & 94.444 & 37.557 & 1.013\\
\hline 
\multicolumn{24}{l}{\textbf{Non-robust method}}\\ 
\hline 
\multicolumn{24}{l}{Model specification: QR--True}\\ 
\hline 
\hspace{2mm} PAPP & 0.291 & 8.528 & 98.148 & 34.586 & 1.033 & & 8.356 & 21.288 & 76.389 & 57.366 & 0.746 & & 9.612 & 12.558 & 78.704 & 31.136 & 0.981 & & 14.580 & 15.439 & 22.222 & 20.750 & 1.04\\
\hline 
\multicolumn{24}{l}{Model specification: QR--False}\\ 
\hline 
\hspace{2mm} PAPP & 59.044 & 59.464 & 0.000 & 27.586 & 0.995 & & 36.477 & 37.351 & 0.000 & 30.541 & 0.968 & & 27.229 & 28.064 & 2.315 & 26.766 & 1.002 & & 41.224 & 41.593 & 0.000 & 21.907 & 1.008\\
\hline 
\multicolumn{24}{l}{\textbf{Doubly robust methods}}\\ 
\hline 
\multicolumn{24}{l}{Model specification: QR--True, PM--True}\\ 
\hline 
\hspace{2mm} GPPP & -1.904 & 6.915 & 97.222 & 32.515 & 1.253 & & -2.381 & 9.423 & 98.148 & 42.778 & 1.203 & & -0.828 & 9.124 & 94.907 & 41.719 & 1.180 & & -0.203 & 5.952 & 95.833 & 24.740 & 1.067\\
\hspace{2mm} LWP & -1.841 & 7.877 & 96.759 & 33.629 & 1.126 & & -1.526 & 10.156 & 96.759 & 45.327 & 1.157 & & 0.746 & 12.656 & 92.593 & 44.049 & 0.898 & & -0.293 & 7.747 & 91.204 & 27.358 & 0.91\\
\hspace{2mm} AIPW & -0.147 & 6.802 & 95.833 & 28.899 & 1.082 & & -0.122 & 8.356 & 93.056 & 32.175 & 0.980 & & 0.891 & 8.193 & 93.056 & 31.509 & 0.985 & & 0.128 & 5.051 & 96.296 & 21.700 & 1.094\\
\hline 
\multicolumn{24}{l}{Model specification: QR--True, PM--False}\\ 
\hline 
\hspace{2mm} GPPP & -1.465 & 6.874 & 98.148 & 33.459 & 1.279 & & 0.499 & 8.277 & 97.685 & 42.119 & 1.307 & & 2.102 & 7.892 & 99.537 & 39.760 & 1.346 & & 0.354 & 5.216 & 96.759 & 21.606 & 1.069\\
\hspace{2mm} LWP & -2.607 & 7.164 & 96.296 & 31.969 & 1.229 & & -0.384 & 8.034 & 98.611 & 39.763 & 1.274 & & 0.531 & 7.418 & 98.611 & 37.732 & 1.308 & & -11.862 & 162.050 & 95.370 & 46.573 & 0.076\\
\hspace{2mm} AIPW & 0.646 & 6.852 & 97.222 & 28.536 & 1.065 & & 0.208 & 8.127 & 94.444 & 31.529 & 0.988 & & 0.865 & 7.761 & 93.056 & 30.408 & 1.004 & & -0.006 & 5.236 & 96.296 & 21.870 & 1.063\\
\hline 
\multicolumn{24}{l}{Model specification: QR--False, PM--True}\\ 
\hline 
\hspace{2mm} GPPP & 0.154 & 6.819 & 99.537 & 35.644 & 1.338 & & 4.130 & 11.390 & 96.759 & 48.346 & 1.172 & & 2.647 & 10.628 & 97.222 & 43.869 & 1.096 & & -0.911 & 7.270 & 94.444 & 27.402 & 0.977\\
\hspace{2mm} LWP & -1.833 & 6.836 & 97.685 & 32.806 & 1.276 & & -0.516 & 10.511 & 98.148 & 46.784 & 1.143 & & 0.254 & 11.133 & 92.593 & 42.761 & 0.988 & & 1.071 & 13.095 & 85.648 & 33.602 & 0.661\\
\hspace{2mm} AIPW & 1.505 & 12.999 & 95.370 & 51.268 & 1.011 & & 2.173 & 41.495 & 65.741 & 101.955 & 0.626 & & 18.273 & 19.986 & 36.574 & 30.913 & 0.972 & & 26.480 & 27.090 & 0.463 & 23.873 & 1.063\\
\hline 
\multicolumn{24}{l}{Model specification: QR--False, PM--False}\\ 
\hline 
\hspace{2mm} GPPP & 53.595 & 54.003 & 0.000 & 36.316 & 1.413 & & 32.656 & 33.628 & 4.630 & 41.353 & 1.322 & & 25.278 & 26.197 & 13.426 & 36.607 & 1.366 & & 40.628 & 41.023 & 0.000 & 30.304 & 1.374\\
\hspace{2mm} LWP & 53.595 & 54.003 & 0.000 & 36.316 & 1.413 & & 32.656 & 33.628 & 4.630 & 41.353 & 1.322 & & 25.278 & 26.197 & 13.426 & 36.607 & 1.366 & & 40.628 & 41.023 & 0.000 & 30.304 & 1.374\\
\hspace{2mm} AIPW & 55.803 & 56.209 & 0.000 & 27.460 & 1.037 & & 34.159 & 35.144 & 1.389 & 30.859 & 0.951 & & 25.913 & 26.866 & 5.556 & 27.008 & 0.969 & & 41.485 & 41.875 & 0.000 & 22.470 & 1.003\\
\bottomrule
\end{tabular}}
 \begin{tablenotes}
 \footnotesize
 \item UW: Unweighted; FW: Fully weighted; GPPP: Gaussian Process of Propensity Prediction; LWP: Linear-in-weight Prediction; AIPW: Augmented Inverse Propensity Weighting.\\
 NOTE: The PAPP and AIPW methods have been implemented through a bootstrap method. 
 \end{tablenotes}
\end{threeparttable}
\end{sidewaystable}

\newpage
\clearpage

\subsection{Supplemental results on SHRP2/NHTS data application}\label{S:6.5}

\begin{table}[hbt!]
\centering\caption{Adjusted police-reportable crash rates per 100M miles and associated 95\% CIs by covariates}\label{tab:6.20}
\begin{threeparttable}
\scriptsize{\begin{tabular}{l l l l l l}
\toprule
\textbf{Covariate} & \textbf{n} & \textbf{Unweighted (95\% CI)} & \textbf{GPPP (95\% CI)} & \textbf{LWP (95\% CI)} & \textbf{AIPW (95\% CI)}\\
\midrule
\textbf{Total} & 2,862 & 1430.59 (1417.66,1443.52) & 461.29 (296,718.88) & 534.2 (270.47,1055.11) & 464.58 (294.06,734)\\
\hline
\textbf{Gender} &  &  &  &  & \\
\hspace{2mm} Male & 1,357 & 1778.61 (1740.38,1816.84) & 457.56 (293.54,713.22) & 543.75 (270.88,1091.48) & 464.59 (293.9,734.42)\\
\hspace{2mm} Female & 1,505 & 1116.79 (1107.96,1125.62) & 465.4 (298.14,726.49) & 524.11 (268.95,1021.35) & 464.45 (293.01,736.18)\\
\hline
\textbf{Age group} &  &  &  &  & \\
\hspace{2mm} 16-19 & 453 & 2621.22 (2565.47,2676.97) & 1532.82 (976.35,2406.45) & 1468.28 (836.11,2578.44) & 1535.73 (970.36,2430.52)\\
\hspace{2mm} 20-24 & 671 & 1357.97 (1334.13,1381.82) & 860.25 (533.43,1387.32) & 829.97 (383.85,1794.61) & 879.91 (534.14,1449.52)\\
\hspace{2mm} 25-29 & 254 & 1058.64 (1017.15,1100.13) & 788.27 (508.57,1221.8) & 854.11 (488.08,1494.64) & 794.45 (499.52,1263.49)\\
\hspace{2mm} 30-39 & 237 & 331.76 (313.61,349.9) & 261.38 (168.1,406.41) & 322.84 (160.73,648.47) & 265 (168.61,416.48)\\
\hspace{2mm} 40-49 & 214 & 290.26 (273.98,306.54) & 358.91 (225.32,571.7) & 444.7 (222.7,887.98) & 361.39 (220.82,591.43)\\
\hspace{2mm} 50-59 & 235 & 4324.69 (3815.74,4833.64) & 399.07 (256.67,620.47) & 509.73 (260.39,997.82) & 402.83 (253.23,640.81)\\
\hspace{2mm} 60-69 & 276 & 529.89 (509.18,550.6) & 561.92 (358.13,881.66) & 626.86 (298.56,1316.14) & 544.86 (343.45,864.39)\\
\hspace{2mm} 70-79 & 345 & 450.48 (433.74,467.23) & 406.44 (264.44,624.69) & 405.47 (210.49,781.06) & 417.89 (272.45,640.96)\\
\hspace{2mm} 80+ & 177 & 1514.88 (1430.84,1598.91) & 1238.85 (750.24,2045.68) & 1204.41 (645.22,2248.24) & 1248.12 (736.97,2113.81)\\
\hline
\textbf{Race} &  &  &  &  & \\
\hspace{2mm} White & 2,530 & 1461.22 (1445.75,1476.7) & 440.54 (281.63,689.11) & 502.06 (252.37,998.76) & 446.8 (282.12,707.59)\\
\hspace{2mm} Black & 150 & 910.16 (860.84,959.49) & 521.84 (334.41,814.31) & 683.96 (342.54,1365.71) & 511.11 (323.81,806.74)\\
\hspace{2mm} Asian & 96 & 2197.74 (2017.8,2377.68) & 521.55 (330.77,822.36) & 560.17 (311.8,1006.42) & 513.48 (313.78,840.28)\\
\hspace{2mm} Other & 86 & 580.72 (517.11,644.34) & 632.26 (420.17,951.43) & 810.56 (443.05,1482.92) & 634.01 (403.55,996.07)\\
\hline
\textbf{Gender} &  &  &  &  & \\
\hspace{2mm} Non-Hisp & 2,754 & 1442.75 (1429.07,1456.44) & 434.32 (277.78,679.07) & 490.06 (247.28,971.23) & 434.68 (274.25,688.97)\\
\hspace{2mm} Hispanic & 108 & 1120.45 (1053.82,1187.08) & 684.24 (472.5,990.85) & 1023.74 (563.45,1860.07) & 716.38 (471.67,1088.05)\\
\hline
\textbf{Ethnicity} &  &  &  &  & \\
\hspace{2mm} $<$High school & 213 & 3659.76 (3497.14,3822.38) & 1169.05 (730.81,1870.09) & 1329.65 (646.48,2734.8) & 1158.09 (703.44,1906.6)\\
HS comp & 279 & 1606.27 (1554.61,1657.93) & 478.5 (304.26,752.52) & 692.02 (356.97,1341.54) & 472.07 (295.39,754.42)\\
\hspace{2mm} College & 837 & 1248.89 (1231.23,1266.54) & 473.56 (304.61,736.21) & 561.22 (272.75,1154.81) & 483.62 (303.5,770.65)\\
\hspace{2mm} Graduate & 1,068 & 603.63 (597.39,609.87) & 370.9 (238.69,576.34) & 385.55 (198.1,750.37) & 374.99 (238.43,589.77)\\
\hspace{2mm} Post-grad & 465 & 2530.49 (2347.36,2713.63) & 475.47 (302.42,747.56) & 509.52 (261.3,993.53) & 473.87 (299.47,749.84)\\
\hline
\textbf{HH income} &  &  &  &  & \\
\hspace{2mm} 0-49 & 1,164 & 1179.27 (1167.12,1191.42) & 499.59 (315.31,791.56) & 594.45 (288.18,1226.2) & 497.9 (308.84,802.71)\\
\hspace{2mm} 150-99 & 1,049 & 709.89 (702.91,716.88) & 375.32 (243.18,579.25) & 421.01 (215.49,822.54) & 376.41 (242.88,583.35)\\
\hspace{2mm} 100-149 & 442 & 1658.96 (1605.49,1712.43) & 442.6 (286.32,684.19) & 506.46 (266.98,960.73) & 455.88 (289.48,717.9)\\
\hspace{2mm} 150+ & 207 & 6008.42 (5391.28,6625.55) & 676.53 (425.78,1074.96) & 824.7 (417.97,1627.23) & 685.64 (410.65,1144.77)\\
\hline
\textbf{HH size} &  &  &  &  & \\
\hspace{2mm} 1 & 598 & 1155.7 (1128.21,1183.18) & 432.75 (273.84,683.87) & 468.57 (231.47,948.5) & 439.9 (272.26,710.74)\\
\hspace{2mm} 2 & 967 & 698.78 (690.22,707.34) & 453.73 (293.66,701.06) & 516.25 (259.7,1026.23) & 442.13 (283.57,689.34)\\
\hspace{2mm} 3 & 510 & 1536.75 (1493.83,1579.67) & 463.86 (295.02,729.33) & 546.31 (281.53,1060.12) & 470.79 (293.94,754.04)\\
\hspace{2mm} 4 & 512 & 3045.14 (2885.96,3204.33) & 481.49 (301.8,768.17) & 561.53 (291.39,1082.1) & 486.84 (300.07,789.85)\\
\hspace{2mm} 5+ & 275 & 1398.78 (1353.14,1444.42) & 483.9 (314.29,745.04) & 626.53 (314.87,1246.67) & 512.19 (323.83,810.11)\\
\hline
\textbf{Vehicle make} &  &  &  &  & \\
\hspace{2mm} American & 1,045 & 2058.42 (2003.06,2113.78) & 407.92 (260.59,638.55) & 496.88 (246.37,1002.13) & 414.98 (260.96,659.91)\\
\hspace{2mm} Asian & 1,745 & 1034.41 (1025.7,1043.12) & 475.22 (306.94,735.75) & 521.23 (268.96,1010.09) & 478.85 (305.11,751.51)\\
\hspace{2mm} European & 72 & 1920.11 (1707.15,2133.08) & 726.7 (448.99,1176.18) & 963.13 (476.84,1945.38) & 690.15 (409.89,1162.04)\\
\hline
\textbf{Vehicle type} &  &  &  &  & \\
\hspace{2mm} Car & 2,061 & 1736.09 (1715.31,1756.86) & 611.19 (392.09,952.73) & 667.49 (342.48,1300.92) & 607.53 (381.97,966.28)\\
\hspace{2mm} Van & 109 & 629.14 (581.73,676.55) & 682.49 (439.71,1059.32) & 835.22 (446.81,1561.3) & 754.96 (480.49,1186.19)\\
\hspace{2mm} SUV & 551 & 724.36 (696.5,752.21) & 320.33 (203.8,503.48) & 376.21 (182.91,773.77) & 325.25 (208.23,508.02)\\
\hspace{2mm} Pickup & 141 & 344.53 (311.67,377.4) & 233.9 (151.03,362.22) & 336.55 (169.77,667.17) & 238.51 (148.96,381.9)\\
\hline
\textbf{Vehicle age} &  &  &  &  & \\
\hspace{2mm} 0-4 & 320 & 2821.77 (2738.31,2905.24) & 511.45 (319.47,818.8) & 631.43 (316.66,1259.1) & 536.52 (324.72,886.47)\\
\hspace{2mm} 5-9 & 742 & 838.31 (826.17,850.46) & 483.2 (313.51,744.74) & 555.03 (293.54,1049.46) & 478.57 (305.12,750.62)\\
\hspace{2mm} 10-14 & 905 & 977.22 (968.18,986.25) & 438.34 (281.64,682.23) & 489.77 (238.5,1005.77) & 442.87 (279.4,701.98)\\
\hspace{2mm} 15-19 & 382 & 607.65 (592.55,622.76) & 412.86 (264.47,644.5) & 480.04 (248.96,925.61) & 404.66 (256.76,637.75)\\
\hspace{2mm} 20-24 & 197 & 5119.19 (4543.74,5694.65) & 433.61 (279.84,671.86) & 478.1 (219.52,1041.29) & 436.13 (281.22,676.37)\\
\hspace{2mm} 25-29 & 108 & 545.61 (515.12,576.09) & 418.01 (257.83,677.71) & 469.24 (214.37,1027.12) & 429.71 (265.65,695.09)\\
\hspace{2mm} 30+ & 178 & 2030.45 (1897.07,2163.84) & 466.84 (277.69,784.84) & 580.56 (283.85,1187.42) & 486.75 (288.21,822.08)\\
\hline
\textbf{Fuel type} &  &  &  &  & \\
\hspace{2mm} Gas/Diesel & 2,641 & 1526.32 (1511.74,1540.9) & 461.79 (296.55,719.13) & 535.61 (270.59,1060.18) & 465.34 (294.5,735.28)\\
\hspace{2mm} Other & 221 & 286.58 (273.29,299.87) & 439.6 (267.26,723.09) & 476.92 (262.85,865.33) & 432.82 (262.46,713.76)\\
\bottomrule
\end{tabular}}
  \begin{tablenotes}
   \small
   \item
  \end{tablenotes}
 \end{threeparttable}
\end{table}

\end{document}